\documentclass{cambridge6A}
\usepackage{rotating}
\usepackage{floatpag}
\rotfloatpagestyle{empty}
\usepackage{amsmath}
\usepackage{amsthm}
\usepackage{graphicx}
\usepackage[margin=1.25in]{geometry}
\usepackage{epsfig}
\usepackage{amsmath}
\usepackage{amssymb}

\def\udots{\mathinner{\mkern1mu\raise1pt\hbox{.}\mkern2mu
    \raise4pt\hbox{.}\mkern2mu\raise7pt\vbox{\kern7pt\hbox{.}}\mkern1mu}}

\def\Toffoli#1#2#3{
  \newcount \n
  \newcount \y
  \n = 1
  \y = #2
  \loop
     \ifnum \n < #3 
        \put(#1,\y){\circle*{4}}
        \put(#1,\y){\line(0,-1){20}}
        \advance \n by 1
        \advance \y by -20  
  \repeat
  \put(#1,\y){\circle{8}}
  \put(#1,\y){\line(0,-1){4}}
}

\newcommand{\cale}{{\cal E}}

\newcommand{\calH}{{\cal H}}

\newcommand{\bfA}{\text{\boldmath $A$}}

\newcommand{\bfB}{\text{\boldmath $B$}}

\newcommand{\bfe}{\text{\boldmath $E$}}
\newcommand{\bfE}{\text{\boldmath $E$}}

\newcommand{\bfg}{\text{\boldmath $G$}}

\newcommand{\bfH}{\text{\boldmath $H$}}

\newcommand{\bfI}{\text{\boldmath $I$}}

\newcommand{\bfK}{\text{\boldmath $K$}}

\newcommand{\bfM}{\text{\boldmath $M$}}

\newcommand{\bfS}{\text{\boldmath $S$}}

\newcommand{\bfu}{\text{\boldmath $U$}}
\newcommand{\bfU}{\text{\boldmath $U$}}
\newcommand{\bfV}{\text{\boldmath $V$}}

\newcommand{\bfW}{\text{\boldmath $W$}}

\newcommand{\bfX}{\mbox{\boldmath $X$}}
\newcommand{\bfY}{\mbox{\boldmath $Y$}}
\newcommand{\bfZ}{\mbox{\boldmath $Z$}}

\newcommand{{\bfsigma}}{\text{\boldmath $\sigma$}}
\newcommand{{\bfSigma}}{\mbox{\boldmath $\Sigma$}}
\newcommand{{\bfDelta}}{\mbox{\boldmath $\Delta$}}
\newcommand{\bftau}{\text{\boldmath $\tau$}}
\newcommand{\bfomega}{\text{\boldmath $\omega$}}
\newcommand{\bfrho}{\text{\boldmath $\rho$}}
\newcommand{\bfpi}{\text{\boldmath $\Pi$}}

\newcommand{\be}{
\renewcommand{\theequation}{\thesection.\arabic{equation}}\begin{equation}}

\begin{document}

\title{Quantum Information\\ Chapter 10. Quantum Shannon Theory}

  \author{John Preskill\\ Institute for Quantum Information and Matter\\ California Institute of Technology \\[2\baselineskip] Updated June 2025\\[2\baselineskip] For further updates and additional chapters, see: https://preskill.caltech.edu/ph219/ \\[2\baselineskip] Please send corrections to preskill@caltech.edu}

\frontmatter
\maketitle
\setcounter{tocdepth}{2}
\tableofcontents

\chapter*{}

\noindent
This article forms one chapter of {\em Quantum Information} which will be first published by Cambridge University Press.

\vskip .2cm  

\noindent
\copyright \, in the Work, John Preskill, 2025

\vskip .2cm
\noindent
NB: The copy of the Work, as displayed on this website, is a draft, pre-publication copy only. The final, published version of the Work can be purchased through Cambridge University Press and other standard distribution channels. This draft copy is made available for personal use only and must not be sold or re-distributed. 

\chapter*{Preface}

This is the 10th and final chapter of my book {\em Quantum Information}, based on the course I have been teaching at Caltech since 1997. An early version of this chapter (originally Chapter 5) has been available on the course website since 1998, but this version is substantially revised and expanded.

The level of detail is uneven, as I've aimed to provide a gentle introduction, but I've also tried to avoid statements that are incorrect or obscure. Generally speaking, I chose to include topics that are both useful to know and relatively easy to explain; I had to leave out a lot of good stuff, but on the other hand the chapter is already quite long. 

My version of Quantum Shannon Theory is no substitute for the more careful treatment in Wilde's book \cite{wilde}, but it may be more suitable for beginners. This chapter contains occasional references to earlier chapters in my book, but I hope it will be intelligible when read independently of other chapters, including the chapter on quantum error-correcting codes. 

This is a working draft of Chapter 10, which I will continue to update. See the URL on the title page for further updates and drafts of other chapters. Please send an email to preskill@caltech.edu if you notice errors.

Eventually, the complete book will be published by Cambridge University Press. I hesitate to predict the publication date --- they have been far too patient with me.

\mainmatter

\setcounter{chapter}{9}

\chapter{Quantum Shannon Theory}

Quantum information science is a synthesis of three great themes of 20th century thought: quantum physics, computer science, and information theory. Up until now, we have given short shrift to the information theory side of this trio, an oversight now to be remedied.

A suitable name for this chapter might have been {\em Quantum Information Theory}, but I prefer for that term to have a broader meaning, encompassing much that has already been presented in this book. Instead I call it {\em Quantum Shannon Theory}, to emphasize that we will mostly be occupied with generalizing and applying Claude Shannon's great (classical) contributions to a quantum setting. Quantum Shannon theory has several major thrusts:

\begin{enumerate}
\item Compressing quantum information.

\item  Transmitting classical and quantum information through noisy quantum channels.

\item Quantifying, characterizing, transforming, and using quantum entanglement.

\end{enumerate}

\noindent A recurring theme unites these topics --- the properties, interpretation, and applications of von Neumann entropy.

My goal is to introduce some of the main ideas and tools of quantum Shannon theory, but there is a lot we won't cover. For example, 
we will mostly consider information theory in an {\em asymptotic setting}, where the same quantum channel or state is used arbitrarily many times, thus focusing on issues of principle rather than more practical questions about devising efficient protocols.

\section{Shannon for Dummies}

Before we can understand von Neumann entropy and its relevance to quantum
information, we should discuss Shannon entropy and its relevance to classical
information.

Claude Shannon established the two core results of classical information theory
in his landmark 1948 paper.  The two central problems that he solved were:

\begin{enumerate}
\item  How much can a message be {\it compressed}; {\it i.e.}, how
redundant is the information?  This question is answered by the ``source coding theorem,'' also called the ``noiseless coding theorem.''

\item  At what {\it rate} can we communicate reliably over a noisy
channel; {\it i.e.}, how much redundancy must be incorporated into a message to
protect against errors?  This question is answered by the ``noisy channel coding theorem.''
\end{enumerate}

\noindent Both questions concern {\it redundancy} -- how {\it unexpected} is
the next letter of the message, on the average.  One of Shannon's key insights
was that {\it entropy} provides a suitable way to quantify redundancy.

I call this section ``Shannon for Dummies'' because I will try to explain
Shannon's ideas quickly, minimizing distracting details.  That way, I can compress classical information
theory to about 13 pages.

\subsection{Shannon entropy and data compression}

A message is a string of letters, where each letter is chosen from an alphabet of $k$ possible letters. We'll consider an idealized setting in which the message is produced by an ``information source'' which picks each letter by sampling from a probability distribution
\begin{equation}
X := \{x,p(x)\};
\end{equation}
that is, the letter has the value
\begin{equation}
x \in \{0,1,2, \dots d{-}1\}
\end{equation}
with probability $p(x)$. If the source emits an $n$-letter message the particular string $x=x_1x_2 \dots x_n$ occurs with probability
\begin{equation}
p(x_1x_2 \dots x_n) = \prod_{i=1}^n p(x_i).
\end{equation}
Since the letters are statistically independent, and each is produced by consulting the same probability distribution $X$, we say that the letters are {\em independent and identically distributed}, abbreviated {\em i.i.d.} We'll use $X^n$ to denote the ensemble of $n$-letter messages in which each letter is generated independently by sampling from $X$, and $\vec x = (x_1x_2\dots x_n)$ to denote a string of letters. 

Now consider long $n$-letter messages, $n \gg 1$.  We ask: is it possible
to compress the message to a shorter string of letters that conveys essentially
the same information? The answer is: Yes, it's possible, unless the distribution X is uniformly random. 

If the alphabet is binary, then each letter is either $0$ with probability $1-p$ or $1$ with probability $p$,  where $0 \leq p \leq 1$.
For $n$  very large, the law of large numbers tells us that typical strings
will contain  about $n(1 - p)~0$'s and about $np~1$'s. The
number of distinct strings of this form is of order the binomial coefficient
$\binom{n}{np}$, and from the Stirling approximation $\log n!=n\log n
- n + O (\log n)$ we obtain
\begin{align}
&\log\binom{n}{np}=\log \left({n!\over (np)! \left(n(1-p)\right)!}\right)\notag\\
&\approx n\log n - n - \left(np\log np - np + n (1 - p) \log n (1 - p) - n (1-p)\right)\notag\\
&= n H(p),
\end{align}
where
\begin{equation}
H(p) = - p\log p - (1 - p) \log (1 - p)
\end{equation}
is the {\it entropy} function. 

In this derivation we used the Stirling approximation in the appropriate form for natural logarithms. But from now on we will prefer to use logarithms with base 2, which is more convenient for expressing a quantity of information in bits; thus if no base is indicated, it will be understood that the base is 2 unless otherwise stated. Adopting this convention in the expression for $H(p)$, the number of typical strings is of
order $2^{nH(p)}$.

To convey essentially all the information carried by a string of $n$ bits, it
suffices to choose a block code that assigns a nonnegative integer to each of the
typical strings.  This block code needs to distinguish about $2^{nH(p)}$ messages (all occurring
with nearly equal {\it a priori} probability), so we may specify any one of the
messages using a binary string with length only slightly longer than $nH(p)$.  Since $0 \leq H(p) \leq 1$
for $0 \leq p \leq 1$, and $H(p) = 1$ only for $p = {1\over 2}$, the block code
shortens the message for any $p\not= {1\over 2}$ (whenever $0$ and $1$ are not
equally probable).  This is Shannon's result.  The key idea is that we do not
need a codeword for every sequence of letters, only for the  {\it typical}
sequences.  The probability that the actual message is atypical becomes
negligible asymptotically, {\it i.e.}, in the limit $n \rightarrow \infty$.

Similar reasoning applies to the case where $X$ samples from a $d$-letter alphabet. In a string
of $n$ letters, $x$ typically occurs about $np(x)$ times, and the number of
typical strings is of order
\begin{equation}
\frac{n!}{\prod\nolimits_x (np(x))!} \simeq 2^{nH (X)},
\end{equation}
where we have again invoked the Stirling approximation and now
\begin{equation}
H(X) = -\sum_x  p(x) \log_2 p(x) =\mathbb{E}_X\left[ \log_2\frac{1}{p(x)}\right].
\end{equation}
is the {\it Shannon} entropy (or simply entropy) of the ensemble $X = \{x,
p(x)\}$, and $\mathbb{E}_X[\,\cdot\,]$ denotes the expectation value defined by this ensemble. Adopting a block code that assigns integers to the typical sequences,
the information in a string of $n$ letters can be compressed to about $nH(X)$ bits.
In this sense a letter $x$ chosen from the ensemble carries, on the average,
$H(X)$ bits of information.

It is useful to restate this reasoning more carefully using the {\em strong law of large numbers}, which asserts that a sample average for a random variable almost certainly converges to its expected value in the limit of many trials. 
If we sample from the distribution $Y=\{y,p(y)\}$ $n$ times, let $y_i, i\in \{1,2,\dots ,n\}$ denote the $i$th sample, and let 
\begin{equation}
\mu[Y] = \mathbb{E}_Y[y] = \sum_y y ~p(y)
\end{equation}
denote the expected value of $y$. Then for any positive $\varepsilon$ and $\delta$ there is a positive integer $N$ such that
\begin{equation}
\left|\frac{1}{n}\sum_{i=1}^n y_i - \mu[Y]\right| \le \delta
\end{equation}
with probability at least $1-\varepsilon$ for all $n \ge N$. We can apply this statement to the random variable $\log_2 p(x)$. Let us say that a sequence of $n$ letters is {\em $\delta$-typical} if 
\begin{equation}
H(X) - \delta \le  -\frac{1}{n}\log_2 p(x_1x_2\dots x_n) \le H(X) + \delta;
\end{equation}
then the strong law of large numbers says that for any $\varepsilon, \delta > 0$ and $n$ sufficiently large, an $n$-letter sequence will be $\delta$-typical with probability $\ge 1-\varepsilon$. 

Since each $\delta$-typical $n$-letter sequence $\vec x$ occurs with probability $p(\vec x)$ satisfying
\begin{equation}\label{typical_P}
p_{\min}=2^{-n(H + \delta)} \le p(\vec x) \le 2^{-n(H-\delta)}= p_{\max},
\end{equation}
we may infer upper and lower bounds on the {\em number} $N_{\rm typ}(\varepsilon, \delta, n)$ of typical sequences:
\begin{equation}
N_{\rm typ}~p_{\min}\le \sum_{{\rm typical}~x}p(x) \le 1, \quad N_{\rm typ}~p_{\max} \ge \sum_{{\rm typical}~x} p(x) \ge 1 -\varepsilon,
\end{equation}
implies
\begin{equation}
2^{n(H + \delta)} \geq N_{\rm typ} (\varepsilon, \delta, n) \geq (1 - \varepsilon)
2^{n(H-\delta)}.
\end{equation}
Therefore, we can encode all typical sequences using a block code with length $n(H+\delta)$ bits. That way, any message emitted by the source can be compressed and decoded successfully as long as the message is typical; the compression procedure achieves a success probability $p_{\rm success} \ge 1 - \varepsilon$, no matter how the atypical sequences are decoded. 

What if we try to compress the message even further, say to $H(X) - \delta'$ bits per letter, where $\delta'$ is a constant independent of the message length $n$? Then we'll run into trouble, because there won't be enough codewords to cover all the typical messages, and we won't be able to decode the compressed message with negligible probability of error. 
The probability $p_{\rm success}$ of successfully
decoding the message will be bounded above by
\begin{equation}
p_{\rm success} \leq 2^{n(H - \delta')} 2^{-n (H - \delta)} +\varepsilon=
2^{-n(\delta' - \delta)} + \varepsilon.
\end{equation}
We can correctly decode only $2^{n(H - \delta')}$ typical messages, each
occurring with probability no higher than $2^{-n(H - \delta)}$; we add $\varepsilon$, an upper bound on the probability of an atypical message, allowing optimistically for the possibility that we somehow manage to decode the atypical
messages correctly.  Since we may choose $\varepsilon$ and $\delta$ as small as we please, this
success probability becomes small as $n \rightarrow \infty$, if $\delta'$ is a positive constant.

The number of bits per letter encoding the compressed message is called the {\em rate} of the compression code, and we say a rate $R$ is {\em achievable} asymptotically (as $n \rightarrow \infty$) if there is a sequence of codes with rate at least $R$ and error probability approaching zero in the limit of large $n$. To summarize our conclusion, we have found that
\begin{align}
{\rm Compression ~Rate} &= H(X) + o(1) ~{\rm is} ~{ achievable},\notag\\
{\rm Compression ~Rate} &= H(X) - \Omega(1) ~{\rm is} ~{not ~ achievable},
\end{align}
where $o(1)$ denotes a positive quantity which may be chosen as small as we please, and $\Omega(1)$ denotes a positive constant.
This is Shannon's source coding theorem.

We have not discussed at all the details of the compression code. We might imagine a huge lookup table which assigns a unique codeword to each message and vice versa, but because such a table has size exponential in $n$ it is quite impractical for compressing and decompressing long messages. It is fascinating to study how to make the coding and decoding efficient while preserving a near optimal rate of compression, and quite important, too, if we really want to compress something. But this practical aspect of classical compression theory is beyond the scope of this book. 

\subsection{Joint typicality, conditional entropy, and mutual information}
\label{subsec:joint-typ}

The Shannon entropy quantifies my {\em ignorance} per letter about the output of an information source. If the source $X$ produces an $n$-letter message, then $n(H(X)+o(1))$ bits suffice to convey the content of the message, while $n(H(X) - \Omega(1))$ bits do not suffice.

Two information sources $X$ and $Y$ can be correlated. Letters drawn from the sources are governed by a joint distribution $XY = \{(x,y), p(x,y)\}$, in which a pair of letters $(x,y)$ appears with probability $p(x,y)$. The sources are independent if $p(x,y) = p(x)p(y)$, but correlated otherwise. If $XY$ is a joint distribution, we use $X$ to denote the marginal distribution, defined as
\begin{equation}
X = \left\{x, p(x) = \sum_y p(x,y)\right\},
\end{equation}
and similarly for $Y$. If $X$ and $Y$ are correlated, then by reading a message generated by $Y^n$ I reduce my ignorance about a message generated by $X^n$, which should make it possible to compress the output of $X$ further than if I did not have access to $Y$.

To make this idea more precise, we use the concept of {\em jointly typical sequences}. Sampling from the distribution $X^nY^n$, that is, sampling $n$ times from the joint distribution $XY$, produces a message $(\vec x,\vec y) = (x_1x_2 \dots x_n,y_1y_2\dots y_n)$ with probability
\begin{align}
p(\vec x,\vec y) = p(x_1,y_1)p(x_2,y_2)\dots p(x_n,y_n).
\end{align}
Let us say that $(\vec x,\vec y)$ drawn from $X^nY^n$ is {\em jointly $\delta$-typical} if 
\begin{align}
2^{-n(H(X) + \delta)} &\le p(\vec x) \le 2^{-n(H(X)-\delta)}, \notag \\
2^{-n(H(Y) + \delta)} &\le p(\vec y) \le 2^{-n(H(Y)-\delta)}, \notag\\
2^{-n(H(XY) + \delta)} &\le p(\vec x,\vec y) \le 2^{-n(H(XY)-\delta)}.
\label{eq:joint-typ-prob}
\end{align}
Then, applying the strong law of large numbers simultaneously to the three distributions $X^n$, $Y^n$, and $X^nY^n$, we infer that for $\varepsilon, \delta > 0$ and $n$ sufficiently large, a sequence drawn from $X^nY^n$ will be jointly $\delta$-typical with probability $\ge 1-\varepsilon$. Using Bayes' rule, we can then obtain upper and lower bounds on the {\em conditional} probability $p(\vec x|\vec y)$ for jointly typical sequences: 
\begin{align}
p(\vec x|\vec y) = \frac{p(\vec x,\vec y)}{p(\vec y)} \ge \frac{2^{-n\left(H(XY) +\delta\right)}}{2^{-n\left(H(Y) -\delta\right)}}= 2^{-n\left(H(X|Y) +2\delta\right)},\notag\\
p(\vec x|\vec y) = \frac{p(\vec x,\vec y)}{p(\vec y)} \le \frac{2^{-n\left(H(XY) -\delta\right)}}{2^{-n\left(H(Y) +\delta\right)}}= 2^{-n\left(H(X|Y) -2\delta\right)}.
\label{eq:joint-typ-conditional}
\end{align}
Here we have introduced the quantity
\begin{equation}
H(X|Y) = H(XY) - H(Y) = \mathbb{E}_{XY}[ -\log p (x,y)+ \log p (y)]= \mathbb{E}_{XY}[ - \log p (x|y)],
\end{equation}
which is called the {\em conditional entropy} of $X$ given $Y$.

The conditional entropy quantifies my {\em remaining} ignorance about $x$ once I know $y$. From eq.(\ref{eq:joint-typ-conditional}) we see that if $(\vec x,\vec y)$ is jointly typical (as is the case with high probability for $n$ large), then the number of possible values for $\vec x$ compatible with the known value of $\vec y$ is no more than $2^{n\left(H(X|Y) +2\delta\right)}$; hence we can convey $\vec x$ with a high success probability using only $H(X|Y) + o(1)$ bits per letter. On the other hand we can't do much better, because if we use only $2^{n\left(H(X|Y) - \delta'\right)}$ codewords, we are limited to conveying reliably no more than a fraction $2^{-n\left(\delta'-2\delta\right)}$ of all the jointly typical messages. To summarize, $H(X|Y)$ is the number of {\it additional} bits per
letter needed to specify {\it both} $\vec x$ and $\vec y$ once $\vec y$ is known.  Similarly, $H(Y|X)$ is the number of additional bits per letter needed to specify both $\vec x$ and $\vec y$ when $\vec x$ is known. 

The information about $X$ that I {\it gain} when I learn $Y$ is quantified by
how much the number of bits per letter needed to specify $X$ is {\it reduced}
when $Y$ is known.  This is
\begin{align}
I(X;Y) &\equiv H(X) - H(X|Y)\notag \\
&= H(X) + H(Y) - H(XY)\notag \\
&= H(Y) - H(Y|X),
\end{align}
which is called the {\em mutual information}. The mutual information $I(X;Y)$ quantifies how $X$ and $Y$ are correlated, and is symmetric under
interchange of $X$ and $Y$: I find out as much about $X$ by learning $Y$ as
about $Y$ by learning $X$.  Learning $Y$ never {\it reduces} my knowledge of
$X$, so $I(X;Y)$ is obviously nonnegative, and indeed the  inequality $H(X) \geq H(X|Y)
\geq 0$ follows easily from the concavity of the log function.

Of course, if $X$ and $Y$ are completely uncorrelated, we have $p(x,y) = p(x)
p(y)$, and
\begin{equation}
I(X;Y) \equiv \left\langle \log {p(x,y)\over p(x)p(y)} \right\rangle = 0;
\end{equation}
we don't find out anything about $X$ by learning $Y$ if there is no
correlation between $X$ and $Y$.

\subsection{Distributed source coding}
\label{subsec:distributed-source}

To sharpen our understanding of the operational meaning of conditional entropy, consider this situation: Suppose that the joint distribution $XY$ is sampled $n$ times, where Alice receives the $n$-letter message $\vec x$ and Bob receives the $n$-letter message $\vec y$. Now Alice is to send a message to Bob which will enable Bob to determine $\vec x$ with high success probability, and Alice wants to send as few bits to Bob as possible. This task is harder than in the scenario considered in \S\ref{subsec:joint-typ}, where we assumed that the encoder and the decoder share full knowledge of $\vec y$, and can choose their code for compressing $\vec x$ accordingly. It turns out, though, that even in this more challenging setting Alice can compress the message she sends to Bob down to $n\left(H(X|Y) + o(1)\right)$ bits, using a method called {\em Slepian-Wolf coding}.

Before receiving $(\vec x,\vec y)$, Alice and Bob agree to sort all the possible $n$-letter messages that Alice might receive into $2^{nR}$ possible bins of equal size, where the choice of bins is known to both Alice and Bob. When Alice receives $\vec x$, she sends $nR$ bits to Bob, identifying the bin that contains $\vec x$. After Bob receives this message, he knows both $\vec y$ and the bin containing $\vec x$. If there is a unique message in that bin which is jointly typical with $\vec y$, Bob decodes accordingly. Otherwise, he decodes arbitrarily. This procedure can fail either because $\vec x$ and $\vec y$ are not jointly typical, or because there is more than one message in the bin which is jointly typical with $\vec y$. Otherwise, Bob is sure to decode correctly. 

Since $\vec x$ and $\vec y$ are jointly typical with high probability, the compression scheme works if it is unlikely for a bin to contain an incorrect message which is jointly typical with $\vec y$. If $\vec y$ is typical, what can we say about the number $N_{{\rm typ}|\vec y}$ of messages $\vec x$ that are jointly typical with $\vec y$? Using eq.(\ref{eq:joint-typ-conditional}), we have 
\begin{equation}
1 \ge \sum_{{\rm typical}~\vec x|\vec y}p(\vec x|\vec y) \ge \quad N_{{\rm typ}|\vec y}~2^{-n\left(H(X|Y) +2\delta\right)},
\end{equation}
and thus
\begin{equation}
N_{{\rm typ}|\vec y} \le 2^{n\left(H(X|Y) +2\delta\right)}.
\end{equation}
Now, to estimate the probability of a decoding error, we need to specify how the bins are chosen. Let's assume the bins are chosen uniformly at random, or equivalently, let's consider averaging uniformly over all codes that divide the length-$n$ strings into $2^{nR}$ bins of equal size. Then the probability that a particular bin contains a message jointly typical with a specified $\vec y$ purely by accident is bounded above by
\begin{equation}
2^{-nR} N_{{\rm typ}|\vec y} \le  2^{-n\left(R- H(X|Y) -2\delta\right)}.
\end{equation}
We conclude that if Alice sends $R$ bits to Bob per each letter of the message $x$, where
\begin{equation}
R = H(X|Y) + o(1),
\end{equation}
then the probability of a decoding error vanishes in the limit $n\rightarrow \infty$, at least when we average uniformly  over all codes. Surely, then, there must exist a particular sequence of codes Alice and Bob can use to achieve the rate $R=H(X|Y) + o(1)$, as we wanted to show. 

In this scenario, Alice and Bob jointly know $(x,y)$, but initially neither Alice nor Bob has access to all their shared information. The goal is to merge all the information on Bob's side with minimal communication from Alice to Bob, and we have found that $H(X|Y)+o(1)$ bits of communication per letter suffice for this purpose. Similarly, the information can be merged on Alice's side using $H(Y|X)+o(1)$ bits of communication per letter from Bob to Alice. Note that this argument, based on averaging over codes, is nonconstructive; it establishes the existence of a code that achieves the stated communication rate without exhibiting any such code. 

\subsection{Relative entropy and hypothesis testing}
\label{subsec:relative-hypothesis}
Here we briefly introduce another useful entropic quantity, the \emph{relative entropy} (also called the Kullback-Leibler divergence), which arises in the context of hypothesis testing. Suppose we sample $n$ times from the distribution $X=\{x,p(x)\}$, where $n$ is very large, but we mistakenly believe that we are sampling instead from a different distribution $Y=\{x,q(x)\}$. With enough samples, we should be able to rule out this incorrect hypothesis with high probability.

To see how this works with minimal distracting details, suppose that $n$ is so large that the sequence drawn from the distribution is $\delta$-typical with probability $1-\varepsilon$ where $\delta$ and $\varepsilon$ are negligibly small. That is, assume that when we sample $n$ times from $X$, the letter $x$ is drawn exactly $np(x)$ times. Under our hypothesis, then, in which $x$ occurs with probability $q(x)$, the observed sequence $\vec x$ has probability
\begin{equation}
    P(\vec x) = \prod_{x=0}^{d-1} q(x)^{np(x)}\implies \log P(\vec x)= n\sum_{x=0}^{d-1} p(x)\log q(x)=-n\mathbb{E}_p[-\log q(x)];
\end{equation}
that is,
\begin{equation}
    P(\vec x) = 2^{-n\mathbb{E}_p[-\log q(x)]}.
\end{equation}
The quantity $\mathbb{E}_p[-\log q(x)]$ is called the \emph{cross entropy} of $q$ relative to $p$.

This expression $P(\vec x)$ gives the probability for any particular typical sequence $\vec x$, and the total number of typical sequences is $2^{nH(X)}$. Therefore, under our hypothesis, the total probability of the observed frequencies, in which the letter $x$ occurs $np(x)$ times, is
\begin{equation}
    \text{Prob} = 2^{nH(X)}P(\vec x) = 2^{-n D(p\|q)},
\end{equation}
where 
\begin{equation}
  D(p\|q) = \sum_x p(x)\left(\log p(x)-\log q(x)\right)  =\mathbb{E}_p\left[\log\frac{p(x)}{q(x)}\right],
\end{equation}
which is called the \emph{relative entropy of $p$ with respect to $q$.}

We see that the relative entropy governs the confidence with which we can rule out the incorrect hypothesis when $n$ is large; in this sense it is a measure of the distinguishability of the two distributions $p(x)$ and $q(x)$. It is not, however, a metric on probability vectors, because it is asymmetric between $p$ and $q$. This asymmetry occurs because $p$ and $q$ have different roles in the hypothesis testing scenario: $p$ is the true distribution and $q$ is the hypothetical one. 

Note that since $\text{Prob}= 2^{-n D(p\|q)}$ is a probability and therefore no larger than 1, it follows that the relative entropy is nonnegative,
\begin{equation}
    D(p\|q)\ge 0.
\end{equation}

\subsection{The noisy channel coding theorem}
\label{subsec:shannon-noisy}

Suppose Alice wants to send a message to Bob, but the communication channel linking Alice and Bob is noisy. Each time they use the channel, Bob receives the letter $y$ with probability $p(y|x)$ if Alice sends the letter $x$. Using the channel $n \gg 1$ times, Alice hopes to transmit a long message to Bob.

Alice and Bob realize that to communicate reliably despite the noise they should use some kind of code. For example, Alice might try sending the same bit $k$ times, with Bob using a majority vote of the $k$ noisy bits he receives to decode what Alice sent. One wonders: for a given channel, is it possible to ensure perfect transmission asymptotically, {\em i.e.}, in the limit where the number of channel uses $n\to \infty$? And what can be said about the {\em rate} of the code; that is, how many bits must be sent per letter of the transmitted message?

Shannon answered these questions. He showed that {\em any} channel can be used for perfectly reliable communication at an asymptotic nonzero rate, as long as there is {\it some} correlation between the channel's input and its output.  Furthermore, he found a useful formula for the optimal rate that can be achieved.  These results are the content of the {\em noisy channel coding theorem}.

\subsubsection{Capacity of the binary symmetric channel.}
To be concrete, suppose we use the binary alphabet $\{0,1\}$, and the {\em binary symmetric channel}; this channel acts on each bit
independently, flipping its value with probability $p$, and leaving it intact
with probability $1 - p$.  Thus the conditional probabilities characterizing the channel are
\begin{equation}
\begin{array}{ll}
p(0|0) = 1 - p, & p(0|1) = p,\\
p(1|0) = p, & p(1|1) = 1 - p.
\end{array}
\end{equation}

We want to construct a family of codes with increasing block size $n$, such that
the probability of a decoding error goes to zero as $n \rightarrow \infty$.  For each $n$, the code contains $2^k$ {\em codewords} among the $2^n$ possible strings of length $n$. The rate $R$ of the code, the number of encoded data bits transmitted per physical bit carried by the channel, is 
\begin{equation}
R = {k\over n}.
\end{equation}

To protect against errors, we should choose the code so that the codewords are as ``far apart'' as possible. For given values of $n$ and $k$, we want to maximize the number of bits that must be flipped to change one codeword to another, the {\em Hamming distance} between the two codewords. For any $n$-bit input message, we expect about $np$ of the bits to flip --- the input diffuses into one of about 
$2^{nH(p)}$ typical output strings, occupying an ``error sphere'' of ``Hamming
radius'' $np$ about the input string.  To decode reliably, we want to choose our input codewords so that the error spheres of two different codewords do not overlap substantially. Otherwise, two different inputs will sometimes yield
the same output, and decoding errors will inevitably occur.  To avoid
such decoding ambiguities, the total number of strings contained in all
$2^k=2^{nR}$ error spheres should not exceed the total number $2^n$ of bits in the
output message; we therefore require
\begin{equation}
2^{nH(p)} 2^{nR} \leq 2^n
\end{equation}
or
\begin{equation}
R \leq 1 - H(p) := C(p).
\end{equation}
If transmission is highly reliable, we cannot expect the rate of the code to
exceed $C(p)$.  But is the rate $R = C(p)$ actually {\it achievable}
asymptotically?

In fact transmission with $R=C-o(1)$ and negligible decoding error probability is possible.  Perhaps Shannon's most ingenious idea
was that this rate can be achieved by an average over
``random codes.''  Though choosing a code at random does not seem like a clever strategy, rather surprisingly it turns out that random coding 
achieves as high a rate as any other coding scheme in the limit $n\to\infty$.  Since $C$ is the optimal rate for reliable transmission of data over
the noisy channel it is called the {\it channel capacity.}

Suppose that $X$ is the uniformly random ensemble for a single bit (either 0 with $p=\frac{1}{2}$ or 1 with $p = \frac{1}{2}$), and that we sample from $X^n$ a total of 
$2^{nR}$ times to generate $2^{nR}$ ``random codewords.'' The resulting code is known by both Alice and Bob. To send $nR$ bits of information, Alice chooses one of the codewords and sends it to Bob by using the channel $n$ times. To decode the $n$-bit message he receives, Bob draws
 a ``Hamming sphere'' with ``radius'' slightly larger than $np$, containing
\begin{equation}
2^{n(H(p) + \delta)}
\end{equation}
strings.  If this sphere contains a unique codeword, Bob decodes the message accordingly. If the sphere contains more than one codeword, or no codewords, Bob decodes arbitrarily. 

How likely is a decoding error?  For any positive $\delta$, Bob's decoding sphere is large enough that it is very likely to contain the codeword sent by Alice when $n$ is sufficiently large. Therefore, we need only worry that the sphere might contain another codeword just by accident. Since there are altogether $2^n$ possible strings, Bob's sphere contains a fraction
\begin{equation}
f = \frac{2^{n(H(p) + \delta)}}{2^n} = 2^{-n(C(p) - \delta)},
\end{equation}
of all the strings. Because the codewords are uniformly random, the probability that Bob's sphere contains any particular codeword aside from the one sent by Alice is $f$, and the probability  that the sphere contains any one of the $2^{nR} -1 $ invalid codewords is no more than
\begin{equation}
2^{nR} f= 2^{-n(C(p) - R - \delta)}.
\end{equation}
Since $\delta$ may be as small as we please, we may choose $R = C(p) -c$ where $c$ is any positive constant, and the decoding error probability will
approach zero as $n \to \infty$.

When we speak of codes chosen at random, we really mean that we are averaging over many possible codes. The argument so far has shown that the 
 {\it average} probability of error is small,
where we average over the choice of random code, and for each specified code
we also average over all codewords. It follows that there must be a particular sequence of codes such that 
the average probability of error (when we average over the codewords) vanishes in the limit $n\to \infty$. 
We would like a stronger result -- that the probability of
error is small for {\it every} codeword.

To establish the stronger result, let $p_i$ denote the probability of a
decoding error when codeword $i$ is sent.  For any positive $\varepsilon$ and sufficiently large $n$, we have demonstrated the existence of a code such that
\begin{equation}
{1\over 2^{nR}} \sum_{i= 1}^{2^{nR}} p_i \le \varepsilon.
\end{equation}
Let $N_{2\varepsilon}$ denote the number of codewords with $p_i \ge
2\varepsilon$.  Then we infer that
\begin{equation}
{1\over 2^{nR}} (N_{2\varepsilon}) 2\varepsilon \le \varepsilon ~{\rm or}~
N_{2\varepsilon} \le 2^{nR-1};
\end{equation}
we see that we can throw away at most half of the codewords, to achieve $p_i \le
2\varepsilon$ for {\it every} codeword.  The new code we have constructed has
\begin{equation}
{\rm Rate} = R - {1\over n},
\end{equation}
which approaches $R$ as $n \rightarrow \infty$. We have seen, then, that the rate $R = C(p) - o(1)$ is asymptotically achievable with negligible probability of error, where $C(p) = 1 - H(p)$.

\subsubsection{Mutual information as an achievable rate.} Now consider how to apply this random coding argument to more general alphabets and
channels.  The channel is characterized by $p(y|x)$, the conditional probability that the letter $y$ is received when the letter $x$ is sent. We fix an ensemble $X=\{x,p(x)\}$ for the input letters, and generate the codewords for a length-$n$ code with rate $R$ by sampling $2^{nR}$ times from the distribution $X^n$; the code is known by both the sender Alice and the receiver Bob. To convey an encoded $nR$-bit message, one of the $2^{nR}$ $n$-letter codewords is selected and sent by using the channel $n$ times. The channel acts independently on the $n$ letters, governed by the same conditional probability distribution $p(y|x)$ each time it is used. The input ensemble $X$, together with the conditional probability characterizing the channel, determines the joint ensemble $XY$ for each letter sent, and therefore the joint ensemble $X^n Y^n$ for the $n$ uses of the channel.

To define a decoding procedure, we use the notion of joint typicality introduced in \S\ref{subsec:joint-typ}. When Bob receives the $n$-letter output message $\vec y$, he determines whether there is an $n$-letter input codeword $\vec x$ jointly typical with $\vec y$. If such $\vec x$ exists and is unique, Bob decodes accordingly. If there is no $\vec x$ jointly typical with $\vec y$, or more than one such $\vec x$, Bob decodes arbitrarily. 

How likely is a decoding error? For any positive $\varepsilon$ and $\delta$, the $(\vec x,\vec y)$ drawn from $X^nY^n$ is jointly $\delta$-typical with probability at least $1-\varepsilon$ if $n$ is sufficiently large. Therefore, we need only worry that there might more than one codeword jointly typical with $\vec y$. 

Suppose that Alice samples $X^n$ to generate a codeword $\vec x$, which she sends to Bob using the channel $n$ times. Then Alice samples $X^n$ a second time, producing another codeword $\vec x'$. With probability close to one, both $\vec y$ and $\vec x'$ are $\delta$-typical. But what is the probability that $\vec x'$ is {\em jointly} $\delta$-typical with $\vec y$? 

Because the samples are independent, the probability of drawing these two codewords factorizes as $p(\vec x',\vec x) = p(\vec x')p(\vec x)$, and likewise the channel output $\vec y$ when the first codeword is sent is independent of the second channel input $\vec x'$, so $p(\vec x',\vec y) = p(\vec x')p(\vec y)$. From eq.(\ref{eq:joint-typ-prob}) we obtain an upper bound on  the number $N_{\rm j.t.}$ of jointly $\delta$-typical $(\vec x,\vec y)$:
\begin{align}
1\ge \sum_{{\rm j.t.} ~(\vec x,\vec y)} p(\vec x,\vec y) \ge N_{\rm j.t.}~ 2^{-n(H(XY)+\delta)} \implies N_{\rm j.t.}\le  2^{n(H(XY)+\delta)}.
\end{align}
We also know that each $\delta$-typical $\vec x'$ occurs with probability $p(\vec x') \le 2^{-n(H(X)-\delta)}$ and that each $\delta$-typical $\vec y$ occurs with probability $p(\vec y) \le 2^{-n(H(Y)-\delta)}$. Therefore, the probability that $\vec x'$ and $\vec y$ are jointly $\delta$-typical is bounded above by
\begin{align}
\sum_{{\rm j.t.}~(\vec x',\vec y)} p(\vec x')p(\vec y)&\le N_{\rm j.t.} ~ 2^{-n(H(X)-\delta)}2^{-n(H(Y)-\delta)}\notag\\
&\le 2^{n(H(XY)+\delta)}2^{-n(H(X)-\delta)}2^{-n(H(Y)-\delta)}\notag\\
&= 2^{-n(I(X;Y) - 3\delta)}.
\end{align}
If there are $2^{nR}$ codewords, all generated independently by sampling $X^n$, then the probability that {\em any} other codeword besides $\vec x$ is jointly typical with $\vec y$ is bounded above by
\begin{align}
2^{nR} 2^{-n(I(X;Y) - 3\delta)}
= 2^{n(R - I(X;Y)+ 3\delta)}.
\end{align}
Since $\varepsilon$ and $\delta$ are as small as we please, we may choose $R = I(X;Y) - c$, where $c$ is any positive constant, and the decoding error probability will
approach zero as $n \to \infty$.

So far we have shown that the error probability is small when we average over codes and over codewords. To complete the argument we use the same reasoning as in our discussion of the capacity of the binary symmetric channel. There must exist a particular sequence of codes with zero error probability in the limit $n\to \infty$, when we average over codewords. And by pruning the codewords, reducing the rate by a negligible amount,  we can ensure that the error probability is small for {\em every} codeword. We conclude that the rate 
\begin{equation}
R = I(X;Y) - o(1)
\end{equation} 
is asymptotically achievable with negligible probability of error. This result provides a concrete operational interpretation for the mutual information $I(X;Y)$; it is the information per letter we can transmit over the channel, supporting the heuristic claim that $I(X;Y)$ quantifies the information we gain about $X$ when we have access to $Y$.

The mutual information $I(X;Y)$ depends not only on the channel's conditional
probability $p(y|x)$ but also on the {\em a priori} probability $p(x)$ defining the codeword ensemble $X$. The achievability argument for random coding applies for any choice of $X$, so we have demonstrated that errorless transmission over the noisy channel is possible for any rate $R$ strictly less than
\begin{equation}
C := \max_X I(X;Y).
\end{equation}
This quantity $C$ is called the {\it channel capacity}; it depends only on the conditional
probabilities $p(y|x)$ that define the channel.

\subsubsection{Upper bound on the capacity.}
We have now shown that any rate $R < C$ is achievable, but can
$R$  exceed $C$ with the error probability still approaching $0$ for large
$n$?  To see that a rate for errorless transmission exceeding $C$ is not possible, we reason as follows.

Consider any code with $2^{nR}$ codewords, and consider the uniform ensemble on the codewords, denoted $\tilde X^n$, in which each
codeword occurs with probability $2^{-nR}$.  Evidently, then,
\begin{equation}
H(\tilde{X}^n) = nR.
\end{equation}
Sending the codewords through $n$ uses of the channel we obtain an ensemble $\tilde Y^n$ of output states, and a joint ensemble $\tilde X^n\tilde Y^n$. 

Because the channel acts on each letter independently, the
conditional probability for $n$ uses of the channel factorizes:
\begin{equation}
p(y_1 y_2 \cdots y_n|x_1 x_2 \cdots x_n) = p(y_1|x_1) p(y_2|x_2) \cdots
p(y_n|x_n),
\end{equation}
and it follows that the conditional entropy satisfies
\begin{align}
H(\tilde{Y}^n|\tilde{X}^n) = \langle -\log p(\vec y|\vec x)\rangle &= \sum_i \langle
-\log p(y_i|x_i)\rangle\notag \\
&= \sum_i H (\tilde{Y}_i|\tilde{X}_i),
\label{eq:additive-conditional-entropy}
\end{align}
where $\tilde{X}_i$ and $\tilde{Y}_i$ are the marginal probability
distributions for the $i$th letter determined by our distribution on the
codewords.  Because Shannon entropy is subadditive, $H(XY) \leq H(X) + H(Y)$, we have
\begin{equation}
H(\tilde{Y}^n) \leq \sum_i H(\tilde{Y}_i),
\end{equation}
and therefore
\begin{align}
I(\tilde{Y}^n; \tilde{X}^n) &= H(\tilde{Y}^n) - H (\tilde{Y}^n |
\tilde{X}^n)\notag \\
&\leq \sum_i (H(\tilde{Y}_i) - H(\tilde{Y}_i | \tilde{X}_i))\notag \\
&= \sum_i I(\tilde{Y}_i; \tilde{X}_i) \leq n C.
\label{eq:capacity-from-subadditivity}
\end{align}
The mutual information of the messages sent and received is bounded above by
the sum of the mutual information per letter, and the mutual information for
each letter is bounded above by the capacity, because $C$ is defined as the
maximum of $I(X;Y)$ over all input ensembles.

Recalling the symmetry of mutual information, we have
\begin{align}\label{mutual_bound}
I(\tilde{X}^n; \tilde{Y}^n) &= H(\tilde{X}^n) -
H(\tilde{X}^n|\tilde{Y}^n)\notag \\
&= nR - H(\tilde{X}^n |\tilde{Y}^n) \leq nC.
\end{align}
Now, if we can decode reliably as $n \rightarrow \infty$, this means that the
input codeword is completely determined by the signal received, or that the
conditional entropy of the input (per letter) must get small
\begin{equation}
{1\over n} H (\tilde{X}^n |\tilde{Y}^n) \rightarrow 0.
\label{eq:conditional-to-zero}
\end{equation}
If errorless transmission is possible, then, eq.~(\ref{mutual_bound}) becomes
\begin{equation}
R  \leq C +o(1),
\end{equation}
in the limit $n \rightarrow \infty$.  The asymptotic rate cannot exceed the capacity.
In Exercise \ref{ex:fano}, you will sharpen the statement eq.(\ref{eq:conditional-to-zero}), showing that 
\begin{align}
{1\over n} H (\tilde{X}^n |\tilde{Y}^n)\le \frac{1}{n} H_2(p_e) + p_e R,
\label{eq:capacity-error}
\end{align}
where $p_e$ denotes the decoding error probability, and $H_2(p_e) = -p_e\log_2 p_e - (1-p_e)\log_2 (1-p_e)$ .


We have now seen that the capacity $C$ is the highest achievable rate of communication
through the noisy channel, where the probability of error
goes to zero as the number of letters in the message goes to infinity.  This is
Shannon's noisy channel coding theorem. What is particularly remarkable is that, although the capacity is achieved by messages that are many letters in length, we have obtained a {\em single-letter formula} for the capacity, expressed in terms of the optimal mutual information $I(X;Y)$ for just a single use of the channel. 

The method we used  to show that $R = C - o(1)$ is achievable, averaging over random codes, is not constructive.  Since a
random code has no structure or pattern, encoding and decoding are
unwieldy, requiring an exponentially large code book.  Nevertheless, the
theorem is important and useful, because it tells us what is achievable, and not achievable, in principle. Furthermore, 
since $I(X;Y)$ is a concave function of $X = \{x, p(x)\}$ (with $\{p(y|x)\}$
fixed), it has a unique local maximum, and $C$ can often be computed (at least
numerically) for channels of interest. Finding codes which can be efficiently encoded and decoded, and come close to achieving the capacity, is a very interesting pursuit, but beyond the scope of our lightning introduction to Shannon theory.

\section{Von Neumann Entropy}

In classical information theory, we often consider a source that prepares
messages of $n$ letters ($n \gg 1$), where each letter is drawn independently
from an ensemble $X = \{x,p(x)\}$.  We have seen that the Shannon entropy
$H(X)$ is the number of incompressible bits of information carried per letter
(asymptotically as $n \rightarrow \infty$).

We may also be interested in correlations among messages.  The correlations
between two ensembles of letters $X$ and $Y$ are characterized by conditional
probabilities $p(y|x)$.  We have seen that the mutual information
\begin{equation}
I(X;Y) = H(X) - H(X|Y) = H(Y) - H(Y|X),
\end{equation}
is the number of bits of information per letter about $X$ that we can acquire
by reading $Y$ (or vice versa).  If the $p(y|x)$'s characterize a noisy
channel, then, $I(X;Y)$ is the amount of information per letter that can be
transmitted through the channel (given the {\it a priori} distribution $X$ for the
channel inputs).

We would like to generalize these considerations to {\it quantum} information.
We may imagine a source that prepares messages of $n$ letters, but where
each letter is chosen from an ensemble of quantum states.  The signal alphabet
consists of a set of quantum states $\{\bfrho(x)\}$, each occurring with a specified
{\it a priori} probability $p(x)$.

As we discussed at length in Chapter 2, the probability of any outcome of any
measurement of a letter chosen from this ensemble, if the observer has no
knowledge about which letter was prepared, can be completely characterized by
the density operator
\begin{equation}
\bfrho = \sum_x p(x) \bfrho(x);
\end{equation}
for a POVM $\bfE =\{\bfE_a\}$, the probability of outcome $a$ is 
\begin{equation}
{\rm Prob}(a) = {\rm tr} (\bfE_a \bfrho).
\end{equation}
For this (or any) density operator, we may define the von Neumann entropy
\begin{equation}
H(\bfrho) = - {\rm tr} (\bfrho\log \bfrho).
\end{equation}
Of course, we may choose an orthonormal basis $\{|a\rangle\}$ that diagonalizes
$\bfrho$,
\begin{equation}
\bfrho = \sum_a \lambda_a |a\rangle \langle a|;
\end{equation}
the vector of eigenvalues $\lambda(\bfrho)$ is a probability distribution, and the von Neumann entropy of $\bfrho$ is just the Shannon entropy of this distribution,
\begin{equation}
H(\bfrho) = H(\lambda(\bfrho)).
\end{equation}
If $\bfrho_A$ is the density operator of system $A$, we will sometimes use the notation
\begin{equation}
H(A) := H(\bfrho_A).
\end{equation}
Our convention is to denote quantum systems with $A, B, C, \dots$ and classical probability distributions with $X, Y, Z, \dots$.

In the case where the signal alphabet $\{|\varphi(x)\rangle, p(x)\}$ consists of mutually orthogonal pure
states, the quantum source reduces to a classical one; all of the signal states
can be perfectly distinguished, and $H(\bfrho) = H(X)$, where $X$ is the classical ensemble $\{x,p(x)\}$. The quantum source is
more interesting when the signal states $\{\bfrho(x)\}$ are not mutually commuting.
We will argue that the von Neumann entropy quantifies the incompressible
information content of the quantum source (in the case where the signal states
are pure) much as the Shannon entropy quantifies the information content of a
classical source.

Indeed, we will find that von Neumann entropy plays multiple roles.  It quantifies
not only the {\it quantum} information content per letter of the pure-state ensemble (the
minimum number of qubits per letter needed to reliably encode the information)
but also its {\it classical} information content (the maximum amount of
information per letter---in bits, not qubits---that we can gain about the
preparation by making the best possible measurement).  And we will see that
von Neumann entropy enters quantum information in yet other ways --- for example,
quantifying the entanglement of a bipartite pure state.  Thus quantum
information theory is largely concerned with the interpretation and uses of von
Neumann entropy, much as classical information theory is largely concerned with
the interpretation and uses of Shannon entropy.

In fact, the mathematical machinery we need to develop quantum information
theory is very similar to Shannon's mathematics (typical sequences, random
coding, \ldots); so similar as to sometimes obscure that the conceptual
context is really quite different.  The central issue in quantum information
theory is that nonorthogonal quantum states cannot be perfectly
distinguished, a feature with no classical analog.

\subsection{Mathematical properties of $H(\bfrho)$}

There are a handful of properties of the von Neumann entropy $H(\bfrho)$ which are frequently useful,
many of which are closely analogous to corresponding properties of the Shannon entropy $H(X)$.  Proofs of some of these are Exercises \ref{ex:rel-entropy}, \ref{ex:von-neumann-entropy}, \ref{ex:rel-entropy-monotonicity}. 
\begin{enumerate}
\item {\bf Pure states}.  A pure state $\bfrho =
|\varphi\rangle\langle\varphi|$ has $H(\bfrho) = 0$.

\item {\bf Unitary invariance}.  The entropy is unchanged by a unitary change of
basis,
\begin{equation}
H(\bfu \bfrho \bfu^{-1}) = H(\bfrho),
\end{equation}
because $H(\bfrho)$ depends only on the eigenvalues of
$\bfrho$.

\item {\bf Maximum}.  If $\bfrho$ has $d$ nonvanishing eigenvalues, then
\begin{equation}
H(\bfrho) \leq \log d,
\end{equation}
with equality when all the nonzero eigenvalues are equal.  The entropy is
maximized when the quantum state is maximally mixed.

\item {\bf Concavity}.  For $\lambda_1, \lambda_2, \cdots , \lambda_n
\geq 0$ and $\lambda_1 + \lambda_2 + \cdots + \lambda_n = 1$,
\begin{equation}
H(\lambda_1 \bfrho_1 + \cdots + \lambda_n \bfrho_n) \geq \lambda_1 H(\bfrho_1)
+ \cdots + \lambda_n H(\bfrho_n).
\end{equation}
The von Neumann entropy is larger if we are {\it more ignorant} about
how the state was prepared.  This property is a consequence of the concavity of
the log function.

\item {\bf Subadditivity}.  Consider a bipartite system $AB$ in the state
$\bfrho_{AB}$.  Then
\begin{equation}
H(AB) \leq H(A) + H(B)
\end{equation}
(where $\bfrho_A = {\rm tr}_B \left(\bfrho_{AB}\right)$ and $\bfrho_B = {\rm tr}_A
\left(\bfrho_{AB}\right)$), with equality only for $\bfrho_{AB} = \bfrho_A \otimes \bfrho_B$.
Thus, entropy is {\it additive} for uncorrelated systems, but otherwise the
entropy of the whole is less than the sum of the entropy of the parts.  This
property is the quantum generalization of subadditivity of Shannon entropy:
\begin{equation}
H(XY) \leq H(X) + H(Y).
\end{equation} 

\item {\bf Bipartite pure states}. If the state $\bfrho_{AB}$ of the bipartite system $AB$ is pure, then
\begin{equation}
H(A) = H(B),
\end{equation}
because $\bfrho_A$ and $\bfrho_B$ have the same nonzero eigenvalues. 

\item {\bf Quantum mutual information}. As in the classical case, we define the mutual information of two quantum systems as
\begin{equation}
I(A;B) = H(A) +H(B) - H(AB), 
\end{equation}
which is nonnegative because of the subadditivity of von Neumann entropy, and zero only for a product state $\bfrho_{AB} = \bfrho_A \otimes \bfrho_B$. 

\item {\bf Triangle inequality (Araki-Lieb inequality)}. For a bipartite system,
\begin{equation}
H(AB) \geq |H(A) - H(B)|.
\label{eq:araki-lieb}
\end{equation}
To derive the triangle inequality, consider the tripartite pure state $|\psi\rangle_{ABC}$ which purifies $\bfrho_{AB}={\rm tr}_C\left(|\psi\rangle\langle \psi|\right)$. Since $|\psi\rangle$ is pure, $H(A) = H(BC)$ and $H(C)=H(AB)$; applying subadditivity to $BC$ yields $H(A) \le H(B) + H(C)= H(B)+H(AB)$. The same inequality applies with $A$ and $B$ interchanged, from which we obtain eq.(\ref{eq:araki-lieb}).
\end{enumerate}

\noindent The triangle inequality contrasts sharply with the analogous property of
Shannon entropy,
\begin{equation}
H(XY) \geq H(X), H(Y).
\end{equation}
The Shannon entropy of just part of a classical bipartite system cannot be greater than the Shannon entropy of the whole system.
Not so for the von Neumann entropy!  For example, in the case of an entangled bipartite
pure quantum state, we have $H(A) = H(B) > 0 $, while $H(AB) = 0$. The entropy of the global system vanishes because our ignorance is minimal --- we know as much about $AB$ as the laws of quantum physics will allow. But we have incomplete knowledge of the parts $A$ and $B$, with our ignorance quantified by $H(A)=H(B)$. For a quantum system, but not for a classical one, information can be encoded in the correlations among the parts of the system, yet be invisible when we look at the parts one at a time. 

Equivalently, a property that holds classically but not quantumly is
\begin{equation}
H(X|Y)= H(XY)-H(Y) \geq 0.
\end{equation}
The Shannon conditional entropy $H(X|Y)$ quantifies our remaining ignorance about $X$ when we know $Y$, and equals zero when knowing $Y$ makes us certain about $X$. On the other hand, the von Neumann conditional entropy,
\begin{equation}
H(A|B) = H(AB) - H(B),
\end{equation}
can be negative; in particular we have $H(A|B) = -H(A) = -H(B) < 0$ if $\bfrho_{AB}$ is an entangled pure state. How can it make sense that ``knowing'' the subsystem $B$ makes us ``more than certain'' about the subsystem $A$? We'll return to this intriguing question in \S\ref{subsec:mother}.

When $X$ and $Y$ are perfectly correlated, then $H(XY) = H(X) = H(Y)$; the conditional entropy is $H(X|Y) = H(Y|X) = 0$ and the mutual information is $I(X;Y) = H(X)$. In contrast, for a bipartite pure state of $AB$, the quantum state for which we may regard $A$ and $B$ as perfectly correlated, the mutual information is $I(A;B) = 2H(A) = 2H(B)$. In this sense the quantum correlations are stronger than classical correlations. 

\subsection{Mixing, measurement, and entropy}
\label{subsec:mixing-measurement-entropy}

The Shannon entropy also has a property called {\em Schur concavity}, which means that if $X= \{x,p(x)\}$ and $Y=\{y,q(y)\}$ are two ensembles such that $p \prec q$, then $H(X)\ge H(Y)$. 
Recall that $p \prec q$ ($q$ {\em majorizes} $p$) means that ``$p$ is at least as random as $q$'' in the sense that $p = Dq$ for some doubly stochastic matrix $D$. Thus Schur concavity of $H$ says that an ensemble with more randomness has higher entropy. 

The von Neumann entropy $H(\bfrho)$ of a density operator is the Shannon entropy of its vector of eigenvalues $\lambda(\bfrho)$. Furthermore, we showed in Exercise 2.6 that if the quantum state ensemble $\mathcal{E}=\{|\varphi(x)\rangle, p(x)\}$  realizes $\bfrho$, then $p \prec \lambda(\bfrho)$; therefore $H(\bfrho)\le H(X)$, where equality holds only for an ensemble of mutually orthogonal states. The decrease in entropy $H(X) - H(\bfrho)$ quantifies how {\it distinguishability is lost} when we mix
nonorthogonal pure states.  As we will soon see, the amount of information we can gain by measuring $\bfrho$ is no more than $H(\bfrho)$ bits, so some of the information about which state was prepared has been irretrievably lost if $H(\bfrho) < H(X)$. 

If we perform an orthogonal measurement on $\bfrho$ by projecting onto the basis $\{|y\rangle\}$, then outcome $y$ occurs with probability 
\begin{equation}
q(y) = \langle y |\bfrho |y\rangle = \sum_a |\langle y|a\rangle|^2 \lambda_a, \quad {\rm where} \quad \bfrho = \sum_a \lambda_a |a\rangle\langle a|
\end{equation}
and $\{|a\rangle\}$ is the basis in which $\bfrho$ is diagonal. Since $D_{ya}=|\langle y|a\rangle|^2$ is a doubly stochastic matrix, $q \prec \lambda(\bfrho)$ and therefore $H(Y) \ge H(\bfrho)$, where equality holds only if the measurement is in the basis $\{|a\rangle\}$. Mathematically, the conclusion is that for a nondiagonal and nonnegative Hermitian matrix, the diagonal elements are more random than the eigenvalues. Speaking more physically, the outcome of an orthogonal measurement is easiest to predict if we measure an observable which commutes with the density operator, and becomes less predictable if we measure in a different basis. 

This majorization property has a further consequence, which will be useful for our discussion of quantum compression. Suppose that $\bfrho$ is a density operator of a $d$-dimensional system, with eigenvalues $\lambda_1\ge \lambda_2 \ge \dots \ge \lambda_d$ and that $\bfE'= \sum_{i=1}^{d'} |e_i\rangle\langle e_i|$ is a projector onto a subspace $\Lambda$ of dimension $d' \le d$ with orthonormal basis $\{|e_i\rangle\}$. Then
\begin{align}
{\rm tr}\left(\bfrho \bfE'\right) = \sum_{i=1}^{d'} \langle e_i|\bfrho|e_i\rangle \le \sum_{a=1}^{d'} \lambda_a,
\end{align}
where the inequality follows because the diagonal elements of $\bfrho$ in the basis $\{|e_i\rangle\}$ are majorized by the eigenvalues of $\bfrho$. In other words, if we perform a two-outcome orthogonal measurement, projecting onto either $\Lambda$ or its orthogonal complement $\Lambda^\perp$, the probability of projecting onto $\Lambda$ is no larger than the sum of the $d'$ largest eigenvalues of $\bfrho$ (the {\em Ky Fan dominance principle}).

\subsection{Strong subadditivity}

In addition to the subadditivity property $I(X;Y) \ge 0$, correlations of classical random variables obey a further  property called {\em strong subadditivity}:
\begin{align}
I(X;YZ) \ge I(X;Y).
\end{align}
This is the eminently reasonable statement that the correlations of $X$ with $YZ$ are at least as strong as the correlations of $X$ with $Y$ alone.

There is another useful way to think about (classical) strong subadditivity. Recalling the definition of mutual information we have
\begin{align}\label{eq:SSA-classical}
I(X;YZ) -I(X;Y)&= -\left\langle \log \frac{p(x)p(y,z)}{p(x,y,z)} + \log \frac{p(x,y)}{p(x)p(y)}\right\rangle\notag\\
&=-\left\langle \log \frac{p(x,y)}{p(y)} ~\frac{p(y,z)}{p(y)}~\frac{p(y)}{p(x,y,z)}\right\rangle\notag\\
&= -\left\langle \log \frac{p(x|y)p(z|y)}{p(x,z|y)}\right\rangle=\sum_y p(y) I(X;Z|y) \ge 0,
\end{align}
where in the last line we used $p(x,y,z)= p(x,z|y)p(y)$.
For each fixed $y$, $p(x,z|y)$ is a normalized probability distribution with nonnegative mutual information; hence $I(X;YZ) -I(X;Y)$ is a convex combination of nonnegative terms and therefore nonnegative. The quantity $I(X;Z|Y):=I(X;YZ) - I(X;Y)$ is called the {\em conditional mutual information}, because it quantifies how strongly $X$ and $Z$ are correlated when $Y$ is known; strong subadditivity can be restated as the nonnegativity of conditional mutual information, 
\begin{align}
I(X;Z|Y) \ge 0.
\end{align}
The identities in eq.(\ref{eq:SSA-classical}) demonstrate that $I(X;Z|Y)$ is symmetric under interchange of $X$ and $Z$, although this is not manifest in the expression $I(X;Z|Y)=I(X;YZ) - I(X;Y)$.

One might ask under what conditions strong subadditivity is satisfied as an equality; that is, when does the conditional mutual information vanish? Since $I(X;Z|Y)$ is a sum of nonnegative terms, each of these terms must vanish if $I(X;Z|Y)=0$. Therefore for each $y$ with $p(y) > 0$, we have $I(X;Z|y)=0$. The mutual information vanishes only for a product distribution; therefore
\begin{align}
p(x,z|y) = p(x|y) p (z|y)\implies p(x,y,z) = p(x|y)p(z|y)p(y).
\end{align}
This means that the correlations between $x$ and $z$ arise solely from their shared correlation with $y$, in which case we say that $x$ and $z$ are {\em conditionally independent}.

Correlations of quantum systems also obey strong subadditivity:
\begin{equation}
I(A;BC) - I(A;B) := I(A;C|B) \ge 0.
\end{equation}
But while the proof is elementary in the classical case, in the quantum setting strong subadditivity is a rather deep result with many important consequences. We will postpone the proof until \S\ref{subsec:operational-ssa}, where we will be able to justify the quantum statement by giving it a clear operational meaning. 

We'll also see in Exercise \ref{ex:rel-entropy-monotonicity} that strong subadditivity follows easily from another deep property, the monotonicity of relative entropy:
\begin{equation}
D(\bfrho_A\|\bfsigma_A) \leq D(\bfrho_{AB} \| \bfsigma_{AB}),
\end{equation}
where
\begin{equation}
D(\bfrho\|\bfsigma):= {\rm tr}~\bfrho\left(\log \bfrho - \log \bfsigma\right).
\end{equation}
We won't discuss the details, but it turns out that the quantum relative entropy of $\bfrho$ with respect to $\bfsigma$ can be interpreted as a measure of the distinguishability of the two states in the context of hypothesis testing, much as the classical relative entropy $D(p\|q)$ is a measure of distinguishability of probability distributions as discussed in \S\ref{subsec:relative-hypothesis}. Hence the monotonicity of quantum relative entropy amounts to the reasonable statement that quantum states on $AB$ become no easier to distinguish when we look at only the subsystem $A$ than when we look at the full system $AB$. 
It also follows (Exercise \ref{ex:rel-entropy-monotonicity}), that the action of a quantum channel $\mathcal{N}$ cannot increase relative entropy:
\begin{equation}
D(\mathcal{N}(\bfrho)\|\mathcal{N}(\bfsigma)) \leq D(\bfrho\| \bfsigma).
\end{equation}

There are a few other ways of formulating strong subadditivity which are helpful to keep in mind. By expressing the quantum mutual information in terms of the von Neumann entropy we find
\begin{equation}
H(ABC) + H(B) \leq H(AB) + H(BC).
\end{equation}
While $A, B, C$ are three disjoint quantum systems, we may view $AB$ and $BC$ as overlapping systems with intersection $B$ and union $ABC$; then strong subadditivity says that the sum of the entropies of two overlapping systems is at least as large as the sum of the entropies of their union and their intersection. In terms of conditional entropy, strong subadditivity becomes
\begin{equation}
H(A|B) \ge H(A|BC); 
\end{equation}
loosely speaking, our ignorance about $A$ when we know only $B$ is no smaller than our ignorance about $A$ when we know both $B$ and $C$, but with the proviso that for quantum information ``ignorance'' can sometimes be negative!

As in the classical case, it is instructive to consider the condition for equality in strong subadditivity. What does it mean for systems to have {\em quantum conditional independence}, $I(A;C|B)=0$? It is easy to formulate a sufficient condition. Suppose that system $B$ has a decomposition as a direct sum of tensor products of Hilbert spaces
\begin{align}
\mathcal{H}_B = \bigoplus_j \mathcal{H}_{B_j} =\bigoplus_j \mathcal{H}_{B_j^L}\otimes \mathcal{H}_{B_j^R},
\end{align}
and that the state of $ABC$ has the block diagonal form
\begin{align}
\bfrho_{ABC} = \bigoplus_j p_j ~\bfrho_{A B_j^L}\otimes \bfrho_{B_j^R C}.
\label{eq:quantum-conditional-independence}
\end{align}
In each block labeled by $j$ the state is a tensor product, with conditional mutual information
\begin{align}
I(A;C|B_j) = I(A;B_j C) - I(A;B_j) =  I(A;B_j^L) - I(A;B_j^L) = 0;
\end{align}
What is less obvious is that the converse is also true --- any state with $I(A;C|B)=0$ has a decomposition as in eq.(\ref{eq:quantum-conditional-independence}). This is a useful fact, though we will not give the proof here. 

\subsection{Monotonicity of mutual information}

Strong subadditivity implies another important property of  quantum mutual information, its {\em monotonicity} --- a quantum channel acting on system $B$ cannot increase the mutual information of $A$ and $B$.  
To derive monotonicity, suppose that a quantum channel $\mathcal{N}^{B\to B'}$ maps $B$ to $B'$. Like any quantum channel, $\mathcal{N}$ has an isometric extension, its Stinespring dilation $\bfU^{B\to B'E}$, mapping $B$ to $B'$ and a suitable environment system $E$. Since the isometry $\bfU$ does not change the eigenvalues of the density operator, it preserves the entropy of $B$ and of $AB$, 
\begin{equation}
H(B) = H(B'E), \quad H(AB) = H(AB'E), 
\end{equation}
which implies
\begin{align}
I(A;B)& = H(A) + H(B) - H(AB) \notag\\
&= H(A) +H(B'E) - H(ABE') = I(A;B'E).
\end{align}
From strong subadditivity, we obtain
\begin{equation}
I(A;B) = I(A;B'E) \ge I(A;B')
\end{equation}
the desired statement of monotonicity.

\subsection{Entropy and thermodynamics}

The concept of entropy first entered science through the study of
thermodynamics, and the mathematical properties of entropy we have enumerated have many interesting thermodynamic implications. Here we will just mention a few ways in which the nonnegativity and monotonicity of quantum relative entropy relate to ideas encountered in thermodynamics. 

There are two distinct ways to approach the foundations of
quantum statistical physics.  In one, we consider the evolution of an
isolated closed quantum system, but ask what we will observe if we have access to only a portion of the full system. Even though the evolution of the full system is unitary, the evolution of a subsystem is not, and the subsystem may be accurately described by a thermal ensemble at late times. Information which is initially encoded locally in an out-of-equilibrium state becomes encoded more and more nonlocally as the system evolves, eventually becoming invisible to an observer confined to the subsystem. 

In the other approach, we consider the evolution of an open system $A$, in contact with an unobserved environment $E$, and track the evolution of $A$ only. From a fundamental perspective this second approach may be regarded as a special case of the first, since $AE$ is closed, with $A$ as a privileged subsystem. In practice, though, it is often more convenient to describe the evolution of an open system using a master equation as in Chapter 3, and to analyze evolution toward thermal equilibrium without explicit reference to the environment. 

\subsubsection{Free energy and the second law.}
Tools of quantum Shannon theory can help us understand why the state of an open system with Hamiltonian $\bfH$ might be expected to be close to the thermal {\em Gibbs state}
\begin{align}
\bfrho_\beta = \frac{e^{-\beta \bfH}}{{\rm tr}\left(e^{-\beta \bfH}\right)},
\end{align}
where $kT= \beta^{-1}$ is the temperature. Here let's observe one noteworthy feature of this state. For an arbitrary density operator $\bfrho$, consider its {\em free energy}
\begin{align}
F(\bfrho) = E(\bfrho) - \beta^{-1}S(\bfrho)
\end{align}
where $E(\bfrho) = \langle \bfH \rangle_{\bfrho}$ denotes the expectation value of the Hamiltonian in this state; for this subsection we respect the conventions of thermodynamics by denoting von Neumann entropy by $S(\bfrho)$ rather than $H(\bfrho)$ (lest $H$ be confused with the Hamiltonian $\bfH$), and by using natural logarithms. Expressing $F(\bfrho)$ and the free energy $F(\bfrho_\beta)$ of the Gibbs state as 
\begin{align}
&F(\bfrho) = {\rm tr} \left(\bfrho \bfH\right)- \beta^{-1} S(\bfrho) = \beta^{-1} {\rm tr}~\bfrho \left(\ln\bfrho +  \beta\bfH  \right),\notag\\ 
&F(\bfrho_\beta)= -\beta^{-1} \ln\left( {\rm tr}~ e^{-\beta \bfH}\right),
\end{align}
we see that the relative entropy of $\bfrho$ and $\bfrho_\beta$ is 
\begin{align}
D(\bfrho\|\bfrho_\beta) &= {\rm tr} \left(\bfrho\ln \bfrho\right) - {\rm tr}\left(\bfrho\ln \bfrho_\beta\right)\notag\\
&= \beta\left(F(\bfrho) - F(\bfrho_\beta)\right)\ge 0,
\label{relative-entropy-gibbs-bound}
\end{align}
with equality only for $\bfrho = \bfrho_\beta$. The nonnegativity of relative entropy implies that at a given temperature $\beta^{-1}$, the Gibbs state $\bfrho_\beta$ has the lowest possible free energy. Our open system, in contact with a thermal reservoir at temperature $\beta^{-1}$, will prefer the Gibbs state if it wishes to minimize its free energy. 

What can we say about the {\em approach} to thermal equilibrium of an open system? We may anticipate that the joint unitary evolution of system and reservoir induces a quantum channel $\mathcal{N}$ acting on the system alone, and we know that relative entropy is monotonic --- if 
\begin{align}
\mathcal{N}: \bfrho\mapsto \bfrho', \quad \mathcal{N}:\bfsigma\mapsto \bfsigma',
\end{align} 
then 
\begin{align}
D(\bfrho'\|\bfsigma') \le D(\bfrho\|\bfsigma).
\end{align}
Furthermore, if the Gibbs state is an {\em equilibrium} state, we expect this channel to preserve the Gibbs state
\begin{align}
\mathcal{N}:\bfrho_\beta \mapsto \bfrho_\beta;
\end{align}
therefore, 
\begin{align}
D(\bfrho'\|\bfrho_\beta) = \beta\left(F(\bfrho') - F(\bfrho_\beta)\right)\le \beta\left(F(\bfrho) - F(\bfrho_\beta)\right) = D(\bfrho\|\bfrho_\beta),
\end{align}
and hence
\begin{align}
F(\bfrho') \le F(\bfrho).
\end{align}
Any channel that preserves the Gibbs state cannot increase the free energy; instead, free energy of an out-of-equilibrium state is monotonically decreasing under open-state evolution.  This statement is a version of the second law of thermodynamics. 

We'll have more to say about how quantum information theory illuminates thermodynamics in \S\ref{subsec:negative-entropy}.

\subsection{Bekenstein's entropy bound.}
\label{subsec:bekenstein}

Similar ideas lead to Bekenstein's bound on entropy in quantum field theory. The field-theoretic details, though interesting, would lead us far afield. The gist is that Bekenstein proposed an inequality relating the energy and the entropy in a bounded spatial region. This bound was motivated by gravitational physics, but can be formulated without reference to gravitation, and follows from properties of relative entropy.

A subtlety is that entropy of a region is infinite in quantum field theory, because of contributions coming from arbitrarily short-wavelength quantum fluctuations near the boundary of the region. Therefore we have to make a subtraction to define a finite quantity. The natural way to do this is to subtract away the entropy of the same region in the vacuum state of the theory, as any finite energy state in a finite volume has the same structure as the vacuum at very short distances. Although the vacuum is a pure state, it, and any other reasonable state, has a marginal state in a finite region which is highly mixed, because of entanglement between the region and its complement. 

For the purpose of our discussion here, we may designate any mixed state $\bfrho_0$ we choose supported in the bounded region as the ``vacuum,'' and define a corresponding ``modular Hamiltonian'' $\bfK$ by
\begin{align}
\bfrho_0 = \frac{e^{-\bfK}}{{\rm tr}\left(e^{-\bfK} \right)}.
\end{align}
That is, we regard the state as the thermal mixed state of $\bfK$, with the temperature arbitrarily set to unity (which is just a normalization convention for $\bfK$). Then by rewriting eq.(\ref{relative-entropy-gibbs-bound}) we see that, for any state $\bfrho$, $D(\bfrho\|\bfrho_0)\ge 0$ implies
\begin{align}
S(\bfrho) - S(\bfrho_0) \le {\rm tr}\left(\bfrho \bfK\right) - {\rm tr}\left(\bfrho_0 \bfK\right)
\end{align}
The left-hand side, the entropy with vacuum entropy subtracted, is not larger than the right-hand side, the (modular) energy with vacuum energy subtracted. This is one version of Bekenstein's bound. Here $\bfK$, which is dimensionless, can be loosely interpreted as $ER$, where $E$ is the energy contained in the region and $R$ is its linear size. 

While the bound follows easily from nonnegativity of relative entropy, the subtle part of the argument is recognizing that the (suitably subtracted) expectation value of the modular Hamiltonian is a reasonable way to define $ER$. The detailed justification for this involves properties of relativistic quantum field theory that we won't go into here. Suffice it to say that, because we constructed $\bfK$ by regarding the marginal state of the vacuum as the Gibbs state associated with the Hamiltonian $\bfK$, we expect $\bfK$ to be linear in the energy, and dimensional analysis then requires inclusion of the factor of $R$ (in units with $\hbar = c = 1$).

Bekenstein was led to conjecture such a bound by thinking about black hole thermodynamics. Leaving out numerical factors, just to get a feel for the orders of magnitude of things, the entropy of a black hole with circumference $\sim R$ is $S \sim R^2/G$, and its mass (energy) is $E\sim R/G$, where $G$ is Newton's gravitational constant; hence $S\sim ER$ for a black hole. Bekenstein realized that unless $S = O(ER)$ for arbitrary states and regions, we could throw extra stuff into the region, making a black hole with lower entropy than the initial state, thus violating the (generalized) second law of thermodynamics. Though black holes provided the motivation, $G$ drops out of the inequality, which holds even in nongravitational relativistic quantum field theories. 


\subsection{Entropic uncertainty relations}
The uncertainty principle asserts that noncommuting observables cannot simultaneously have definite values. To translate this statement into mathematics, recall that a Hermitian observable $\bfA$ has spectral representation
\begin{align}
\bfA = \sum_x |x\rangle a(x) \langle x|
\end{align}
where $\{|x\rangle\}$ is the orthonormal basis of eigenvectors of $\bfA$ and $\{a(x)\}$ is the corresponding vector of eigenvalues; if $\bfA$ is measured in the state $\bfrho$, the outcome $a(x)$ occurs with probability $p(x) = \langle x|\bfrho |x\rangle$. Thus $\bfA$ has expectation value ${\rm tr}(\bfrho \bfA)$ and variance
\begin{align}
\left(\Delta A\right)^2 = {\rm tr} \left(\bfrho \bfA^2\right) - \left({\rm tr}\bfrho \bfA \right)^2.
\end{align}
Using the Cauchy-Schwarz inequality, we can show that if $\bfA$ and $\bfB$ are two Hermitian observables and $\bfrho = |\psi\rangle\langle \psi|$ is a pure state, then
\begin{align}
\Delta A \Delta B \ge \frac{1}{2} |\langle \psi|[\bfA,\bfB]|\psi\rangle |.
\label{eq:uncertain-cs}
\end{align}

Eq.(\ref{eq:uncertain-cs}) is a useful statement of the uncertainty principle, but has drawbacks. It depends on the state $|\psi\rangle$ and for that reason does not fully capture the incompatibility of the two observables. Furthermore, the variance does not characterize very well the unpredictability of the measurement outcomes; entropy would be a more informative measure. 

In fact there are {\em entropic uncertainty relations} which  do not suffer from these deficiencies. If we measure a state $\bfrho$ by projecting onto the orthonormal basis $\{|x\rangle\}$, the outcomes define a classical ensemble
\begin{align}
X= \{x, p(x) = \langle x |\bfrho| x\rangle\};
\end{align}
that is, a probability vector whose entries are the diagonal elements of $\bfrho$ in the $x$-basis. The Shannon entropy $H(X)$ quantifies how uncertain we are about the outcome before we perform the measurement. If $\{|z\rangle\}$ is another orthonormal basis, there is a corresponding classical ensemble $Z$ describing the probability distribution of outcomes when we measure the same state $\bfrho$ in the $z$-basis. If the two bases are incompatible, there is a tradeoff between our uncertainty about $X$ and about $Z$, captured by the inequality
\begin{align}
H(X) + H(Z) \ge \log\left(\frac{1}{c}\right) + H(\bfrho),
\label{eq:entropic-uncertain}
\end{align}
where
\begin{align}
c = \max_{x,z} |\langle x |z\rangle|^2.
\end{align}
The second term on the right-hand side, which  vanishes if $\bfrho$ is a pure state, reminds us that our uncertainty increases when the state is mixed. Like many good things in quantum information theory, this entropic uncertainty relation follows from the monotonicity of the quantum relative entropy. 

For each measurement there is a corresponding quantum channel, realized by performing the measurement and printing the outcome in a classical register,
\begin{align}
&\mathcal{M}_X: \bfrho \mapsto \sum_x |x\rangle\langle x|\bfrho |x\rangle\langle x|=: \bfrho_X,\notag\\
&\mathcal{M}_Z: \bfrho \mapsto \sum_z |z\rangle\langle z|\bfrho |z\rangle\langle z|=: \bfrho_Z.
\end{align}
The Shannon entropy of the measurement outcome distribution is also the von Neumann entropy of the corresponding channel's output state,
\begin{align}
H(X) = H(\bfrho_X),\quad H(Z) = H(\bfrho_Z);
\end{align}
the entropy of this output state can be expressed in terms of the relative entropy of input and output, and the entropy of the channel input, as in
\begin{align}
H(X) =-{\rm tr} \bfrho_X\log \bfrho_X = -{\rm tr}\bfrho \log \bfrho_X = D(\bfrho\|\bfrho_X) +H(\bfrho).
\label{eq:entropic-uncertain1}
\end{align}

Using the monotonicity of relative entropy under the action of the channel $\mathcal{M}_Z$, we have
\begin{align}
D(\bfrho\|\bfrho_X) \ge D(\bfrho_Z\|\mathcal{M}_Z(\bfrho_X)),
\label{eq:entropic-uncertain2}
\end{align}
where
\begin{align}
D(\bfrho_Z\|\mathcal{M}_Z(\bfrho_X)) = - H(\bfrho_Z)  - {\rm tr} \bfrho_Z \log \mathcal{M}_Z(\bfrho_X),
\label{eq:entropic-uncertain3}
\end{align}
and 
\begin{align}
\mathcal{M}_Z(\bfrho_X) = \sum_{x,z} |z\rangle\langle z|x\rangle\langle x|\bfrho|x\rangle\langle x|z\rangle\langle z|.
\end{align}
Writing
\begin{align}
\log \mathcal{M}_Z(\bfrho_X) = \sum_{z} |z\rangle\log \left(\sum_x\langle z|x\rangle\langle x|\bfrho|x\rangle\langle x|z\rangle\right) \langle z|,
\end{align}
we see that 
\begin{align}
-{\rm tr}\bfrho_Z\log \mathcal{M}_Z(\bfrho_X) = -\sum_{z}\langle z|\bfrho|z\rangle \log \left(\sum_x\langle z|x\rangle\langle x|\bfrho|x\rangle\langle x|z\rangle\right).
\end{align}

Now, because $ - \log (\cdot)$ is a monotonically decreasing function, we have
\begin{align}
-\log \left(\sum_x\langle z|x\rangle\langle x|\bfrho|x\rangle\langle x|z\rangle\right) &\ge -\log\left(\max_{x,z} |\langle x |z\rangle |^2 \sum_x \langle x|\bfrho|x\rangle \right)\notag\\
&=\log\left(\frac{1}{c}\right),
\end{align}
and therefore
\begin{align}
-{\rm tr}\bfrho_Z\log \mathcal{M}_Z(\bfrho_X)\ge \log\left(\frac{1}{c}\right).
\label{eq:entropic-uncertain4}
\end{align}
Finally, putting together eq.(\ref{eq:entropic-uncertain1}), (\ref{eq:entropic-uncertain2}) (\ref{eq:entropic-uncertain3}), (\ref{eq:entropic-uncertain4}), we find
\begin{align}
&H(X) -H(\bfrho)=D(\bfrho\|\bfrho_X)  \ge D(\bfrho_Z\|\mathcal{M}_Z(\bfrho_X)) \notag\\
&= - H(Z)  - {\rm tr} \bfrho_Z \log \mathcal{M}_Z(\bfrho_X)\ge -H(Z)+\log\left(\frac{1}{c}\right),
\end{align}
which is equivalent to eq.(\ref{eq:entropic-uncertain}).

We say that two different bases $\{|x\rangle\}$, $\{|z\rangle\}$ for a $d$-dimensional Hilbert space are {\em mutually unbiased} if for all $x, z$
\begin{align}
|\langle x |z\rangle|^2 = \frac{1}{d};
\end{align}
thus, if we measure any $x$-basis state $|x\rangle$ in the $z$-basis, all $d$ outcomes are equally probable. For measurements in two mutually unbiased bases performed on a pure state, the entropic uncertainty relation becomes
\begin{align}
H(X) + H(Z) \ge \log d. 
\end{align}
Clearly this inequality is tight, as it is saturated by $x$-basis (or $z$-basis) states, for which $H(X)=0$ and $H(Z) = \log d$. 


\section{Quantum Source Coding}

What is the quantum analog of Shannon's source coding theorem?

Let's consider a long message consisting of $n$ letters, where each letter is a pure quantum state chosen by sampling from the ensemble
\begin{equation}
\{|\varphi(x)\rangle, p(x)\}.
\end{equation}
If the states of this ensemble are mutually orthogonal, then the message might as well be classical; the interesting quantum case is where the states are not orthogonal and therefore not perfectly distinguishable. 
The density operator realized by this ensemble is 
\begin{equation}
\bfrho = \sum_x p(x) |\varphi(x) \rangle \langle \varphi(x)|,
\end{equation}
and the entire $n$-letter message has the density operator
\begin{equation}
\bfrho^{\otimes n} = \bfrho \otimes \cdots \otimes \bfrho.
\end{equation}

How {\it redundant} is the quantum information in this message?  We would like to
devise a {\it quantum code} allowing us to compress the message to a
smaller Hilbert space, but without much compromising the fidelity of the message.
Perhaps we have a quantum memory device, and we know the {\it statistical} properties of the
recorded data; specifically, we know $\bfrho$.  We want to conserve space on our (very expensive) quantum hard drive by compressing the data.

The optimal compression that can be achieved was found by Schumacher.  As you might guess, the message can be compressed to a Hilbert 
space $\calH$ with
\begin{equation}
\dim \calH = 2^{n (H(\bfrho)+o(1))}
\end{equation}
with negligible loss of fidelity as $n\to \infty$, while errorless compression to dimension $2^{n (H(\bfrho)-\Omega(1))}$ is not possible. 
In this sense, the von Neumann entropy is the number of {\it qubits} of quantum
information carried per letter of the message. Compression is always possible unless $\bfrho$ is maximally mixed, just as we can always compress a classical message unless the information source is uniformly random. 
This result provides a precise operational interpretation for von Neumann entropy. 

Once Shannon's results are known and understood, the proof of Schumacher's compression
theorem is not difficult, as the mathematical ideas needed are very similar to those used by Shannon. But conceptually quantum compression is very different from its classical counterpart, as the imperfect distinguishability of nonorthogonal quantum states is the central idea. 

\subsection{Quantum compression: an example}

Before discussing Schumacher's quantum compression protocol in full
generality, it is helpful to consider a simple example.  Suppose that each
letter is a  single qubit drawn from the ensemble
\begin{eqnarray}
|\uparrow_z\rangle &= \begin{pmatrix}1 \\ 0\end{pmatrix}, &\quad p = {1\over 2},\\
|\uparrow_x\rangle &= \begin{pmatrix}\frac{1}{\sqrt{2}} \\ \frac{1}{\sqrt{2}}\end{pmatrix}, &\quad p = {1\over
2}, 
\end{eqnarray}
so that the density operator of each letter is
\begin{align}
\bfrho &= {1\over 2} |\uparrow_z\rangle \langle \uparrow_z| + {1\over 2}
|\uparrow_x\rangle\langle \uparrow_x|\notag \\
&= {1\over 2} \begin{pmatrix}1&0\\0&0\end{pmatrix} + {1\over 2} \begin{pmatrix}{1\over 2}&
{1\over 2}\\ {1\over 2} &{1\over 2}\end{pmatrix} = \begin{pmatrix}{3\over 4} &
{1\over4}\\{1\over 4} &{1\over 4}\end{pmatrix}.
\end{align}
As is obvious from symmetry, the eigenstates of $\bfrho$ are qubits oriented up
and down along the axis $\hat n = {1\over \sqrt{2}} (\hat x + \hat z)$,
\begin{align}
|0'\rangle &\equiv |\uparrow_{\hat n} \rangle = \begin{pmatrix}\cos {\pi\over 8}\\
\sin{\pi\over 8}\end{pmatrix},\notag \\
|1'\rangle &\equiv |\downarrow_{\hat n} \rangle = \begin{pmatrix}\sin {\pi\over8}\\-\cos {\pi\over 8}\end{pmatrix};
\end{align}
the eigenvalues are
\begin{align}
\lambda(0') & = {1\over 2} + {1\over 2\sqrt{2}} = \cos^2 {\pi\over 8},\notag
\\
\lambda(1') & = {1\over 2} - {1\over 2\sqrt{2}} = \sin^2 {\pi\over 8};
\end{align}
evidently $\lambda(0') + \lambda(1') = 1$ and $\lambda(0') \lambda (1') =
{1\over 8} = {\rm det} \bfrho$.  The eigenstate $|0'\rangle$ has equal (and
relatively large) overlap with both signal states
\begin{equation}
|\langle 0'|\uparrow_z\rangle|^2 = |\langle 0'|\uparrow_x\rangle|^2 = \cos^2
{\pi\over 8} = .8535,
\end{equation}
while $|1'\rangle$ has equal (and relatively small) overlap with both,
\begin{equation}
|\langle 1'|\uparrow_z\rangle|^2 = |\langle 1'|\uparrow_x\rangle|^2 = \sin^2
{\pi\over 8} = .1465.
\end{equation}
Thus if we don't know whether $|\uparrow_z\rangle$  or $|\uparrow_x\rangle$ was
sent, the best guess we can make  is $|\psi\rangle = |0'\rangle$.  This guess
has the maximal {\it fidelity} with $\bfrho$
\begin{equation}
F = {1\over 2} |\langle\uparrow_z|\psi\rangle|^2 + {1\over 2}
|\langle\uparrow_x|\psi\rangle|^2,
\end{equation}
among all possible single-qubit states $|\psi\rangle$ ($F$ = .8535).

Now imagine that Alice needs to send three letters to Bob, but she can afford
to send only two qubits.  Still, she
wants Bob to reconstruct her state with the highest possible fidelity.
She could send Bob two of her three letters, and ask Bob to guess $|0'\rangle$
for the third.  Then Bob receives two letters with perfect fidelity, and his guess has $F =
.8535$ for the third; hence $F = .8535$ overall.  But is there a more clever
procedure that achieves higher fidelity?

Yes, there is.  By diagonalizing $\bfrho$, we decomposed
the Hilbert space of a single qubit into a ``likely'' one-dimensional subspace
(spanned by $|0'\rangle$) and an ``unlikely'' one-dimensional subspace (spanned
by $|1'\rangle$).  In a similar way we can decompose the Hilbert space of three
qubits into likely and unlikely subspaces.  If
$|\psi\rangle=|\psi_1\rangle\otimes |\psi_2\rangle\otimes |\psi_3\rangle$ is any signal state,
where the state of each qubit is either $|\uparrow_z\rangle$ or
$|\uparrow_x\rangle$, we have
\begin{align}
|\langle 0'0'0'|\psi\rangle|^2 &= \cos^6 \left({\pi\over 8}\right) =
.6219,\notag \\
|\langle 0'0'1'|\psi\rangle|^2 &= |\langle 0'1'0'|\psi\rangle|^2 = |\langle
1'0'0'|\psi\rangle|^2 = \cos^4 \left({\pi\over 8}\right) \sin^2 \left({\pi\over
8}\right) = .1067,\notag \\
|\langle 0'1'1'|\psi\rangle|^2 &= |\langle 1'0'1'|\psi\rangle|^2 = |\langle
1'1'0'|\psi\rangle|^2 = \cos^2 \left({\pi\over 8}\right) \sin^4 \left({\pi\over
8}\right) = .0183,\notag \\
|\langle 1'1'1'|\psi\rangle|^2 &= \sin^6 \left({\pi\over 8}\right) = .0031.
\end{align}
Thus, we may decompose the space into the likely subspace $\Lambda$ spanned by
$\{|0'0'0'\rangle,|0'0'1'\rangle,|0'1'0'\rangle,|1'0'0'\rangle\}$, and its
orthogonal complement $\Lambda^\perp$.  If we make an incomplete orthogonal measurement
that projects a signal state onto $\Lambda$ or $\Lambda^\perp$, the probability
of projecting onto the likely subspace $\Lambda$ is
\begin{equation}
p_{\rm likely} = .6219 + 3 (.1067) = . 9419,
\end{equation}
while the probability of projecting onto the unlikely subspace is
\begin{equation}
p_{\rm unlikely} = 3 (.0183) + .0031 = .0581.
\end{equation}

To perform this measurement, Alice could, for example, first apply a
unitary transformation $\bfu$ that rotates the four high-probability basis
states to
\begin{equation}
|\cdot\rangle\otimes |\cdot \rangle\otimes |0\rangle,
\end{equation}
and the four low-probability basis states to
\begin{equation}
|\cdot\rangle \otimes |\cdot\rangle \otimes |1\rangle;
\end{equation}
then Alice measures the third qubit to perform the projection.  If the
outcome is $|0\rangle$, then Alice's input state has in effect been projected
onto $\Lambda$.  She sends the remaining two unmeasured qubits to Bob.  When
Bob receives this compressed two-qubit state $|\psi_{\rm comp}\rangle$, he
decompresses it by appending $|0\rangle$ and applying $\bfu^{-1}$, obtaining
\begin{equation}
|\psi'\rangle = \bfu^{-1} (|\psi_{\rm comp}\rangle \otimes|0\rangle).
\end{equation}
If Alice's measurement of the third qubit yields $|1\rangle$, she has projected
her input state onto the low-probability subspace $\Lambda^\perp$.  In this
event, the best thing she can do is send the state that Bob will decompress to
the most likely state $|0'0'0'\rangle$ -- that is, she sends the state
$|\psi_{\rm comp}\rangle$ such that
\begin{equation}
|\psi'\rangle = \bfu^{-1} (|\psi_{\rm comp}\rangle\otimes|0\rangle) =
|0'0'0'\rangle.
\end{equation}
Thus, if Alice encodes the three-qubit signal state $|\psi\rangle$, sends two
qubits to Bob, and Bob decodes as just described, then Bob obtains the state
$\bfrho'$
\begin{equation}
|\psi\rangle\langle\psi| \rightarrow
\bfrho' = \bfe |\psi\rangle\langle \psi|\bfe + |0'0'0'\rangle \langle \psi|(\bfI -
\bfe) |\psi\rangle\langle 0'0'0'|,
\end{equation}
where $\bfe$ is the projection onto $\Lambda$.  The fidelity achieved by this
procedure is
\begin{align}
F &= \langle\psi|\bfrho'|\psi\rangle = (\langle\psi| \bfe | \psi\rangle)^2 +
(\langle \psi| (\bfI- \bfe) |\psi\rangle) (\langle \psi |0'0'0'\rangle)^2\notag
\\
&= (.9419)^2 + (.0581)(.6219) = .9234.
\end{align}
This is indeed better than the naive procedure of sending two of the three
qubits each with perfect fidelity.

As we consider longer messages with more letters, the fidelity of the
compression improves, as long as we don't try to compress too much.  The Von-Neumann entropy of the one-qubit ensemble is
\begin{equation}
H(\bfrho) = H \left(\cos^2 {\pi\over 8}\right) = .60088 \ldots
\end{equation}
Therefore, according to Schumacher's theorem, we can shorten a long message by
the factor, say, .6009, and still achieve very good fidelity.

\subsection{Schumacher compression in general}
\label{subsec:schumacher-compression}

The key to Shannon's noiseless coding theorem is that we can code the typical
sequences and ignore the rest, without much loss of fidelity.  To quantify the
compressibility of quantum information, we promote the notion of a typical {\it
sequence} to that of a typical {\it subspace}.  The key to Schumacher's
noiseless quantum coding theorem is that we can code the typical subspace and
ignore its orthogonal complement, without much loss of fidelity.

We consider a message of $n$ letters where each letter is a pure quantum state
drawn from the ensemble $\{|\varphi(x)\rangle, p(x)\}$, so that the density
operator of a single letter is
\begin{equation}
\bfrho = \sum_x p(x) |\varphi(x)\rangle\langle \varphi(x)|.
\end{equation}
Since the letters are drawn independently,  the density operator of
the entire message is
\begin{equation}
\bfrho^{\otimes n} \equiv \bfrho \otimes \cdots \otimes \bfrho.
\end{equation}
We claim that, for $n$ large, this density matrix has nearly all of its
support on a subspace of the full Hilbert space of the messages, where the
dimension of this subspace asymptotically approaches $2^{nH(\bfrho)}$.

This claim follows directly from the corresponding classical statement, for we may consider $\bfrho$ to be realized by an ensemble of orthonormal pure states, its eigenstates, where the probability assigned to each eigenstate is the corresponding eigenvalue.  In this basis our source of quantum information is effectively
classical, producing messages which are tensor products of $\bfrho$ eigenstates,
each with a probability given by the product of the corresponding eigenvalues.
For a specified $n$ and $\delta$, define the {\em $\delta$-typical subspace} $\Lambda$ as the
space spanned by the eigenvectors of $\bfrho^{\otimes n}$ with eigenvalues $\lambda$
satisfying
\begin{equation}
2^{-n(H-\delta)} \geq \lambda \geq 2^{-n(H +\delta)}.
\end{equation}
Borrowing directly from Shannon's argument, we infer that for any $\delta, \varepsilon
> 0$ and $n$ sufficiently large, the sum of the eigenvalues of $\bfrho^{\otimes n}$ that
obey this condition satisfies
\begin{equation}
{\rm tr} (\bfrho^{\otimes n} \bfe) \ge 1 - \varepsilon,
\end{equation}
where $\bfe$ denotes the projection onto the typical subspace $\Lambda$, and the number
$\dim (\Lambda)$ of such eigenvalues satisfies
\begin{equation}
2^{n(H + \delta)} \geq \dim (\Lambda) \geq (1 - \varepsilon) 2^{n(H -
\delta)}.
\end{equation}
Our coding strategy is to send states in the typical subspace faithfully. We can make a measurement that projects the input message onto
either $\Lambda$ or $\Lambda^\perp$; the outcome will be $\Lambda$ with
probability $p_\Lambda = {\rm tr} (\bfrho^{\otimes n} \bfe) \ge 1 - \varepsilon$.  In that
event, the projected state is coded and sent.  Asymptotically, the probability
of the other outcome becomes negligible, so it matters little what we do in
that case.

The coding of the projected state merely packages it so it can be carried by a
minimal number of qubits.  For example, we apply a unitary change of basis
$\bfu$ that takes each state $|\psi_{\rm typ} \rangle$ in $\Lambda$ to a state
of the form
\begin{equation}
\bfU |\psi_{\rm typ} \rangle = |\psi_{\rm comp}\rangle \otimes |0_{\rm rest}\rangle,
\label{eq:U-compression}
\end{equation}
where $|\psi_{\rm comp}\rangle$ is a state of $n(H + \delta)$ qubits, and
$|0_{\rm rest}\rangle$ denotes the state $|0\rangle \otimes \ldots \otimes
|0\rangle$ of the remaining qubits.  Alice sends $|\psi_{\rm comp}\rangle$ to
Bob, who decodes by appending $|0_{\rm rest}\rangle$ and applying $\bfu^{-1}$.

Suppose that
\begin{equation}
|\varphi(\vec x)\rangle = |\varphi(x_1)\rangle\otimes  \ldots \otimes
|\varphi(x_n)\rangle,
\end{equation}
denotes any one of the $n$-letter pure state messages that might be sent.
After coding, transmission, and decoding are carried out as just described, Bob
has reconstructed a state
\begin{align}
|\varphi(\vec x)\rangle\langle \varphi(\vec x)|\mapsto \bfrho'(\vec x) &= \bfe
|\varphi(\vec x)\rangle\langle \varphi(\vec x)|\bfe\notag \\
&+ \bfrho_{\rm Junk}(\vec x) \langle\varphi(\vec x)|({\bfI} - \bfe)|\varphi(\vec x)\rangle,
\end{align}
where $\bfrho_{\rm Junk}(\vec x)$ is the state we choose to send if the
measurement yields the outcome $\Lambda^\perp$.  What can we say about the
fidelity of this procedure?

The fidelity varies from message to message, so we consider the fidelity averaged over the ensemble of
possible messages:
\begin{align}
\bar F &= \sum_{\vec x} p({\vec x}) \langle\varphi({\vec x}) | \bfrho'({\vec x}) |\varphi({\vec x})\rangle\notag \\
&= \sum_{\vec x} p({\vec x}) \langle \varphi({\vec x})|\bfe |\varphi({\vec x})\rangle \langle \varphi({\vec x})|\bfe |
\varphi({\vec x})\rangle \notag\\
&+ \sum_{\vec x} p({\vec x}) \langle\varphi({\vec x})| \bfrho_{\rm
Junk}({\vec x})|\varphi({\vec x})\rangle\langle\varphi({\vec x})|{\bfI} - \bfe |\varphi({\vec x})\rangle\notag
\\
&\geq \sum_{\vec x} p({\vec x}) \langle\varphi({\vec x})|\bfe |\varphi({\vec x})\rangle^2,
\end{align}
where the last inequality holds because the ``Junk'' term is nonnegative.
Since any real number $z$ satisfies
\begin{equation}
(z - 1)^2 \geq 0, ~~{\rm or}~~ z^2 \geq 2z - 1,
\end{equation}
we have (setting $z = \langle \varphi({\vec x})|\bfe |\varphi({\vec x})\rangle$)
\begin{equation}
\langle \varphi({\vec x})|\bfe |\varphi({\vec x})\rangle^2 \geq 2 \langle\varphi({\vec x}) |\bfe
|\varphi({\vec x})\rangle - 1,
\end{equation}
and hence
\begin{align}
\bar F &\geq \sum_{\vec x} p({\vec x}) (2 \langle \varphi({\vec x})|\bfe|\varphi({\vec x})\rangle - 1)\notag \\
&= 2 ~{\rm tr} (\bfrho^{\otimes n} \bfe) - 1 \ge 2 (1 - \varepsilon) - 1 = 1 -
2\varepsilon.
\end{align}
Since $\varepsilon$ and $\delta$ can be as small as we please, we have shown  that it is possible to compress the message to 
$n (H + o(1))$ qubits, while achieving an average fidelity that becomes
arbitrarily good as $n$ gets large.

Is further compression possible?
Let us suppose that Bob will decode the message $\bfrho_{\rm comp}({\vec x})$ that he
receives by appending qubits and applying a unitary transformation $\bfu^{-1}$,
obtaining
\begin{equation}
\bfrho'({\vec x}) = \bfu^{-1} (\bfrho_{\rm comp}({\vec x}) \otimes |0\rangle\langle0|)\bfu
\end{equation}
(``unitary decoding''), and suppose that $\bfrho_{\rm comp}({\vec x})$ has been compressed
to $n(H -\delta')$ qubits.  Then, no matter how the input messages have been
encoded, the decoded messages are all contained in a subspace $\Lambda'$ of
Bob's Hilbert space with ${\rm dim}(\Lambda')= 2^{n(H-\delta')}$.  

If the input message is $|\varphi({\vec x})\rangle$, then the density operator reconstructed by Bob can be diagonalized as
\begin{equation}
\bfrho'({\vec x})= \sum_{a_{{\vec x}}} |a_{\vec x}\rangle \lambda_{a_{\vec x}} \langle a_{\vec x}|,
\end{equation}
where the $|a_{\vec x}\rangle$'s are mutually orthogonal states in $\Lambda'$.  The
fidelity of the reconstructed message is
\begin{align}
F({\vec x}) &= \langle \varphi({\vec x})|\bfrho'({\vec x})|\varphi({\vec x})\rangle\notag \\
&= \sum_{a_{{\vec x}}} \lambda_{a_{{\vec x}}} \langle\varphi({\vec x})|a_{\vec x}\rangle\langle
a_{\vec x}|\varphi({\vec x})\rangle\notag \\
&\leq \sum_{a_{{\vec x}}} \langle\varphi({\vec x})|a_{\vec x}\rangle\langle a_{\vec x}|\varphi({\vec x})\rangle \leq
\langle\varphi({\vec x})|\bfe'|\varphi({\vec x})\rangle,
\end{align}
where $\bfe'$ denotes the orthogonal projection onto the subspace $\Lambda'$.
The average fidelity therefore obeys
\begin{equation}
\bar F = \sum_{\vec x} p({\vec x}) F({\vec x}) \leq \sum_{\vec x} p({\vec x}) \langle \varphi({\vec x})|\bfe' |\varphi({\vec x})\rangle =
{\rm tr} (\bfrho^{\otimes n} \bfe').
\end{equation}
But, according to the Ky Fan dominance principle discussed in \S\ref{subsec:mixing-measurement-entropy}, since $\bfe'$ projects onto a space of dimension $2^{n(H-\delta')}, {\rm
tr}(\bfrho^{\otimes n} \bfe')$ can be no larger than the sum of the $2^{n(H-\delta')}$
largest eigenvalues of $\bfrho^{\otimes n}$. The $\delta$-typical eigenvalues of $\bfrho^{\otimes n}$ are no smaller than $2^{-n(H - \delta)}$, so the sum of the $2^{n(H-\delta')}$ largest eigenvalues can be bounded above:
\begin{align}
{\rm tr} (\bfrho^{\otimes n} \bfE')\le 2^{n(H-\delta')}2^{-n(H - \delta)} + \varepsilon = 2^{-n(\delta'-\delta)} + \varepsilon,
\end{align}
where the $+~\varepsilon$ accounts for the contribution from the atypical eigenvalues. Since we may choose $\varepsilon$ and $\delta$ as small as we please for sufficiently large $n$, we conclude that the average fidelity $\bar F$ gets small as $n\to\infty$ if we compress to $H(\bfrho) - \Omega(1)$ qubits per letter. 
We find, then, that $H(\bfrho)$ qubits per letter is the optimal compression
of the quantum information that can be achieved if we are to obtain good
fidelity as $n$ goes to infinity.  This is Schumacher's quantum source
coding theorem.

The above argument applies to any conceivable encoding scheme, but only to a
restricted class of decoding schemes, unitary decodings.  The extension of the argument to general decoding schemes is sketched in \S\ref{subsec:monotonicity-chi}. The conclusion is
the same.  The point is that $n(H-\delta)$ qubits are too few to faithfully
encode the typical subspace.

There is another useful way to think about Schumacher's quantum compression protocol. Suppose that Alice's density operator $\bfrho_A^{\otimes n}$ has a {\em purification} $|\psi\rangle_{RA}$ which Alice shares with Robert. Alice wants to convey her share of $|\psi\rangle_{RA}$ to Bob with high fidelity, sending as few qubits to Bob as possible. To accomplish this task, Alice can use the same procedure as described above, attempting to compress the state of $A$ by projecting onto its typical subspace $\Lambda$. Alice's projection succeeds with probability
\begin{align} 
P(\bfE) = \langle \psi|\bfI\otimes \bfE|\psi\rangle = {\rm tr}\left(\bfrho^{\otimes n} \bfE\right)\ge 1-\varepsilon,
\end{align}
where $\bfE$ projects onto $\Lambda$, and when successful prepares the state
\begin{align}
\frac{\left(\bfI\otimes \bfE\right) |\psi\rangle}{\sqrt{P(\bfE)}}.
\end{align}
Therefore, after Bob decompresses, the state he shares with Robert has fidelity $F_e$ with $|\psi\rangle$ satisfying
\begin{align}
F_e 
\ge \langle \psi|\bfI\otimes \bfE |\psi\rangle \langle \psi|\bfI\otimes \bfE|\psi\rangle = \left({\rm tr}\left(\bfrho^{\otimes n} \bfE\right)\right)^2= P(\bfE)^2 \ge (1-\varepsilon)^2\ge 1-2\varepsilon.
\end{align}
We conclude that Alice can transfer her share of the pure state $|\psi\rangle_{RA}$ to Bob by sending $nH(\bfrho) +o(n)$ qubits, achieving arbitrarily good {\em entanglement fidelity} $F_e$ as $n\to \infty$. In \S\ref{subsec:mother} we'll derive a more general version of this result.

To summarize, there is a close analogy between Shannon's classical source coding
theorem and Schumacher's quantum source coding theorem.  In the classical
case, nearly all long messages are typical sequences, so we can code only these
and still have a small probability of error.  In the quantum case, nearly all
long messages have nearly perfect overlap with the typical subspace, so we can
code only the typical subspace and still achieve good fidelity.

Alternatively, Alice could send classical information to Bob, the string
$x_1x_2\cdots x_n$, and Bob
could follow these classical instructions to reconstruct Alice's state $|\varphi(x_1)\rangle\otimes  \ldots \otimes|\varphi(x_n)\rangle$.
By this means, they could achieve high-fidelity compression to $H(X)+o(1)$ bits --- or
qubits --- per letter, where $X$ is the classical ensemble $\{x,p(x)\}$. But if $\{|\varphi(x)\rangle, p(x)\}$ is an ensemble of {\it
nonorthogonal} pure states, this classically achievable amount of compression is not optimal; some of
the classical information about the preparation of the state is
redundant, because the nonorthogonal states cannot be perfectly distinguished.
Schumacher coding goes  further, achieving optimal compression to
$H(\bfrho)+o(1)$ qubits per letter.
Quantum compression packages the message more efficiently than classical compression, but at a price --- Bob receives the quantum state Alice intended to send, but Bob doesn't know what he has.  In contrast to
the classical case, Bob can't fully decipher Alice's quantum message accurately.
An attempt to read the message will unavoidably
disturb it.

\section{Entanglement Concentration and Dilution}
\label{sec:concentration-dilution}

Any bipartite pure state that is not a product state is entangled. But {\em how} entangled? Can we compare two states and say that one is more entangled than the other?

For example, consider the two bipartite states 
\begin{align}
&|\phi^+\rangle = {1\over\sqrt{2}} (|00\rangle + | 11\rangle),\notag\\
&|\psi\rangle = \sqrt{\frac{2}{3}}~ |00\rangle + {1\over \sqrt{6}} ~|11\rangle +
{1\over \sqrt{6}} ~|22\rangle.
\end{align}
$|\phi^+\rangle$ is a maximally entangled state of two qubits, while $|\psi\rangle$ is a  {\it partially} entangled state of two {\it qutrits}.
Which is more entangled?

It is not immediately clear that the question has a meaningful answer.  Why
should it be possible to find an unambiguous way of ordering all bipartite pure states  according to their degree of
entanglement?  Can we compare a pair of qutrits with a pair of qubits any more
than we can compare apples and oranges?

A crucial feature of entanglement is that it cannot be created by local
operations and classical communication (LOCC).  In particular, if Alice and Bob share a bipartite pure state, its Schmidt number does not increase if Alice or Bob performs a unitary transformation on her/his share of the state, nor if Alice or Bob measures her/his share, even if Alice and Bob exchange classical
messages about their actions and measurement outcomes.  Therefore, any quantitative measure of entanglement should have the property that LOCC cannot increase it, and it should also vanish for an unentangled product state. An obvious candidate is the Schmidt number, but on reflection it
does not seem very satisfactory.  Consider
\begin{equation}
|\psi_\varepsilon \rangle = \sqrt{1-2|\varepsilon|^2}~|00\rangle + \varepsilon |11\rangle + \varepsilon
|22\rangle,
\end{equation}
which has Schmidt number 3 for any $|\varepsilon| > 0$.  Do we really want to say that
$|\psi_\varepsilon \rangle$ is ``more entangled'' than $|\phi^+\rangle$?
Entanglement, after all, can be regarded as a resource --- we might plan to use
it for teleportation, for example --- and it seems clear that $|\psi_\varepsilon\rangle$ (for $|\varepsilon|\ll 1$) is
a less valuable resource than $|\phi^+\rangle$.

It turns out, though, that there is a natural and useful way to quantify the
entanglement of any bipartite pure state.  To compare two states, we use LOCC to convert both states to a common currency that can be
compared directly.  The common currency is {\em maximal} entanglement, and the amount of shared entanglement can be expressed in units of Bell pairs (maximally entangled two-qubit states), also called {\em ebits} of entanglement.

To quantify the entanglement of a particular bipartite pure state,
$|\psi\rangle_{AB}$, imagine preparing $n$ identical
copies of that state.  Alice and Bob share a large supply of maximally entangled
{\it Bell pairs}.  Using LOCC, they are to convert $k$ 
Bell pairs ($|\phi^+\rangle_{AB})^{\otimes k}$) to $n$ high-fidelity copies of the desired state
($|\psi\rangle_{AB})^{\otimes n}$).  What is the minimum number $k_{\rm min}$ of Bell
pairs with which they can perform this task? 

To obtain a precise answer, we consider the {\em asymptotic} setting, requiring arbitrarily high-fidelity conversion in the limit of large $n$. We say that a rate $R$ of conversion from $|\phi^+\rangle$ to $|\psi\rangle$ is asymptotically achievable if for any $\varepsilon, \delta > 0$, there is an LOCC protocol with
\begin{align} 
\frac{k}{n} \le R + \delta,
\end{align}
which prepares the target state $|\psi^+\rangle^{\otimes n}$ with fidelity $F\ge 1-\varepsilon$. We define the {\em entanglement cost} $E_C$ of $|\psi\rangle$ as the infimum of achievable conversion rates:
\begin{equation}
E_C(|\psi\rangle) :=\inf\left\{{\rm achievable} ~ {\rm rate}~{\rm for ~creating}~ |\psi\rangle~{\rm from~Bell~pairs}\right\}.
\end{equation}
Asymptotically, we can create many copies of $|\psi\rangle$ by consuming $E_C$ Bell pairs per copy. 

Now imagine that $n$ copies of $|\psi\rangle_{AB}$ are already shared by Alice and Bob. Using LOCC, Alice and Bob are to convert $(|\psi\rangle_{AB})^{\otimes n}$ back to the standard currency: $k'$ Bell pairs $|\phi^+\rangle_{AB}^{\otimes k'}$.
What is the maximum number $k'_{\rm max}$ of Bell pairs they can extract from $|\psi\rangle_{AB}^{\otimes n}$? In this case we say that a rate $R'$ of conversion from $|\psi\rangle$ to $|\phi^+\rangle$ is asymptotically achievable if for any $\varepsilon, \delta > 0$, there is an LOCC protocol with
\begin{align} 
\frac{k'}{n} \ge R' - \delta,
\end{align}
which prepares the target state $|\phi^+\rangle^{\otimes k'}$ with fidelity $F\ge 1-\varepsilon$. We define the {\em distillable entanglement} $E_D$ of $|\psi\rangle$ as the supremum of achievable conversion rates:
\begin{equation}
E_D(|\psi\rangle) :=\sup\left\{{\rm achievable} ~ {\rm rate}~ {\rm for~distilling~Bell~pairs~from~} |\psi\rangle\right\}.
\end{equation}
Asymptotically, we can convert many copies of $|\psi\rangle$ to Bell pairs, obtaining $E_D$ Bell pairs per copy of $|\psi\rangle$ consumed. 

Since it is an inviolable principle that LOCC cannot create
entanglement, it is certain that
\begin{equation}
E_D(|\psi\rangle) \le E_C(|\psi\rangle);
\end{equation}
otherwise Alice and Bob could increase their number of shared Bell pairs by converting them to copies of $|\psi\rangle$ and then back to Bell pairs. In fact the entanglement cost and distillable entanglement are {\em equal} for bipartite pure states. (The story is more complicated for bipartite mixed states; see \S\ref{sec:quant-mixed-entanglement}.) Therefore, for pure states at least we may drop the subscript, using $E(|\psi\rangle)$ to denote the {\em entanglement} of $|\psi\rangle$. We don't need to distinguish between entanglement cost and distillable entanglement because conversion of entanglement from one form to another is an asymptotically {\em reversible} process. $E$ quantifies both what we have to pay in Bell pairs to create $|\psi\rangle$, and the value of $|\psi\rangle$ in Bell pairs for performing tasks like quantum teleportation which consume entanglement. 

But what is the value of $E(|\psi\rangle_{AB})$? Perhaps you can guess --- it is
\begin{equation}
E(|\psi\rangle_{AB}) = H(\bfrho_A) = H(\bfrho_B),
\end{equation}
the von Neumann entropy of Alice's density operator $\bfrho_A$
(or equivalently Bob's density operator $\bfrho_B$).  This is clearly the right answer in the
case where $|\psi\rangle_{AB}$ is a product of $k$ Bell pairs.  In that case
$\bfrho_A$ (or $\bfrho_B$) is ${1\over 2} {\bfI}$ for each qubit in Alice's
possession
\begin{equation}
\bfrho_A = \left(\frac{1}{2} {\bfI}\right)^{\otimes k},
\end{equation}
and
\begin{equation}
H(\bfrho_A) = k~H \left({1\over 2} {\bfI}\right) = k.
\end{equation}
How do we see that $E = H(\bfrho_A)$ is the right answer for any bipartite
pure state?

Though it is perfectly fine to use Bell pairs as the common currency for comparing bipartite entangled states, in the asymptotic setting it is simpler and more natural to allow fractions of a Bell pair, which is what we'll do here. That is, we'll consider a maximally entangled state of two $d$-dimensional systems to be $\log_2 d$ Bell pairs, even if $d$ is not a power of two. So our goal will be to show that Alice and Bob can use LOCC to convert shared maximal entanglement of systems with dimension $d = 2^{n(H(\bfrho_A) + \delta)}$ into $n$ copies of $|\psi\rangle$, for any positive $\delta$ and with arbitrarily good fidelity as $n\to \infty$, and conversely that Alice and Bob can use LOCC to convert $n$ copies of $|\psi\rangle$ into a shared maximally entangled state of $d$-dimensional systems with arbitrarily good fidelity, where $d = 2^{n(H(\bfrho_A) - \delta)}$. This suffices to demonstrate that $E_C(|\psi\rangle) = E_D(|\psi\rangle) = H(\bfrho_A$).

First let's see that if Alice and Bob share $k = n (H(\bfrho_A) + \delta)$
Bell pairs, then they can prepare $|\psi\rangle_{AB}^{\otimes n}$
with high fidelity using LOCC.  They perform this task, called {\em entanglement dilution}, by combining quantum
teleportation with Schumacher compression. To get started, Alice locally creates $n$ copies of 
$|\psi\rangle_{AC}$, where $A$ and $C$ are systems she controls in her laboratory. Next she wishes to teleport the $C^n$ share of these copies to Bob, but to minimize the consumption of Bell pairs, she should compress $C^n$ before teleporting it.

If $A$ and $C$ are $d$-dimensional, then the bipartite state $|\psi\rangle_{AC}$ can be  expressed in terms of its Schmidt basis as
\begin{equation}
|\psi\rangle_{AC} = \sqrt{p_0}~ |00\rangle + \sqrt{p_1}~|11\rangle +
\ldots + \sqrt{p_{d{-}1}}~ |d{-}1,d{-}1\rangle,
\end{equation}
and $n$ copies of the state can be expressed as
\begin{align}
|\psi\rangle_{AC}^{\otimes n}&= \sum_{x_1, \dots, x_n=0}^{d-1} \sqrt{p(x_1)\dots p(x_n)}~ |x_1x_2\dots x_n\rangle_{A^n}\otimes |x_1x_2\dots x_n\rangle_{C^n}\notag\\
 &= \sum_{\vec x} \sqrt{p(\vec x)}~|\vec x\rangle_{A^n}\otimes |\vec x\rangle_{C^n},
\end{align}
where $\sum_{\vec x}p(\vec x) = 1$.
If Alice attempts to project onto the $\delta$-typical subspace of $C^n$, she succeeds with high probability
\begin{equation}
P = \sum_{\delta\text{-}{\rm typical}~\vec x} p(\vec x) \ge 1 -\varepsilon
\end{equation} 
and when successful prepares the post-measurement state
\begin{align}
|\Psi\rangle_{A^nC^n} &= P^{-1/2}\sum_{\delta\text{-}{\rm typical}~\vec x} \sqrt{p(\vec x)}~|\vec x\rangle_{A^n}\otimes |\vec x\rangle_{C^n} ,
\end{align}
such that 
\begin{equation}
\langle \Psi| \psi^{\otimes n}\rangle = P^{-1/2} \sum_{\delta\text{-}{\rm typical}~\vec x} p(\vec x) = \sqrt{P} \ge \sqrt{1 -\varepsilon}.
\end{equation}
Since the typical subspace has dimension at most $2^{n(H(\bfrho) + \delta)}$, Alice can teleport the $C^n$ half of $|\Psi\rangle$ to Bob with perfect fidelity using no more than $n(H(\bfrho) + \delta)$ Bell pairs shared by Alice and Bob. The teleportation uses LOCC: Alice's entangled measurement, classical communication from Alice to Bob to convey the measurement outcome, and Bob's unitary transformation conditioned on the outcome. Finally, after the teleportation, Bob decompresses, so that Alice and Bob share a state which has high fidelity with $|\psi\rangle_{AB}^{\otimes n}$. This protocol demonstrates that the entanglement cost $E_C$ of $|\psi\rangle$ is not more than $H(\bfrho_A)$.


Now consider the distillable entanglement $E_D$. Suppose Alice and Bob share the state
$|\psi\rangle_{AB}^{\otimes n}$.  Since $|\psi\rangle_{AB}$ is, in general, a {\it
partially} entangled state, the entanglement that Alice and Bob share is in a diluted
form.  They wish to {\it concentrate} their shared entanglement, squeezing it
down to the smallest possible Hilbert space; that is, they want to convert it
to maximally-entangled pairs.  We will show that Alice and Bob can ``distill''
at least
\begin{equation}
k' = n (H(\bfrho_A) - \delta)
\end{equation}
Bell pairs from $|\psi\rangle_{AB}^{\otimes n}$, with high likelihood of success.

To illustrate the concentration of entanglement, imagine that Alice and Bob
have $n$ copies of the two-qubit state $|\psi\rangle$, which is 
\begin{equation}
|\psi(p)\rangle = \sqrt{1-p}~ |00\rangle + \sqrt{p} ~|11\rangle,
\end{equation}
where $0\le p \le 1$, when expressed in its Schmidt basis.
That is, Alice and Bob
share the state
\begin{equation}\label{n_conc}
|\psi(p)\rangle^{\otimes n} = (\sqrt{1-p} ~|00\rangle + \sqrt{p}~ |11\rangle)^{\otimes n}.
\end{equation}
When we expand this state in the $\{|0\rangle, |1\rangle \}$ basis, we find $2^n$ terms, in each of which Alice and Bob hold exactly the same binary string of length $n$.

Now suppose Alice (or Bob) performs a local measurement on her (his) $n$ qubits, measuring 
the {\it total} spin along the $z$-axis
\begin{equation}
\bfsigma_{3}^{{\rm total}} = \sum_{i = 1}^n \bfsigma_{3}^{(i)}.
\end{equation}
Equivalently, the measurement determines the {\em Hamming weight} of Alice's $n$ qubits, the number of $|1\rangle$'s in Alice's $n$-bit string; that is,  the number of spins pointing up. 

In the expansion of $|\psi(p)\rangle^{\otimes n}$ there are $\binom{n}{m}$ terms in which Alice's string has Hamming weight $m$, each occurring with the same amplitude:  $\left(1-p\right)^{(n-m)/2}p^{m/2}$. Hence the probability that Alice's measurement finds Hamming weight $m$ is
\begin{equation}
p(m) = \binom{n}{m} (1-p)^{n-m} p^m.
\end{equation}
Furthermore, because Alice is careful not to acquire any additional information besides the Hamming weight when she conducts the measurement, by measuring the Hamming weight $m$ she prepares a
{\it uniform} superposition of all $\binom{n}{m}$ strings with $m$ up spins. Because Alice and Bob have perfectly correlated strings, if Bob were to measure the Hamming weight of his qubits he would find the same outcome as Alice. 
Alternatively, Alice could report
her outcome to Bob in a classical message, saving Bob the trouble of doing
the measurement himself. Thus, Alice and Bob share a maximally entangled state 
\begin{equation}
\sum_{i=1}^D|i\rangle_A\otimes |i\rangle_B,
\end{equation}
where the sum runs over the $D = \binom{n}{m}$ strings with Hamming weight $m$. 


For $n$ large the binomial distribution $\{p(m)\}$ approaches a sharply peaked function of $m$ with mean $\mu = np$ and variance $\sigma^2 = n p (1-p)$. Hence the probability of a large deviation from the mean,
\begin{equation}
|m - np| = \Omega(n) ,
\end{equation}
is $\exp\left(- \Omega(n)\right)$. Using Stirling's approximation, it then follows that 
\begin{equation}
2^{n(H(p) - o(1))} \le D \le 2^{n(H(p) + o(1))}.
\end{equation}
with probability approaching one as $n\to\infty$, 
where $H(p) = - p\log_2 p - (1 - p)\log_2 (1 - p)$ is the entropy function.  Thus with high probability Alice and Bob share a maximally entangled state of Hilbert spaces $\mathcal{H}_A$ and $\mathcal{H}_B$ with ${\rm dim}(\mathcal{H}_A)= {\rm dim}(\mathcal{H}_B) = D$ and $\log_2 D \ge n(H(p) - \delta)$. In this sense Alice and Bob can distill $H(p) - \delta$ Bell pairs per copy of $|\psi\rangle_{AB}$.

Though the number $m$ of up spins that Alice (or Bob) finds in her
(his) measurement is typically close to $np$, it can fluctuate about
this value.  Sometimes Alice and Bob will be lucky, and then will manage to
distill more than $H(p)$ Bell pairs per copy of
$|\psi(p)\rangle_{AB}$.  But the probability of doing substantially better
becomes negligible as $n \rightarrow \infty$.


The same idea applies to bipartite pure states in larger
Hilbert spaces.  If $A$ and $B$ are $d$-dimensional systems, then $|\psi\rangle_{AB}$ has the Schmidt decomposition
\begin{equation}
|\psi(X)\rangle_{AB} = \sum_{i = 0}^{d{-}1} \sqrt{p(x)} ~|x\rangle_A\otimes |x\rangle_{B},
\end{equation}
where $X$ is the classical ensemble $\{x, p(x)\}$, and $H(\bfrho_A) = H(\bfrho_B) = H(X)$. 
The Schmidt decomposition of $n$ copies of $\psi\rangle$ is
\begin{equation}
\sum_{x_1, x_2, \dots ,x_n=0}^{d{-}1} \sqrt{p(x_1)p(x_2)\dots p(x_n)}~ |x_1x_2\dots x_n\rangle_{A^n}\otimes |x_1x_2\dots x_n\rangle_{B^n}.
\end{equation}
Now Alice (or Bob) can measure the total
number of $|0\rangle$'s, the total number of $|1\rangle$'s, etc. in her (his)
possession.  If she finds $m_0 |0\rangle$'s, $m_1 |1\rangle$'s, etc., then her
measurement prepares a maximally entangled state with Schmidt number
\begin{equation}
D(m_0,m_1, \dots, m_{d{-}1})= \frac{n!}{m_0!m_1!\dots m_{d{-}1}!}
\end{equation}
and this outcome occurs with probability
\begin{equation}
p(m) = D(m_0,m_1, \dots, m_{d{-}1}) p(0)^{m_0}p(1)^{m_1}\dots p(d{-}1)^{m_{d-1}}.
\end{equation}
For $n$ large, Alice will typically find $m_x\approx n p(x)$, and again the probability of a large deviation is small, so that, from Stirling's approximation 
\begin{equation}
2^{n(H(X) - o(1))} \le D \le 2^{n(H(X) + o(1))}
\end{equation}
with high probability.
Thus, asymptotically for $n \rightarrow \infty$, $n(H(\bfrho_A) - o(1))$ high-fidelity Bell
pairs can be distilled from $n$ copies of $|\psi\rangle$, establishing that $E_D(|\psi\rangle) \ge H(\bfrho_A)$, and therefore $E_D(|\psi\rangle) = E_C(|\psi\rangle)=E(|\psi\rangle)$. 

This entanglement concentration protocol uses local operations but does not require any classical communication. When Alice and Bob do the same measurement they always get the same outcome, so there is no need for them to communicate. Classical communication really is necessary, though, to perform entanglement dilution. The protocol we described here, based on teleportation, requires two bits of classical one-way communication per Bell pair consumed; in a more clever protocol this can be reduced to $O(\sqrt{n})$ bits, but no further. Since the classical communication cost is sublinear in $n$, the number of bits of classical communication needed per copy of $|\psi\rangle$ becomes negligible in the limit $n\to \infty$.

Here we have discussed the entanglement cost and distillable entanglement for bipartite pure states. An achievable rate for distilling Bell pairs from bipartite mixed states will be derived in \S\ref{subsec:mother}.

\section{Quantifying Mixed-State Entanglement}
\label{sec:quant-mixed-entanglement}

\subsection{Asymptotic irreversibility under LOCC}

The entanglement cost $E_C$ and the distillable entanglement $E_D$ are natural and operationally meaningful ways to quantify entanglement. It's quite satisfying to find that, because entanglement dilution and concentration are asymptotically reversible for pure states, these two measures of pure-state bipartite entanglement agree, and provide another operational role for the von Neumann entropy of a marginal quantum state. 

We can define $E_C$ and $E_D$ for bipartite mixed states just as we did for pure states, but the story is more complicated --- when we prepare many copies of a mixed state shared by Alice and Bob, the dilution of Bell pairs is not in general reversible, even asymptotically, and the distillable entanglement can be strictly less than the entanglement cost, though it can never be larger. There are even bipartite mixed states with nonzero entanglement cost and zero distillable entanglement, a phenomenon called {\em bound entanglement}. This irreversibility is not shocking; any bipartite operation which maps many copies of the pure state $|\phi^+\rangle_{AB}$ to many copies of the mixed state $\bfrho_{AB}$ necessarily discards some information to the environment, and we don't normally expect a process that forgets information to be reversible. 

This separation between $E_C$ and $E_D$ raises the question, what is the preferred way to quantify the amount of entanglement when two parties share a mixed quantum state? The answer is: it depends. Many different measures of bipartite mixed-state entanglement have been proposed, each with its own distinctive advantages and disadvantages. Even though they do not always agree, both $E_C$ and $E_D$ are certainly valid measures. A further distinction can be made between the rate $E_{D1}$ at which entanglement can be distilled with one-way communication between the parties, and the rate $E_{D}$ with two-way communication. There are bipartite mixed states for which $E_{D} > E_{D1}$, and even states for which $E_{D}$ is nonzero while $E_{D1}$ is zero. In contrast to the pure-state case, we don't have nice formulas for the values of the various entanglement measures, though there are useful upper and lower bounds. We will derive a lower bound on $E_{D1}$ in \S\ref{subsec:mother} (the {\em hashing inequality}).

There are certain properties that any reasonable measure of bipartite quantum entanglement should have. The most important is that it must not increase under local operations and classical communication, because quantum entanglement cannot be created by LOCC alone. A function on bipartite states that is nonincreasing under LOCC is called an {\em entanglement monotone}. Note that an entanglement monotone will also be {\em invariant} under local unitary operations $U_{AB}=U_A\otimes U_B$, for if $U_{AB}$ can reduce the entanglement for any state, its inverse can increase entanglement. 

A second important property is that a bipartite entanglement measure must {\em vanish for separable states}. Recall from Chapter 4 that a bipartite mixed state is separable if it can be expressed as a convex combination of product states,
\begin{align}
\bfrho_{AB} = \sum_x p(x)~ |\alpha(x)\rangle\langle \alpha(x)|_A ~\otimes ~|\beta(x)\rangle \langle \beta(x)|_B.
\end{align}
A separable state is not entangled, as it can be created using LOCC. Via classical communication, Alice and Bob can establish a shared source of randomness, the distribution $X = \{x,p(x)\}$. Then they may jointly sample from $X$; if the outcome is $x$, Alice prepares $|\alpha(x)\rangle$ while Bob prepares $|\beta(x)\rangle$. 

A third desirable property for a bipartite entanglement measure is that it should agree with $E = E_C = E_D$ for bipartite pure states. Both the entanglement cost and the distillable entanglement respect all three of these properties. 

We remark in passing that, despite the irreversibility of entanglement dilution under LOCC, there is a mathematically viable way to formulate a reversible theory of bipartite entanglement which applies even to mixed states. In this formulation, we allow Alice and Bob to perform arbitrary bipartite operations that are incapable of creating entanglement; these include LOCC as well as additional operations which cannot be realized using LOCC. In this framework, dilution and concentration of entanglement become asymptotically reversible even for mixed states, and a unique measure of entanglement can be formulated characterizing the optimal rate of conversion between copies of $\bfrho_{AB}$ and  Bell pairs using these non-entangling operations.

Irreversible bipartite entanglement theory under LOCC, and also the reversible theory under non-entangling bipartite operations, are both examples of {\em resource theories}. In the resource theory framework, one or more parties are able to perform some restricted class of operations, and they are capable of preparing a certain restricted class of states using these operations. In addition, the parties may also have access to  {\em resource states}, which are outside the class they can prepare on their own. Using their restricted operations, they can transform resource states from one form to another, or consume resource states to perform operations beyond what they could achieve with their restricted operations alone. The name ``resource state'' conveys that such states are valuable because they may be consumed to do useful things.

In a two-party setting, where LOCC is allowed or more general non-entangling operations are allowed, bipartite entangled states may be regarded as a valuable resource.  Resource theory also applies if the allowed operations are required to obey certain symmetries; then states breaking this symmetry become a resource. In thermodynamics, states deviating from thermal equilibrium are a resource. Entanglement theory, as a particularly well developed resource theory, provides guidance and tools which are broadly applicable to many different interesting situations.

\subsection{Squashed entanglement}
As an example of an alternative bipartite entanglement measure, consider the {\em squashed entanglement} $E_{\rm sq}$, defined by
\begin{align}
E_{\rm sq}(\bfrho_{AB}) = \inf \left\{\frac{1}{2} I(A;B|C): \bfrho_{AB} = {\rm tr}_C \left(\bfrho_{ABC}\right)\right\}
\end{align}
The squashed entanglement of $\bfrho_{AB}$ is the greatest lower bound on the quantum conditional mutual information of all possible extensions of $\bfrho_{AB}$ to a tripartite state $\bfrho_{ABC}$; it can be shown to be an entanglement monotone. The locution ``squashed'' conveys that choosing an optimal conditioning system $C$ squashes out the non-quantum correlations between $A$ and $B$.

For pure states the extension is superfluous, so that 
\begin{align}
E_{\rm sq}(|\psi\rangle_{AB}) = \frac{1}{2} I(A;B) = H(A) = H(B)= E(|\psi\rangle_{AB}). 
\end{align}
For a separable state, we may choose the extension
\begin{align}
\bfrho_{ABC} = \sum_x p(x)~ |\alpha(x)\rangle\langle \alpha(x)|_A ~\otimes ~|\beta(x)\rangle \langle \beta(x)|_B ~\otimes ~|x\rangle\langle x|_C.
\end{align}
where $\{|x\rangle_C\}$ is an orthonormal set; the state $\bfrho_{ABC}$ has the block-diagonal form eq.(\ref{eq:quantum-conditional-independence}) and hence $I(A;B|C)=0$. Conversely, if $\bfrho_{AB}$ has any extension $\bfrho_{ABC}$ with $I(A;B|C) = 0$, then $\bfrho_{ABC}$ has the form eq.(\ref{eq:quantum-conditional-independence}) and therefore $\bfrho_{AB}$ is separable. 

$E_{\rm sq}$ is difficult to compute, because the infimum is to be evaluated over all possible extensions, where the system $C$ may have arbitrarily high dimension. This property also raises the logical possibility that there are nonseparable states for which the infimum vanishes; conceivably, though a nonseparable $\bfrho_{AB}$ can have no finite-dimensional extension for which $I(A;B|C)=0$, perhaps $I(A;B|C)$ can approach zero as the dimension of $C$ increases. Fortunately, though this is not easy to show, it turns out that  $E_{\rm sq}$ is strictly positive for any nonseparable state. In this sense, then,  it is a faithful entanglement measure, strictly positive if and only if the state is nonseparable. 

One desirable property of $E_{\rm sq}$, not shared by $E_C$ and $E_D$, is its additivity on tensor products (Exercise \ref{ex:squashed-additivity}), 
\begin{align}
E_{\rm sq}(\bfrho_{AB}\otimes \bfrho_{A'B'}) = E_{\rm sq}(\bfrho_{AB}) + E_{\rm sq}(\bfrho_{A'B'}) .
\end{align}
Though, unlike $E_C$ and $E_D$, squashed entanglement does not have an obvious operational meaning, any additive entanglement monotone which matches $E$ for bipartite pure states is bounded above and below by $E_C$ and $E_D$ respectively, 
\begin{align}
E_C \ge E_{\rm sq} \ge E_D.
\label{eq:squashed-upper-lower}
\end{align}

\subsection{Entanglement monogamy}

Classical correlations are {\em polyamorous}; they can be shared among many parties. If Alice and Bob read the same newspaper, then they have information in common and become correlated. Nothing prevents Claire from reading the same newspaper; then Claire is just as strongly correlated with Alice and with Bob as Alice and Bob are with one another. Furthermore, David, Edith, and all their friends can read the newspaper and join the party as well.

Quantum correlations are not like that; they are harder to share. If Bob's state is pure, then the tripartite quantum state is a product $\bfrho_B\otimes \bfrho_{AC}$, and Bob is completely uncorrelated with Alice and Claire. If Bob's state is mixed, then he can be entangled with other parties. But if Bob is fully entangled with Alice (shares a pure state with Alice), then the state is a product $\bfrho_{AB}\otimes \bfrho_C$; Bob has used up all his ability to entangle by sharing with Alice, and Bob cannot be correlated with Claire at all. Conversely, if Bob shares a pure state with Claire, the state is $\bfrho_A\otimes \bfrho_{BC}$, and Bob is uncorrelated with Alice. Thus we say that quantum entanglement is {\em monogamous}.

Entanglement measures obey monogamy inequalities which reflect this tradeoff between Bob's entanglement with Alice and with Claire in a three-party state. Squashed entanglement, in particular, obeys a monogamy relation following easily from its definition, which was our primary motivation for introducing this quantity; we have
\begin{align}
E_{\rm sq}(A;B) +E_{\rm sq}(A;C) \le E_{\rm sq}(A;BC).
\label{eq:squashed-monogamy}
\end{align}
In particular, in the case of a pure tripartite state, $E_{\rm sq} = H(A)$ is the (pure-state) entanglement shared between $A$ and $BC$. The inequality is saturated if Alice's system is divided into subsystems $A_1$ and $A_2$ such that  the tripartite pure state is
\begin{align}
|\psi\rangle_{ABC} = |\psi_1\rangle_{A_1B}\otimes |\psi_2\rangle_{A_2C}.
\end{align}
In general, combining eq.(\ref{eq:squashed-upper-lower}) with eq.(\ref{eq:squashed-monogamy}) yields
\begin{align}
E_D(A;B) +E_D(A;C) \le E_C(A;BC);
\end{align}
loosely speaking, the entanglement cost $E_C(A;BC)$ imposes a ceiling on Alice's ability to entangle with Bob and Claire individually, requiring her to trade in some distillable entanglement with Bob to increase her distillable entanglement with Claire.

To prove the monogamy relation eq.(\ref{eq:squashed-monogamy}), we note that mutual information obeys a {\em chain rule} which is really just a restatement of the definition of conditional mutual information:
\begin{align}
I(A;BC) = I(A;C) + I(A;B|C).
\label{eq:chain-rule-mutual}
\end{align}
A similar equation follows directly from the definition if we condition on a fourth system $D$, 
\begin{align}
I(A;BC|D) = I(A;C|D) + I(A;B|CD).
\label{eq:chain-rule-mutual-conditioned}
\end{align}
Now, $E_{\rm sq}(A;BC)$ is the infimum of $I(A;BC|D)$ over all possible extensions of $\bfrho_{ABC}$ to $\bfrho_{ABCD}$. But since $\bfrho_{ABCD}$ is also an extension of $\bfrho_{AB}$ and $\bfrho_{AC}$, we have
\begin{align}
I(A;BC|D) \ge  E_{\rm sq}(A;C) + E_{\rm sq}(A;B)
\end{align}
for any such extension. Taking the infimum over all $\bfrho_{ABCD}$ yields eq.(\ref{eq:squashed-monogamy}).

A further aspect of monogamy arises when we consider extending a quantum state to more parties. We say that the bipartite state $\bfrho_{AB}$ of systems $A$ and $B$ is $k$-{\em extendable} if there is a $(k{+}1)$-part state $\bfrho_{AB_1 \dots B_k}$ whose marginal state on $A B_j$ matches $\bfrho_{AB}$ for each $j=1,2,\dots k$, and such that $\bfrho_{AB_1 \dots B_k}$ is invariant under permutations of the $k$ systems $B_1, B_2 \dots B_k$. Separable states are $k$-extendable for every $k$, and entangled pure states are not even $2$-extendable. Every entangled mixed state fails to be $k$-extendable for some finite $k$, and we may regard the maximal value $k_{\rm max}$ for which such a symmetric extension exists as a rough measure of how entangled the state is --- bipartite entangled states with larger and larger $k_{\rm max}$ are closer and closer to being separable. 


\section{Accessible Information}

\subsection{How much can we learn from a measurement?}
\label{subsec:how-much-measurement}

Consider a game played by Alice and Bob. Alice prepares a quantum state drawn from the ensemble $\mathcal{E} = \{\bfrho(x), p(x)\}$ and sends the state to Bob. Bob knows this ensemble, but not the particular state that Alice chose to send. After receiving the state, Bob performs a POVM with elements $\{\bfE(y)\}\equiv \bfE$, hoping to find out as much as he can about what Alice sent. The conditional probability that Bob obtains outcome $y$ if Alice sent $\bfrho(x)$ is $p(y|x) = {\rm tr}\left(\bfE(y)\bfrho(x)\right)$, and the joint distribution governing Alice's preparation and Bob's measurement is $p(x,y) = p(y|x) p(x)$. 

Before he measures, Bob's ignorance about Alice's state is quantified by $H(X)$, the number of ``bits per letter'' needed to specify $x$; after he measures his ignorance is reduced to $H(X|Y) = H(XY) - H(Y)$. The improvement in Bob's knowledge achieved by the measurement is Bob's {\em information gain}, the mutual information
\begin{equation}
I(X;Y) = H(X) - H(X|Y).
\end{equation}
Bob's best strategy (his {\em optimal measurement}) maximizes this information gain. The best information gain Bob can achieve, 
\begin{equation}
{\rm Acc}(\mathcal{E}) = \max_{\bfE} I(X;Y),
\end{equation}
is a property of the ensemble $\mathcal{E}$ called the {\em accessible information} of $\mathcal{E}$.

If the states $\{\bfrho(x)\}$ are mutually orthogonal they are perfectly distinguishable. Bob can identify Alice's state with certainty by choosing $\bfE(x)$ to be the projector onto the support of $\bfrho(x)$; Then $p(y|x) = \delta_{x,y}= p(x|y)$, hence $H(X|Y) = \langle-\log p(x|y) \rangle=0$ and ${\rm Acc}(\mathcal{E}) = H(X)$. Bob's task is more challenging if Alice's states are not orthogonal. Then no measurement will identify the state perfectly, so $H(X|Y)$ is necessarily positive and ${\rm Acc}(\mathcal{E}) < H(X)$.

Though there is no simple general formula for the accessible information of an ensemble, we can derive a useful upper bound, called the {\em Holevo bound}. For the special case of an ensemble of pure states $\mathcal{E} = \{|\varphi(x)\rangle, p(x)\}$, the Holevo bound becomes
\begin{equation}
{\rm Acc}(\mathcal{E}) \le H(\bfrho), \quad {\rm where} \quad \bfrho= \sum_x p(x) |\varphi(x)\rangle\langle \varphi(x)|,
\end{equation}
and a sharper statement is possible for an ensemble of mixed states, as we will see. Since the entropy for a quantum system with dimension $d$ can be no larger than $\log d$, the Holevo bound asserts that Alice, by sending $n$ qubits to Bob ($d=2^n$) can convey no more than $n$ bits of information. This is true even if Bob performs a sophisticated collective measurement on all the qubits at once, rather than measuring them one at a time. 

Therefore, if Alice wants to convey classical information to Bob by sending qubits, she can do no better than treating the qubits as though they were classical, sending each qubit in one of the two orthogonal states $\{|0\rangle, |1\rangle\}$ to transmit one bit. This statement is not so obvious. Alice might try to stuff more classical information into a single qubit by sending a state chosen from a large alphabet of pure single-qubit signal states, distributed uniformly on the Bloch sphere. But the enlarged alphabet is to no avail, because as the number of possible signals increases the signals also become less distinguishable, and Bob is not able to extract the extra information Alice hoped to deposit in the qubit. 

If we can send information more efficiently by using an alphabet of mutually orthogonal states, why should we be interested in the accessible information for an ensemble of non-orthogonal states? There are many possible reasons. Perhaps Alice finds it easier to send signals, like coherent states, which are imperfectly distinguishable rather than mutually orthogonal. Or perhaps Alice sends signals to Bob through a noisy channel, so that signals which are orthogonal when they enter the channel are imperfectly distinguishable by the time they reach Bob. 

The accessible information game also arises when an experimental physicist tries to measure an unknown classical force using a quantum system as a probe. For example, to measure the $z$-component of a magnetic field, we may prepare a spin-$\frac{1}{2}$ particle pointing in the $x$-direction; the spin precesses for time $t$ in the unknown field, producing an ensemble of possible final states (which will be an ensemble of  mixed states if the initial preparation is  imperfect, or if decoherence occurs during the experiment). The more information we can gain about the final state of the spin, the more accurately we can determine the value of the magnetic field. 

\subsection{Holevo bound}

Recall that quantum mutual information obeys monotonicity --- if a quantum channel maps $B$ to $B'$, then $I(A;B) \ge I(A;B')$. We derive the Holevo bound by applying monotonicity of mutual information to the accessible information game. We will suppose that Alice records her chosen state in a classical register $X$ and Bob likewise records his measurement outcome in another register $Y$, so that Bob's information gain is the mutual information $I(X;Y)$ of the two registers. After Alice's preparation of her system $A$, the joint state of $XA$ is
\begin{equation}
\bfrho_{XA} = \sum_x p(x) |x\rangle \langle x | \otimes \bfrho(x).
\end{equation}
Bob's measurement is a quantum channel mapping $A$ to $AY$ according to
\begin{equation}
\bfrho(x) \mapsto \sum_y \bfM(y) \bfrho(x) \bfM(y)^\dagger \otimes |y\rangle \langle y|, 
\end{equation}
where $\bfM(y)^\dagger \bfM(y) = \bfE(y)$, yielding the state for $XAY$
\begin{equation}
 \bfrho'_{XAY} = \sum_{x,y} p(x) |x\rangle \langle x | \otimes \bfM(y)\bfrho(x)\bfM(y)^\dagger \otimes |y\rangle\langle y|.
\end{equation}
Now we have
\begin{equation}
I(X;Y)_{ \bfrho'} \le I(X;AY)_{\bfrho'} \le I(X;A)_{ \bfrho},
\end{equation}
where the subscript indicates the state in which the mutual information is evaluated; the first inequality uses strong subadditivity in the state $\bfrho'$, and the second uses monotonicity under the channel mapping $\bfrho$ to $\bfrho'$.

The quantity $I(X;A)$ is an intrinsic property of the ensemble $\mathcal{E}$; it is denoted $\chi(\mathcal{E})$ and called the {\em Holevo chi} of the ensemble. We have shown that however Bob chooses his measurement his information gain is bounded above by the Holevo chi; therefore, 
\begin{equation}
{\rm Acc}(\mathcal{E}) \le \chi(\mathcal{E}) := I(X;A)_\bfrho.
\end{equation}
This is the Holevo bound. 

Now let's calculate $I(X;A)_{\bfrho}$ explicitly. We note that
\begin{align}
H(XA) &= -{\rm tr}_{XA}\left(\sum_x p(x)|x\rangle\langle x|\otimes \bfrho(x) \log \left(\sum_{x'} p(x') |x'\rangle \langle x'|\otimes \bfrho(x')\right)\right)\notag\\
&= - \sum_x {\rm tr}_A~p(x)\bfrho(x) \left(\log p(x) + \log\bfrho(x)\right)\notag\\
& = H(X) + \sum_x p(x) H(\bfrho(x)),
\end{align}
and therefore
\begin{align}
H(A|X) = H(XA) - H(X)= \sum_x p(x) H(\bfrho(x)).
\end{align}
Using $I(X;A) = H(A) - H(A|X)$, we then find
\begin{equation}
\chi(\mathcal{E}) = I(X;A) = H(\bfrho_A) - \sum_x p(x) H(\bfrho_A(x)) \equiv H(A)_{\mathcal{E}} - \langle H(A)\rangle_{\mathcal{E}}
\label{eq:chi-as-concave}
\end{equation}
For an ensemble of pure states, $\chi$ is just the entropy of the density operator arising from the ensemble, but for an ensemble $\mathcal{E}$ of mixed states it is a strictly smaller quantity --  the difference between the entropy $H(\bfrho_{\mathcal{E}})$ of the convex sum of signal states and the convex sum $\langle H\rangle_{\mathcal{E}}$ of the signal state entropies; this difference is always nonnegative because of the concavity of the entropy function (or because mutual information is nonnegative).

\subsection{Monotonicity of Holevo chi}
\label{subsec:monotonicity-chi}

Since Holevo chi is the mutual information $I(X;A)$ of the classical register $X$ and the quantum system $A$, the monotonicity of mutual information also implies the monotonicity of $\chi$. If $\mathcal{N}: A\rightarrow A'$ is a quantum channel, then $I(X;A')\le I(X;A)$ and therefore 
\begin{equation}
\chi(\mathcal{E}')\le \chi(\mathcal{E}),
\end{equation}
where
\begin{align}
\mathcal{E}=\{\bfrho(x)), p(x)\} \quad {\rm and} \quad \mathcal{E'} = \{\bfrho'(x)=\mathcal{N}(\bfrho(x)), p(x)\}.
\end{align}
A channel cannot increase the Holevo chi of an ensemble. 

Its monotonicity provides a further indication that $\chi(\mathcal{E})$ is a useful measure of the information encoded in an ensemble of quantum states; the decoherence described by a quantum channel can reduce this quantity, but never increases it. In contrast, the von Neumann entropy may either increase or decrease under the action of a channel. Mapping pure states to mixed states can increase $H$, but a channel might instead map the mixed states in an ensemble to a fixed pure state $|0\rangle\langle 0|$, decreasing $H$ and improving the purity of each signal state, but without improving the distinguishability of the states.

We discussed the asymptotic limit $H(\bfrho)$ on quantum compression per letter in \S\ref{subsec:schumacher-compression}. There we considered unitary decoding; invoking the monotonicity of Holevo chi clarifies why more general decoders cannot do better. Suppose we compress and decompress the ensemble $\mathcal{E}^{\otimes n}$ using an encoder $\mathcal{N}_e$ and a decoder $\mathcal{N}_d$, where both maps are quantum channels: 
\begin{align}
\mathcal{E}^{\otimes n} \mathrel{\mathop{\longrightarrow}^{\mathcal{N}_e}} \mathcal{\tilde E}^{(n)}\mathrel{\mathop{\longrightarrow}^{\mathcal{N}_d}} \mathcal{\tilde E}^{\prime (n)}\approx\mathcal{E}^{\otimes n}
\end{align}
The Holevo chi of the input pure-state product ensemble is additive, $\chi(\mathcal{E}^{\otimes n}) =H(\bfrho^{\otimes n}) = nH(\bfrho)$, and  $\chi$ of a $d$-dimensional system is no larger than $\log_2 d$; therefore if the ensemble $\mathcal{\tilde E}^{(n)}$ is compressed to $q$ qubits per letter, then because of the monotonicity of $\chi$ the decompressed ensemble $\mathcal{\tilde E}^{\prime (n)}$ has Holevo chi per letter $\frac{1}{n} \chi(\mathcal{\tilde E}^{\prime (n)})\le q$. If the decompressed output ensemble has high fidelity with the input ensemble, its $\chi$ per letter should nearly match the $\chi$ per letter of the input ensemble, hence 
\begin{align}
q\ge \frac{1}{n}\chi(\mathcal{\tilde E}^{\prime (n)})\ge H(\bfrho) - \delta
\end{align}
for any positive $\delta$ and sufficiently large $n$. We conclude that high-fidelity compression to fewer than $H(\bfrho)$ qubits per letter is impossible asymptotically, even when the compression and decompression  maps are arbitrary channels. 

\subsection{Improved distinguishability through coding: an example}
\label{subsec:peres-wootters}

To better acquaint ourselves with the concept of accessible information, let's
consider a single-qubit example.  Alice prepares one of the three possible pure
states
\begin{align}
|\varphi_1\rangle &= |\uparrow_{\hat n_1}\rangle = \begin{pmatrix}1\\
0\end{pmatrix},\notag
\\
|\varphi_2\rangle &= |\uparrow_{\hat n_2}\rangle = \begin{pmatrix}-{1\over 2}\\
{\sqrt{3}\over 2}\end{pmatrix},\notag \\
|\varphi_3\rangle &= |\uparrow_{\hat n_3}\rangle = \begin{pmatrix}-{1\over 2}\\
-{\sqrt{3}\over 2}\end{pmatrix};
\label{eq:qubit-trine}
\end{align}
a spin-${1\over 2}$ object points in one of three directions that are
symmetrically distributed in the $xz$-plane.  Each state has {\it a priori}
probability ${1\over 3}$.  Evidently, Alice's signal states are
nonorthogonal:
\begin{equation}\label{pw_nonorthog}
\langle\varphi_1|\varphi_2\rangle = \langle\varphi_1| \varphi_3\rangle =
\langle\varphi_2 |\varphi_3\rangle = - {1\over 2}.
\end{equation}

Bob's task is to find out as much as he can about what Alice prepared by making
a suitable measurement.  The density matrix of Alice's ensemble is
\begin{equation}
\bfrho = {1\over 3} (|\varphi_1\rangle\langle\varphi_1| +
|\varphi_2\rangle\langle\varphi_2| + |\varphi_3\rangle\langle \varphi_3|) =
{1\over 2} {\bfI},
\end{equation}
which has $H(\bfrho) = 1$.  Therefore, the Holevo bound tells us that the
mutual information of Alice's preparation and Bob's measurement outcome cannot
exceed $1$ bit.

In fact, though, the accessible information is considerably less than the one
bit allowed by the Holevo bound.  In this case, Alice's ensemble has enough
symmetry that it is not hard to guess the optimal measurement.  Bob may choose
a POVM with three outcomes, where
\begin{equation}\label{pwpovm}
\bfE_{ a} = {2\over 3} ({\bfI} - |\varphi_a\rangle\langle \varphi_a|),
\quad a =
1,2,3;
\end{equation}
we see that
\begin{equation}
p(a|b) = \langle\varphi_b|\bfE_{ a} |\varphi_b\rangle =
\left\{\begin{array}{ll}0 & a = b,\\{1\over 2} & a\not= b.\end{array}\right.
\end{equation}
The measurement outcome $a$ {\it excludes} the possibility that
Alice prepared $a$, but leaves equal {\it a posteriori} probabilities $\left(p
= {1\over 2}\right)$ for the other two states.  Bob's information gain is
\begin{equation}
I = H(X) - H(X|Y) = \log_2 3 - 1 = .58496.
\end{equation}
To show that this measurement is really optimal, we may appeal to a variation
on a theorem of Davies, which assures us that an optimal POVM can be chosen
with three $\bfE_a$'s that share the same three-fold symmetry as the three
states in the input ensemble.  This result restricts the possible POVM's enough
so that we can check that eq.~(\ref{pwpovm}) is optimal with an explicit
calculation.  Hence we have found that the ensemble $\cale =
\{|\varphi_a\rangle, p_a = {1\over 3}\}$ has accessible information.
\begin{equation}
{\rm Acc} (\cale) = \log_2 \left({3\over 2}\right) = .58496 ...
\end{equation}
The Holevo bound is not saturated.

Now suppose that Alice has enough cash so that she can afford to send two
qubits to Bob, where again each qubit is drawn from the ensemble $\cale$.  The
obvious thing for Alice to do is prepare one of the {\it nine} states
\begin{equation}
|\varphi_a\rangle \otimes|\varphi_b\rangle, \quad a,b = 1,2,3,
\end{equation}
each with $p_{ab} = 1/9$.  Then Bob's best strategy is to perform the POVM
eq.~(\ref{pwpovm}) on each of the two qubits, achieving a mutual information of
.58496 bits per qubit, as before.

But, determined to do better, 
Alice and Bob decide on a different strategy.  Alice will
prepare one of {\it three} two-qubit states
\begin{equation}\label{pw_twos}
|\Phi_a\rangle = |\varphi_a\rangle\otimes |\varphi_a\rangle, \quad a = 1,2,3,
\end{equation}
each occurring with {\it a priori} probability $p_a = 1/3$.  Considered
one-qubit at a time, Alice's choice is governed by the ensemble $\cale$, but
now her two qubits have (classical) correlations -- both are prepared the same
way.

The three $|\Phi_a\rangle$'s are linearly independent, and so span a
three-dimensional subspace of the four-dimensional two-qubit Hilbert space.  In
Exercise \ref{ex:peres-wootters}, you will show that the density operator
\begin{equation}
\bfrho = {1\over 3} \left(\sum_{a=1}^3 |\Phi_a\rangle\langle\Phi_a|\right),
\end{equation}
has the nonzero eigenvalues $1/2, 1/4, 1/4$, so that
\begin{equation}
H(\bfrho) = - {1\over 2} \log_2 {1\over 2} - 2 \left({1\over 4} \log_2 {1\over
4}\right) = {3\over 2}.
\end{equation}
The Holevo bound requires that the accessible information {\it per qubit} is
no more than $3/4$ bit, which is at least consistent with the possibility
that we can exceed the .58496 bits per qubit attained by the nine-state method.

Naively, it may seem that Alice won't be able to convey as much classical
information to Bob, if she chooses to send one of only three possible states
instead of nine.  But on further reflection, this conclusion is not obvious.
True, Alice has fewer signals to choose from, but the signals are {\it more
distinguishable;} we have
\begin{equation}
\langle\Phi_a | \Phi_b\rangle = {1\over 4}, \quad a\not= b,
\end{equation}
instead of eq.~(\ref{pw_nonorthog}).  It is up to Bob to exploit this improved
distinguishability in his choice of measurement.  In particular, Bob will find
it advantageous to perform {\it collective} measurements on the two qubits
instead of measuring them one at a time.

It is no longer obvious what Bob's optimal measurement will be.  But Bob can
invoke a general procedure that, while not guaranteed optimal, is usually at
least pretty good.  We'll call the POVM constructed by this procedure a
``pretty good measurement'' (or PGM).

Consider some collection of vectors $|\tilde{\Phi}_a\rangle$ that are not
assumed to be orthogonal or normalized.  We want to devise a POVM that can
distinguish these vectors reasonably well.  Let us first construct
\begin{equation}
\bfg = \sum_a |\tilde{\Phi}_a\rangle\langle\tilde{\Phi}_a|;
\end{equation}
This is a positive operator on the space spanned by the
$|\tilde{\Phi}_a\rangle$'s.  Therefore, on that subspace, $\bfg$ has an
inverse, $\bfg^{-1}$ and that inverse has a positive square root $\bfg^{-1/2}$.
 Now we define
\begin{equation}
\bfE_a = \bfg^{-1/2} |\tilde{\Phi}_a\rangle\langle\tilde{\Phi}_a|\bfg^{-1/2},
\label{eq:PGM-POVM}
\end{equation}
and we see that
\begin{align}
\sum_a \bfE_a &= \bfg^{-1/2} \left(\sum_a
|\tilde{\Phi}_a\rangle\langle\tilde{\Phi}_a|\right) \bfg^{-1/2}\notag \\
&= \bfg^{-1/2} \bfg \bfg^{-1/2} = {\bfI},
\end{align}
on the span of the $|\tilde{\Phi}_a\rangle$'s.  If necessary, we can augment
these $\bfE_a$'s with one more positive operator, the projection $\bfE_0$ onto
the orthogonal complement of the span of the $|\tilde{\Phi}_a\rangle$'s, and so
construct a POVM.  This POVM is the PGM associated with the vectors
$|\tilde{\Phi}_a\rangle$.

In the special case where the $|\tilde{\Phi}_a\rangle$'s are orthogonal,
\begin{equation}
|\tilde{\Phi}_a\rangle = \sqrt{\lambda_a} | \phi_a\rangle,
\end{equation}
(where the $|\phi_a\rangle$'s are orthonormal), we have
\begin{align}
\bfE_a &= \sum_{b,c} (|\phi_b\rangle \lambda_b^{-1/2} \langle\phi_b|)
(|\phi_a\rangle\lambda_a \langle\phi_a|) (|\phi_c\rangle \lambda_c^{-1/2}
\langle\phi_c|)\notag \\
&= |\phi_a \rangle\langle\phi_a|;
\end{align}
this is the orthogonal measurement that perfectly distinguishes the
$|\phi_a\rangle$'s and so clearly is optimal.  If the
$|\tilde{\Phi}_a\rangle$'s are linearly independent but not orthogonal, then
the PGM is again an orthogonal measurement (because $n$ one-dimensional
operators in an $n$-dimensional space can constitute a POVM only if mutually
orthogonal --- see Exercise 3.11), but in that case the measurement may not be optimal.

In Exercise \ref{ex:peres-wootters}, you'll construct the PGM for the vectors $|\Phi_a\rangle$ in
eq. ~(\ref{pw_twos}), and you'll show that
\begin{align}
p(a|a) &= \langle\Phi_a| \bfE_a |\Phi_a\rangle = {1\over 3} \left( 1 +
{1\over\sqrt{2}}\right)^2 = .971405,\notag \\
p(b|a) &= \langle\Phi_a| \bfE_b |\Phi_a\rangle = {1\over 6} \left( 1 -
{1\over\sqrt{2}}\right)^2 = .0142977
\end{align}
(for $b\not= a$).  It follows that the conditional entropy of the input is
\begin{equation}
H(X|Y) = .215894,
\end{equation}
and since $H(X) = \log_2 3 = 1.58496$, the information gain is
\begin{equation}
I(X;Y) = H(X) - H (X|Y) = 1.369068,
\end{equation}
a mutual information of .684534 bits per qubit.  Thus, the improved
distinguishability of Alice's signals has indeed paid off -- we have exceeded
the .58496 bits that can be extracted from a single qubit.  We still didn't
saturate the Holevo bound ($I \le 1.5$ in this case), but we came a lot closer
than before.

This example, first described by Peres and Wootters, teaches some useful
lessons.  First, Alice is able to convey more information to Bob by ``pruning''
her set of codewords.  She is better off choosing among fewer signals that are
more distinguishable than more signals that are less distinguishable.  An
alphabet of three letters encodes more than an alphabet of nine letters.

Second, Bob is able to read more of the information if he performs a collective
measurement instead of measuring each qubit separately.  His optimal orthogonal
measurement projects Alice's signal onto a basis of {\it entangled} states.


\subsection{Classical capacity of a quantum channel}

This example illustrates how coding and collective measurement can enhance accessible information, but while using the code narrowed the gap between the accessible information and the Holevo chi of the ensemble, it did not close the gap completely. As is often the case in information theory, we can characterize the accessible information more precisely by considering an asymptotic i.i.d.~setting. 
To be specific, we'll consider the task of sending classical information reliably through a noisy quantum channel $\mathcal{N}^{A\to B}$. 

An ensemble of input signal states $\mathcal{E}= \{\bfrho(x), p(x)\}$ prepared by Alice is mapped by the channel to an ensemble of output signals $\mathcal{E}'=\{\mathcal{N}(\bfrho(x)), p(x)\}$. If Bob measures the output his optimal information gain
\begin{align}
{\rm Acc}(\mathcal{E}') \le I(X;B) = \chi(\mathcal{E}')
\label{eq:Acc-E-prime}
\end{align}
is bounded above by the Holevo chi of the output ensemble $\mathcal{E}'$. To convey as much information through the channel as possible, Alice and Bob may choose the input ensemble $\mathcal{E}$ that maximizes the Holevo chi of the output ensemble $\mathcal{E'}$. The maximum value
\begin{align}
\chi(\mathcal{N}) := \max_{\mathcal{E}} \chi(\mathcal{E}') = \max_{\mathcal{E}}I(X;B)
\end{align}
of $\chi(\mathcal{E}')$ is a property of the channel, which we will call the Holevo chi of $\mathcal{N}$.

As we've seen, Bob's actual optimal information gain in this {\em single-shot} setting may fall short of $\chi(\mathcal{E}')$ in general.
But instead of using the channel just once, suppose that Alice and Bob use the channel $n\gg 1$ times, where Alice sends signal states chosen from a code, and Bob performs an optimal measurement to decode the signals he receives. Then an information gain of  $\chi(\mathcal{N})$ bits per letter really can be achieved asymptotically as $n\to \infty$. 

Let's denote Alice's ensemble of encoded $n$-letter signal states by $\mathcal{\tilde E}^{(n)}$, denote the ensemble of classical labels carried by the signals by $\tilde X^n$, and denote Bob's ensemble of measurement outcomes by $\tilde Y^n$. Let's say that the code has rate $R$ if Alice may choose from among $2^{nR}$ possible signals to send. If classical information can be sent through the channel with rate $R-o(1)$ such that Bob can decode the signal with  negligible error probability as $n\to \infty$, then we say the rate $R$ is {\em achievable}. The classical capacity $C(\mathcal{N})$ of the quantum channel $\mathcal{N}^{A\to B}$ is the supremum of all achievable rates. 

As in our discussion of the  capacity of a classical channel in \S\ref{subsec:shannon-noisy}, we suppose that $\tilde X^n$ is the uniform ensemble over the $2^{nR}$ possible messages, so that $H(\tilde X^n)= nR$. Furthermore, the conditional entropy per letter ${1\over n} H (\tilde{X}^n |\tilde{Y}^n))$ approaches zero as $n\to\infty$ if the error probability is asymptotically negligible; therefore,
\begin{align} 
R & \le  \frac{1}{n}\left(I(\tilde X^n; \tilde Y^n)+o(1) \right)\notag\\
&\le \frac{1}{n}\left( \max_{\mathcal{ E}^{(n)} }I(X^n; B^n)+ o(1)\right) = \frac{1}{n} \left(\chi(\mathcal{N}^{\otimes n})+o(1)\right),
\label{eq:rate-inequality-tilde}
\end{align}
where we obtain the first inequality as in eq.(\ref{mutual_bound}) and  the second inequality by invoking the Holevo bound, optimized over all possible $n$-letter input ensembles. We therefore infer that
\begin{align}
C(\mathcal{N}) \le \lim_{n\to\infty} \frac{1}{n} \chi\left(\mathcal{N}^{\otimes n}\right);
\end{align}
the classical capacity is bounded above by the asymptotic Holevo chi per letter of the product channel $\mathcal{N}^{\otimes n}$.

In fact this upper bound is actually an achievable rate, and hence equal to the classical capacity $C(\mathcal{N})$. However, this formula for the classical capacity is not very useful as it stands, because it requires that we optimize the Holevo chi over message ensembles of arbitrary length; we say that the formula for capacity is {\em regularized} if, as in this case, it involves taking a limit in which the number of channel uses tends to infinity. It would be far preferable to reduce our expression for $C(\mathcal{N})$ to a {\em single-letter formula} involving just one use of the channel. In the case of a classical channel, the reduction of the regularized expression to a single-letter formula was possible, because the conditional entropy for $n$ uses of the channel is additive as in eq.(\ref{eq:additive-conditional-entropy}).

For quantum channels the situation is more complicated, as channels are known to exist such that the Holevo chi is strictly superadditive:
\begin{align}
\chi\left(\mathcal{N}_1\otimes \mathcal {N}_2\right) > \chi\left(\mathcal{N}_1\right)+ \chi\left(\mathcal {N}_2\right).
\end{align}
Therefore, at least for some channels, we are stuck with the not-very-useful regularized formula for the classical capacity. But we can obtain a single-letter formula for the optimal achievable communication rate if we put a restriction on the code used by Alice and Bob. In general, Alice is entitled to choose input codewords which are entangled across the many uses of the channel, and when such entangled codes are permitted the computation of the classical channel capacity may be difficult. But suppose we demand that all of Alice's codewords are product states. With that proviso the Holevo chi becomes subadditive, and we may express the optimal rate as
\begin{align}
C_1\left(\mathcal{N}\right) = \chi(\mathcal{N}).
\label{eq:prod-state-capacity}
\end{align}
$C_1(\mathcal{N})$ is called the {\em product-state capacity} of the channel.

Let's verify the subadditivity of $\chi$ for product-state codes. The product channel $\mathcal{N}^{\otimes n}$ maps product states to product states; hence if Alice's input signals are product states then so are Bob's output signals, and we can express Bob's $n$-letter ensemble as 
\begin{align}
\mathcal{E}^{(n)}= \{\bfrho(x_1)\otimes \bfrho(x_2)\otimes \dots\otimes \bfrho(x_n) ,~p(x_1x_2 \dots x_n )\},
\label{eq:prod-state-ensemble-holevo}
\end{align}
which has Holevo chi
\begin{align}
\chi(\mathcal{E}^{(n)}) = I(X^n;B^n) = H(B^n) - H(B^n|X^n).
\label{eq:holevo-n-letters}
\end{align}
(Here $\mathcal{E}^{(n)}$ is the output ensemble received by Bob when Alice sends product-state codewords, but to simplify the notation we have dropped the prime (indicating output) and tilde (indicating codewords) used earlier, e.g. in eq.(\ref{eq:Acc-E-prime}) and eq.(\ref{eq:rate-inequality-tilde}). We do not assume that the probability distribution $p(x_1x_2 \dots x_n )$ factorizes; hence the $n$ letters received by Bob, though unentangled, can be classically correlated.)
While the von Neumann entropy is subadditive,
\begin{align}
H(B^n) \le \sum_{i=1}^n H(B_i);
\label{eq:n-subadditive}
\end{align}
the (negated) conditional entropy
\begin{align}
-H(B^n|X^n) = -\sum_{\vec x} p(\vec x)~ H\left(\bfrho(\vec x)\right)
\end{align}
(see eq.(\ref{eq:chi-as-concave})) is not subadditive in general.
But for the product-state ensemble eq.(\ref{eq:prod-state-ensemble-holevo}), since the entropy of a product is additive, we have
\begin{align}
H(B^n|X^n) &= \sum_{x_1,x_2, \dots , x_n} p(x_1x_2, \dots x_n) \left(\sum_{i=1}^n H\left(\bfrho(x_i)\right)\right) \notag\\
&= \sum_{i=1}^n \sum_{x_i} p_i(x_i) H(\bfrho(x_i)) = \sum_{i=1}^n H(B_i|X_i)
\label{eq:conditional-holevo-bound}
\end{align}
where $X_i = \{x_i, p_i(x_i)\}$ is the marginal probability distribution for the $i$th letter. Eq.(\ref{eq:conditional-holevo-bound}) is a quantum analog of eq.(\ref{eq:additive-conditional-entropy}), which holds for product-state ensembles but not in general for entangled ensembles. Combining eq.(\ref{eq:holevo-n-letters}), (\ref{eq:n-subadditive}), (\ref{eq:conditional-holevo-bound}), we have
\begin{align}
I(X^n;B^n) \le \sum_{i=1}^n \left(H(B_i) - H(B_i|X_i) \right)=\sum_i I(X_i; B_i)\le  n \chi(\mathcal{N}).
\end{align}
Therefore the Holevo chi of a channel is subadditive when restricted to product-state codewords, as we wanted to show. 

We won't give a careful argument here that $C_1(\mathcal{N})$ is an asymptotically achievable rate using product-state codewords; we'll just give a rough sketch of the idea.  We demonstrate achievability with a random coding argument similar to Shannon's. Alice fixes an input ensemble $\mathcal{E}= \{\bfrho(x), p(x)\}$, and samples from the product ensemble $\mathcal{E}^{\otimes n}$ to generate a codeword; that is, the codeword
\begin{align}
\bfrho(\vec x) = \bfrho(x_1)\otimes \bfrho(x_2)\otimes \cdots\otimes \bfrho(x_n)
\end{align}
is selected with probability $p(\vec x)=p(x_1)p(x_2)\dots p(x_n)$. (In fact Alice should choose each $\bfrho(\vec x)$ to be pure to optimize the communication rate.) This codeword is sent via $n$ uses of the channel $\mathcal{N}$, and Bob receives the product state
\begin{align}
\mathcal{N}^{\otimes n}\left(\bfrho(\vec x)\right)=\mathcal{N}(\bfrho(x_1))\otimes \mathcal{N}(\bfrho(x_2))\otimes \cdots\otimes \mathcal{N}(\bfrho(x_n)).
\end{align}
Averaged over codewords, the joint state of Alice's classical register $X^n$ and Bob's system $B^n$ is
\begin{align}
\bfrho_{X^nB^n} = \sum_{\vec x}  p(\vec x)~ |\vec x\rangle \langle \vec x | \otimes \mathcal{N}^{\otimes n}(\bfrho(\vec x)).
\end{align}

To decode, Bob performs a POVM designed to distinguish the codewords effectively; a variant of the pretty good measurement described in \S\ref{subsec:peres-wootters} does the job well enough. 
The state Bob receives is mostly supported on a typical subspace with dimension $2^{n(H(B)+o(1))}$, and for each typical codeword that Alice sends, what Bob receives is mostly supported on a much smaller typical subspace with dimension $2^{n(H(B|X)+o(1))}$. 
The key point is that ratio of the dimensions of these spaces is exponential in the mutual information of $X$ and $B$:
\begin{align}
\frac{2^{n(H(B|X)+o(1))}}{2^{n(H(B)-o(1))}}= 2^{-n(I(X;B) - o(1))}
\label{eq:decod-subspace-ratio}
\end{align}
Each of Bob's POVM elements has support on the typical subspace arising from a particular one of Alice's codewords. The probability that any codeword is mapped purely by accident to the decoding subspace of a different codeword is suppressed by the ratio eq.(\ref{eq:decod-subspace-ratio}). Therefore, the probability of a decoding error remains small even when there are $2^{nR}$ codewords to distinguish, for $R = I(X;B) - o(1)$. 

We complete the argument with standard Shannonisms. Since the probability of a decoding error is small when we average over codes, it must also be small, averaged over codewords, for a particular sequence of codes. Then by pruning half of the codewords, reducing the rate by a negligible amount, we can ensure that the decoding errors are improbable for every codeword in the code. Therefore $I(X;B)$ is an achievable rate for classical communication. Optimizing over all product-state input ensembles, we obtain eq.(\ref{eq:prod-state-capacity}).

To turn this into an honest argument, we would need to specify Bob's decoding measurement more explicitly and do a careful error analysis. This gets a bit technical, so we'll skip the details. Somewhat surprisingly, though, it turns out to be easier to prove capacity theorems when quantum channels are used for other tasks besides sending classical information. We'll turn to that in \S\ref{sec:quantum-channel-decoupling}. 




\subsection{Entanglement-breaking channels}
Though Holevo chi is superadditive for some quantum channels, there are classes of channels for which chi is additive, and for any such channel $\mathcal{N}$ the classical capacity is $C= \chi(\mathcal{N})$ without any need for regularization. For example, consider {\em entanglement-breaking channels}. We say that $\mathcal{N}^{A\to B}$ is entanglement breaking if for any input state $\bfrho_{RA}$, $I \otimes \mathcal{N}(\bfrho_{RA}) $ is a separable state on $RA$ --- the action of $\mathcal{N}$ on $A$ always breaks its entanglement with $R$. We claim that if $\mathcal{N}_1$ is entanglement breaking, and $\mathcal{N}_2$ is an arbitrary channel, then 
\begin{align}
\chi\left(\mathcal{N}_1\otimes \mathcal{N}_2\right) \le \chi(\mathcal{N}_1) + \chi(\mathcal{N}_2).
\label{eq:ent-break-chi}
\end{align}

To bound the chi of the product channel, consider an input ensemble
\begin{align}
\bfrho_{XA_1A_2} = \sum_x p(x)~|x\rangle\langle x| \otimes \bfrho(x)_{A_1A_2},
\end{align}
which is mapped by $\mathcal{N}_1\otimes \mathcal{N}_2$ to
\begin{align}
\bfrho'_{XB_1B_2} = \left(I\otimes \mathcal{N}_1\otimes \mathcal{N}_2\right)\left(\bfrho_{XA_1A_2}\right).
\end{align}
By tracing out the second output system we find
\begin{align}
\bfrho'_{XB_1} = I\otimes\mathcal{N}_1\left(\bfrho_{XA_1}\right) = \sum_x p(x)~|x\rangle\langle x| \otimes \mathcal{N}_1\left(\bfrho(x)_{A_1}\right),
\end{align}
which, by the definition of $\chi(\mathcal{N}_1)$ implies
\begin{align}
I(X;B_1)_{\bfrho'} \le \chi(\mathcal{N}_1).
\end{align}

We can easily check that, for any three systems $A$, $B$, and $C$, 
\begin{align}
I(A;BC) = I(A;B) +I(AB;C) - I(C;B) \le I(A;B) +I(AB;C),
\end{align}
so that in particular
\begin{align}
I(X;B_1B_2)_{\bfrho'}  \le \chi(\mathcal{N}_1) +I(XB_1;B_2)_{\bfrho'}.
\label{eq:mutualXB1B2}
\end{align}
Eq.(\ref{eq:mutualXB1B2}) holds for any channels $\mathcal{N}_1$ and $\mathcal{N}_2$; now to obtain eq.(\ref{eq:ent-break-chi}) it suffices to show that 
\begin{align}
I(XB_1;B_2)_{\bfrho'} \le \chi(\mathcal{N}_2)
\label{eq:mutualXB1B2-chi}
\end{align}
for entanglement breaking $\mathcal{N}_1$. 

If $\mathcal{N}_1$ is entanglement breaking, then $\bfrho(x)_{A_1A_2} $ is mapped by $\mathcal{N}_1$ to a separable state:
\begin{align}
\mathcal{N}_1\otimes I: \bfrho(x)_{A_1A_2}\mapsto \sum_y p(y|x)~\bfsigma(x,y)_{B_1}\otimes  \bftau(x,y)_{A_2}.
\end{align}
Therefore,
\begin{align}
\bfrho'_{XB_1B_2} = \sum_{x,y}  p(x)p(y|x) |x\rangle\langle x| \otimes \bfsigma(x,y)_{B_1}\otimes \left[\mathcal{N}_2\left(\bftau(x,y)\right)\right]_{B_2}
\end{align}
may be regarded as the marginal state (after tracing out $Y$) of
\begin{align}
\bfomega'_{XYB_1B_2} = \sum_{x,y}  p(x,y) |x,y\rangle\langle x,y| \otimes \bfsigma(x,y)\otimes \mathcal{N}_2\left(\bftau(x,y)\right).
\end{align}
Furthermore,
because $\bfomega'$ becomes a product state when conditioned on $(x,y)$, we find
\begin{align}
I(B_1;B_2|XY)_{\bfomega'} = 0,
\end{align}
and using strong subadditivity together with the definition of conditional mutual information we obtain
\begin{align}
I(XB_1;B_2)_{\bfrho'}& =I(XB_1;B_2)_{\bfomega'} \le I(XYB_1;B_2)_{\bfomega'} \notag\\
&= I(XY;B_2)_{\bfomega'} + I(B_1;B_2|XY)_{\bfomega'} = I(XY;B_2)_{\bfomega'}.
\end{align}
Finally, noting that 
\begin{align}
{\rm tr}_{B_1} ~\bfomega'_{XYB_1B_2} = \sum_{x,y} p(x,y) |x,y\rangle\langle x,y|\otimes \mathcal{N}_2\left(\bftau(x,y)\right)
\end{align}
and recalling the definition of $\chi(\mathcal{N}_2)$, we see that $I(XY;B_2)_{\bfomega'} \le \chi(\mathcal{N}_2)$, establishing eq.(\ref{eq:mutualXB1B2-chi}), and therefore eq.(\ref{eq:ent-break-chi}).

An example of an entanglement-breaking channel is a {\em classical-quantum channel}, also called a {\em c-q channel}, which acts according to
\begin{align}
\mathcal{N}^{A\to B}: \bfrho_A\mapsto \sum_x \langle x|\bfrho_A|x\rangle \bfsigma(x)_B,
\end{align}
where $\{|x\rangle\}$ is an orthonormal basis. In effect, the channel performs a complete orthogonal measurement on the input state and then prepares an output state conditioned on the measurement outcome. The measurement breaks the entanglement between system $A$ and any other system with which it was initially entangled. Therefore, c-q channels are entanglement breaking and have additive Holevo chi. 

\section{Quantum Channel Capacities and Decoupling}
\label{sec:quantum-channel-decoupling}

\subsection{Coherent information and the quantum channel capacity}
\label{subsec:coherent-information}

As we have already emphasized, it's marvelous that the capacity for a classical channel can be expressed in terms of the optimal correlation between input and output for a {\em single use} of the channel, 
\begin{equation}
C := \max_X I(X;Y).
\end{equation}
Another pleasing feature of this formula is its {\em robustness}. For example, the capacity does not increase if we allow the sender and receiver to share randomness, or if we allow feedback from receiver to sender. But for quantum channels the story is more complicated. We've seen already that no simple single-letter formula is known for the classical capacity of a quantum channel, if we allow entanglement among the channel inputs, and we'll soon see that the same is true for the quantum capacity. In addition, it turns out that entanglement shared between sender and receiver can boost the classical and quantum capacities of some channels, and so can ``backward'' communication from receiver to sender. There are a variety of different notions of capacity for quantum channels, all reasonably natural, and all with different achievable rates. 

While Shannon's theory of classical communication over noisy classical channels is pristine and elegant, the same cannot be said for the theory of communication over noisy quantum channels, at least not in its current state. It's still a work in progress. Perhaps some day another genius like Shannon will construct a beautiful theory of quantum capacities. For now, at least there are a lot of interesting things we can say about achievable rates. Furthermore, the tools that have been developed to address questions about quantum capacities have other applications beyond communication theory.

The most direct analog of the classical capacity of a classical channel is the quantum capacity of a quantum channel, unassisted by shared entanglement or feedback. The quantum channel $\mathcal{N}^{A\to B}$ is a TPCP map from $\mathcal{H}_A$ to $\mathcal{H}_B$, and Alice is to use the channel $n$ times to convey a quantum state to Bob with high fidelity. She prepares her state $|\psi\rangle$ in a code subspace
\begin{align}
\mathcal{H}^{(n)} \subseteq \mathcal{H}_A^{\otimes n}
\end{align}
and sends it to Bob, who applies a decoding map, attempting to recover $|\psi\rangle$. The rate $\bar R$ of the code is the number of encoded qubits sent per channel use,
\begin{align}
\bar R = \frac{1}{n}\log_2 {\rm dim}\left(\mathcal{H}^{(n)}\right).
\end{align}
(Here, deviating from our earlier practice, we have used $\bar R$ rather than $R$ to denote the communication rate; from now on we we will use $R$ to denote the reference system, which is introduced in the next paragraph.)
We say that the rate $\bar R$ is {\em achievable} if there is a sequence of codes with increasing $n$ such that for any $\varepsilon, \delta >0$ and for sufficiently large $n$ the rate is at least $\bar R-\delta$ and Bob's recovered state $\bfrho$ has fidelity $F = \langle \psi|\bfrho|\psi\rangle \ge 1 - \varepsilon$. The {\em quantum channel capacity} $Q(\mathcal{N})$ is the supremum of all achievable rates. 

There is a regularized formula for $Q(\mathcal{N})$. To understand the formula we first need to recall that any channel $\mathcal{N}^{A\to B}$ has an isometric Stinespring dilation $\bfU^{A\to BE}$ where $E$ is the channel's ``environment.'' Furthermore, any input density operator $\bfrho_A$ has a purification; if we introduce a {\em reference system} $R$, for any $\bfrho_A$ there is a pure state $\psi_{RA}$ such that $\bfrho_A = {\rm tr}_R \left(|\psi\rangle\langle \psi|\right)$. (I will sometimes use $\psi$ rather than the Dirac ket $|\psi\rangle$ to denote a pure state vector, when the context makes the meaning clear and the ket notation seems unnecessarily cumbersome.) Applying the channel's dilation to $\psi_{RA}$, we obtain an output pure state $\phi_{RBE}$, which we represent graphically as:

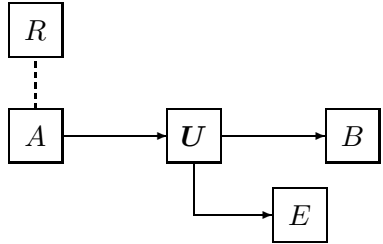
\begin{center}
\begin{picture} (140,90)
\put(0,70){\framebox(20,20){$R$}}
\put(0,30){\framebox(20,20){$A$}}
\put(60,30){\framebox(20,20){$\bfU$}}
\put(120,30){\framebox(20,20){$B$}}
\put(100,0){\framebox(20,20){$E$}}
\put(20,40){\vector(1,0){40}}
\put(80,40){\vector(1,0){40}}
\put(70,10){\vector(1,0){30}}
\put(70,10){\line(0,1){20}}
\multiput(10,50)(0,4){5}{\line(0,1){2}}
\end{picture}
\end{center}

\noindent
We then define the {\em one-shot quantum capacity} of the channel $\mathcal{N}$ by
\begin{align}
Q_1(\mathcal{N}) := \max_A \left(- H(R|B)_{\phi_{RBE}}\right).
\end{align}
Here the maximum is taken over all possible input density operators $\{\bfrho_A\}$, and $H(R|B)$ is the quantum conditional entropy 
\begin{align}
H(R|B) = H(RB) - H(B) = H(E) - H(B),
\end{align}
where in the last equality we used $H(RB) = H(E)$ in a pure state of $RBE$. The quantity $-H(R|B)$ has such a pivotal role in quantum communication theory that it deserves to have its own special name. We call it the {\em coherent information} from $R$ to $B$ and denote it
\begin{align}
I_c(R~\rangle B)_\phi = - H(R|B)_\phi = H(B)_\phi - H(E)_\phi .
\end{align}
This quantity does not depend on how the purification $\phi$ of the density operator $\bfrho_A$ is chosen; any one purification can be obtained from any other by a unitary transformation acting on $R$ alone, which does not alter $H(B)$ or $H(E)$. Indeed, since the expression $H(B) - H(E)$ only depends on the marginal state of $BE$, for the purpose of computing this quantity we could just as well consider the input to the channel to be the mixed state $\bfrho_A$ obtained from $\psi_{RA}$ by tracing out the reference system $R$. Furthermore, the coherent information does not depend on how we choose the dilation of the quantum channel; given a purification of the input density operator $\bfrho_A$, $I_c(R~\rangle B)_\phi = H(B) - H(RB)$ is determined by the output density operator of $RB$.

For a classical channel, $H(R|B)$ is always nonnegative and the coherent information is never positive. In the quantum setting,  $I_c(R~\rangle B)$ is positive if the reference system $R$ is more strongly correlated with the channel output $B$ than with the environment $E$. Indeed, an alternative way to express the coherent information is
\begin{align}
I_c(R~\rangle B) = \frac{1}{2} \left(I(R;B) - I(R;E) \right) = H(B) - H(E),
\end{align}
where we note that (because $\phi_{RBE}$ is pure)
\begin{align}
I(R;B) & = H(R)+H(B) - H(RB) = H(R) +H(B) -H(E),\notag\\
I(R;E) & = H(R) +H(E) -H(RE) = H(R) +H(E) - H(B).
\end{align}

Now we can state the regularized formula for the quantum channel capacity --- it is the optimal asymptotic coherent information per letter
\begin{align}
Q(\mathcal{N}^{A\to B}) = \lim_{n\to\infty}\max_{A^n} \frac{1}{n} I_c(R^n\rangle B^n)_{\phi_{R^nB^nE^n}},
\label{eq:quantum-channel-capacity-regularized}
\end{align}
where the input density operator $\bfrho_{A^n}$ is allowed to be entangled across the $n$ channel uses.
If coherent information were subadditive, we could reduce this expression to a single-letter quantity, the one-shot capacity $Q_1(\mathcal{N})$. But, unfortunately, for some channels the coherent information can be superadditive, in which case the regularized formula is not very informative. At least we can say that $Q_1(\mathcal{N})$ is an achievable rate, and therefore a lower bound on the capacity.

\subsection{The decoupling principle}
\label{subsec:decoupling}
Before we address achievability, let's understand why eq.(\ref{eq:quantum-channel-capacity-regularized}) is an upper bound on the capacity. First we note that the monotonicity of mutual information implies a corresponding monotonicity property for the coherent information. Suppose that the channel $\mathcal{N}_1^{A\to B}$ is followed by a channel $\mathcal{N}_2^{B\to C}$. Because mutual information is monotonic we have
\begin{align}
I(R;A) \ge I(R;B) \ge I(R;C), 
\end{align}
which can also be expressed as
\begin{align}
H(R) - H(R|A) \ge H(R) - H(R|B) \ge H(R) - H(R|C), 
\end{align}
and hence
\begin{align}
I_c(R~\rangle A) \ge I_c(R~\rangle B) \ge I_c(R~\rangle C).
\end{align}
A quantum channel cannot increase the coherent information, which has been called the {\em quantum data-processing inequality}. 

Suppose now that $\bfrho_A$ is a quantum code state, and that the two channels acting in succession are a noisy channel $\mathcal{N}^{A\to B}$ and the decoding map $\mathcal{D}^{B\to \hat B}$ applied  by Bob to the channel output in order to recover the channel input. Consider the action of the dilation $\bfU^{A\to BE}$ of $\mathcal{N}$ followed by the dilation $\bfV^{B\to \hat B B'}$ of $\mathcal{D}$ on the input purification $\psi_{RA}$, under the assumption that Bob is able to recover {\em perfectly}:
\begin{align}
\psi_{RA} \mathrel{\mathop{\longrightarrow}^{\bfU}} \phi_{RBE}\mathrel{\mathop{\longrightarrow}^{\bfV}} \tilde\psi_{R\hat B B' E} =\psi_{R\hat B}\otimes \chi_{B'E}.
\end{align}
If the decoding is perfect, then after decoding Bob holds in system $\hat B$ the purification of the state of $R$, so that
\begin{align}
H(R) = I_c(R~\rangle A)_\psi = I_c(R~\rangle \hat B)_{\tilde \psi}.
\end{align}
Since the initial and final states have the same coherent information, the quantum data processing inequality implies that the same must be true for the intermediate state $\phi_{RBE}$:
\begin{align}
H(R) &= I_c(R~\rangle B) = H(B) - H(E)\notag\\
&\implies H(B) = H(RE) = H(R) + H(E).
\end{align}
Thus the state of $RE$ is a product state. We have found that if Bob is able to recover perfectly from the action of the channel dilation $\bfU^{A\to BE}$ on the pure state $\psi_{RA}$, then, in the resulting channel output pure state $\phi_{RBE}$, the marginal state $\bfrho_{RE}$ must be the product $\bfrho_R\otimes \bfrho_E$. Recall that we encountered this criterion for recoverability earlier, when discussing quantum error-correcting codes in Chapter 7.

Conversely, suppose that $\psi_{RA}$ is an entangled pure state, and Alice wishes to transfer the purification of $R$ to Bob by sending it through the noisy channel $\bfU^{A\to BE}$. And suppose that in the resulting tripartite pure state $\phi_{RBE}$, the marginal state of $RE$ factorizes as $\bfrho_{RE}=\bfrho_R\otimes \bfrho_E$. Then $B$ decomposes into subsystems $B= B_1B_2$ such that
\begin{align}
\phi_{RBE} = \bfW_B \left(\tilde \psi_{R B_1}\otimes \chi_{B_2E}\right).
\end{align}
where $\bfW_B$ is some unitary change of basis in $B$; this holds because the most general purification of $\bfrho_{RE}$ can be obtained from a particular purification by applying a unitary transformation to $B$ alone.
Now Bob can construct an isometric decoder $\bfV^{B_1\to \hat B}\bfW_B^\dagger$, which extracts the purification of $R$ into Bob's preferred subsystem $\hat B$. Since all purifications of $R$ differ by an isometry on Bob's side, Bob can choose his decoding map to output the state $\psi_{R\hat B}$; then the input state of $RA$ is successfully transmitted to $R\hat B$ as desired. Furthermore, we may choose the initial state to be a maximally entangled state $\Phi_{RA}$ of the reference system with the code space of a quantum code; if the marginal state of $RE$ factorizes in the resulting output pure state $\phi_{RBE}$, then by the relative state method of Chapter 3 we conclude that {\em any} state in the code space can be sent through the channel and decoded with perfect fidelity by Bob. 

We have found that purified quantum information transmitted through the noisy channel is exactly correctable if and only if the reference system is completely uncorrelated with the channel's environment, or as we sometimes say, {\em decoupled} from the environment. This is the {\em decoupling principle}, a powerful notion underlying many of the key results in the theory of quantum channels. 

So far we have shown that exact correctability corresponds to exact decoupling. But we can likewise see that approximate correctability corresponds to approximate decoupling. Suppose for example that the state of $RE$ is close to a product state in the $L^1$ norm:
\begin{align}
\|\bfrho_{RE} - \bfrho_R\otimes \bfrho_E\|_1\le \varepsilon.
\end{align}
As we learned in Chapter 2, if two density operators are close together in this norm, that means they also have fidelity close to one and hence purifications with a large overlap. Any purification of the product state $\bfrho_R\otimes \bfrho_E$ has the form
\begin{align}
\tilde \phi_{RBE} = \bfW_B\left(\tilde \psi_{RB_1}\otimes \chi_{B_2E}\right),
\end{align}
and since all purifications of $\bfrho_{RE}$ can be transformed to one another by an isometry acting on the purifying system $B$, there is a way to choose $\bfW_B$ such that
\begin{align}
F(\bfrho_{RE},\bfrho_R\otimes \bfrho_E) = \left|\langle  \phi_{RBE}|\tilde \phi_{RBE}\rangle\right|^2\ge 1-\|\bfrho_{RE}-\bfrho_R\otimes \bfrho_E\|_1 \ge 1-\varepsilon.
\end{align}
Here we have used Uhlmann's theorem, which asserts that the fidelity of two density operators is equal to the maximum overlap of the purifications of the density operators. 
Furthermore, because fidelity is monotonic, both under tracing out $E$ and under the action of Bob's decoding map, and because Bob can decode $\tilde \phi_{RBE}$ perfectly, we conclude that
\begin{align}
F\left (\mathcal{D}^{B\to \hat B}\left(\bfrho_{RB}\right),\psi_{R\hat B} \right)\ge 1 - \varepsilon
\end{align}
if Bob chooses the proper decoding map $\mathcal{D}$. Thus approximate decoupling in the $L^1$ norm implies high-fidelity correctability. It is convenient to note that a similar argument still works if $\bfrho_{RE}$ is close in the $L^1$ norm to $\tilde \bfrho_R \otimes \tilde \bfrho_E$, where $\tilde \bfrho_R$ is not necessarily ${\rm tr}_E\left(\bfrho_{RE}\right)$ and $\tilde \bfrho_E$ is not necessarily ${\rm tr}_R\left(\bfrho_{RE}\right)$. 

On the other hand, if (approximate) decoupling fails, the fidelity of Bob's decoded state will be seriously compromised. Suppose that in the state $\phi_{RBE}$ we have
\begin{align}
I(R;E) = H(R) + H(E) - H(RE) \ge \varepsilon > 0.
\end{align}
Then the coherent information of $\phi$ is
\begin{align}
I_c(R~\rangle B)_\phi = H(B)_\phi - H(E)_\phi  = H(RE)_\phi - H(E)_\phi \le H(R)_\phi - \varepsilon.
\end{align}
By the quantum data processing inequality, we know that the coherent information of Bob's decoded state $\tilde\psi_{R\hat B}$ is no larger; hence 
\begin{align}
I_c(R~\rangle \hat B)_{\tilde \psi} = H(R)_{\tilde\psi} - H(R\hat B)_{\tilde \psi} \le H(R)_{\tilde \psi} - \varepsilon,
\end{align}
and therefore
\begin{align}
H(R\hat B)_{\tilde \psi} \ge \varepsilon
\end{align}
The deviation from perfect decoupling means that the decoded state of $R\hat B$ has some residual entanglement with the environment $E$, and is therefore impure. 

Now we have the tools to derive an upper bound on the quantum channel capacity $Q(\mathcal{N})$. For $n$ channel uses, let $\psi^{(n)}$ be a maximally entangled state of a reference system $\mathcal{H}_R^{(n)}\subseteq \mathcal{H}_R^{\otimes n}$ with a code space $\mathcal{H}_A^{(n)}\subseteq \mathcal{H}_A^{\otimes n}$, where ${\rm dim} ~\mathcal{H}_A^{(n)} = 2^{n\bar R}$, so that
\begin{align}
I_c(R^n\rangle A^n)_{\psi^{(n)}} = H(R^n)_{\psi^{(n)}} = n\bar R.
\end{align}
(Here $\bar R$ denotes the communication rate; we are now using $R$ to denote the reference system.)
Now $A^n$ is transmitted to $B^n$ through $\left(\bfU^{A\to BE}\right)^{\otimes n}$, yielding the pure state $\phi^{(n)}$ of $R^nB^nE^n$. If Bob can decode with high fidelity, then his decoded state must have coherent information $H(R^n)_{\psi^{(n)}} - o(n)$, and the quantum data processing inequality then implies that
\begin{align}
I_c(R^n \rangle B^n)_{\phi^{(n)}} 
= H(R^n)_{\psi^{(n)}}- o(n) = n\bar R - o(n) 
\label{eq:coherent-bound-loose}
\end{align}
and hence
\begin{align}
\bar R = \frac{1}{n} I_c(R^n \rangle B^n)_{\phi^{(n)}} + o(1).
\end{align}
Taking the limit $n\to\infty$ we see that the expression for $Q(\mathcal{N})$ in eq.(\ref{eq:quantum-channel-capacity-regularized}) is an upper bound on the quantum channel capacity. In Exercise \ref{ex:quantum-fano}, you will sharpen the statement eq.(\ref{eq:coherent-bound-loose}), showing that
\begin{align}
H(R^n) - I_c(R^n\rangle B^n) \le 2H_2(\varepsilon) + 4\varepsilon n\bar R 
\end{align}
if there is a decoding map that achieves fidelity $F\ge 1-\varepsilon$.

To show that $Q(\mathcal{N})$ is an achievable rate, rather than just an upper bound, we will need to formulate a quantum version of Shannon's random coding argument. Our strategy (see \S\ref{subsec:proof-father}) will be to demonstrate the existence of codes that achieve approximate decoupling of $E^n$ from $R^n$.


\subsection{Degradable channels}
\label{subsec:degradable-channels}

Though coherent information can be superadditive in some cases, there are classes of channels for which the coherent information is additive, and therefore the quantum channel capacity matches the one-shot capacity, for which there is a single-letter formula. One such class is the class of {\em degradable channels}.

To understand what a degradable channel is, we first need the concept of a {\em complementary channel}. Any channel $\mathcal{N}^{A\to B}$ has a Stinespring dilation $\bfU^{A\to BE}$, from which we obtain $\mathcal{N}^{A\to B}$ by tracing out the environment $E$. Alternatively we obtain the channel $\mathcal{N}^{A\to E}_c$ complementary to $\mathcal{N}^{A\to B}$ by tracing out $B$ instead. Since we have the freedom to compose $\bfU^{A\to BE}$ with an isometry $\bfV^{E\to E}$ without changing $\mathcal{N}^{A\to B}$, the complementary channel is defined only up to an isometry acting on $E$. This lack of uniqueness need not trouble us, because the properties of interest for the complementary channel are invariant under such isometries. 

We say that the channel $\mathcal{N}^{A\to B}$ is degradable if we can obtain its complementary channel by composing $\mathcal{N}^{A\to B}$ with a channel mapping $B$ to $E$:
\begin{align}
\mathcal{N}^{A\to E}_c = \mathcal{T}^{B\to E} \circ \mathcal{N}^{A\to B}.
\end{align}
In this sense, when Alice sends a state through the channel, Bob, who holds system $B$, receives a less noisy copy than Eve, who holds system $E$. 

Now suppose that $\bfU_1^{A_1\to B_1E_1}$ and $\bfU_2^{A_2\to B_2E_2}$ are dilations of the degradable channels $\mathcal{N}_1$ and $\mathcal{N}_2$. Alice  introduces a reference system $R$ and prepares an input pure state $\psi_{RA_1A_2}$, then sends the state to Bob via $\mathcal{N}_1\otimes \mathcal{N}_2$, preparing the output pure state $\phi_{RB_1B_2E_1E_2}$. We would like to evaluate the coherent information $I_c(R~\rangle B_1B_2)_\phi$ in this state. 

The key point is that because both channels are degradable, there is a product channel $\mathcal{T}_1\otimes \mathcal{T}_2$ mapping $B_1B_2$ to $E_1E_2$, and the monotonicity of mutual information therefore implies
\begin{align}
I(B_1;B_2) \ge I(E_1;E_2).
\end{align}
Therefore, the coherent information satisfies
\begin{align}
&I_c(R~\rangle B_1B_2) = H(B_1B_2) - H(E_1 E_2) \notag\\
&= H(B_1) + H(B_2) - I(B_1;B_2) - H(E_1) - H(E_2) +I(E_1;E_2) \notag\\
&\le H(B_1) - H(E_1) + H(B_2) - H(E_2).
\end{align}
These quantities are all evaluated in the state $\phi_{RB_1B_2E_1E_2}$. But notice that for the evaluation of $H(B_1) - H(E_1)$, the isometry $\bfU_2^{A_2\to B_2E_2}$ is irrelevant. This quantity is really the same as the coherent information $I_c(RA_2\rangle B_1)$, where now we regard $A_2$ as part of the reference system for the input to channel $\mathcal{N}_1$. Similarly $H(B_2) - H(E_2) = I_c(RA_1\rangle B_2)$, and therefore,
\begin{align}
I_c(R~\rangle B_1 B_2) \le I_c(RA_2\rangle B_1)+ I_c(RA_1\rangle B_2) \le Q_1(\mathcal{N}_1) + Q_1(\mathcal{N}_2),
\end{align}
where in the last inequality we use the definition of the one-shot capacity as coherent information maximized over all inputs. Since $Q_1(\mathcal{N}_1\otimes \mathcal{N}_2)$ is likewise defined by maximizing the coherent information $I_c(R~\rangle B_1B_2)$, we find that
\begin{align}
Q_1(\mathcal{N}_1\otimes \mathcal{N}_2) \le Q_1(\mathcal{N}_1) +Q_1(\mathcal{N}_2)
\label{eq:degradable-channel-bound}
\end{align}
if $\mathcal{N}_1$ and $\mathcal{N}_2$ are degradable. 

The regularized formula for the capacity of $\mathcal{N}$ is
\begin{align}
Q(\mathcal{N}) = \lim_{n\to \infty} \frac{1}{n} Q_1(\mathcal{N}^{\otimes n}) \le Q_1(\mathcal{N}),
\end{align}
where the last inequality follows from eq.(\ref{eq:degradable-channel-bound}) assuming that $\mathcal{N}$ is degradable. We'll see that $Q_1(\mathcal{N}) $ is actually an achievable rate, and therefore a single-letter formula for the quantum capacity of a degradable channel. 

As a concrete example of a degradable channel, consider the {\em generalized dephasing channel} with dilation
\begin{equation}
U^{A\to BE}: |x\rangle_A \mapsto |x\rangle_B\otimes |\alpha_x\rangle_E,
\end{equation}
where $\{|x\rangle_A\}$, $\{|x\rangle_B\}$ are orthonormal bases for $\mathcal{H}_A$, $\mathcal{H}_B$ respectively, and the states  $\{|\alpha_x\rangle_E\}$ of the environment are normalized but not necessarily orthogonal. (We discussed the special case where $A$ and $B$ are qubits in \S3.4.2.) The corresponding channel is
\begin{align}
\mathcal{N}^{A\to B}:\bfrho\mapsto \sum_{x,x'} |x\rangle\langle x|\bfrho|x'\rangle\langle x'| \langle \alpha_{x'}|\alpha_x\rangle ,
\label{eq:A-to-B-channel-dephasing}
\end{align}
which has the complementary channel
\begin{align}
\mathcal{N}_c^{A\to E}: \bfrho\mapsto \sum_x |\alpha_x\rangle\langle x |\bfrho|x\rangle \langle \alpha_x|.
\label{eq:A-to-E-channel-dephasing}
\end{align}
In the special case where the states $\{|\alpha_x\rangle_E = |x\rangle_E\}$ {\em are} orthonormal, we obtain the {\em completely dephasing channel}
\begin{align}
\Delta^{A\to B}: \bfrho\mapsto \sum_x |x\rangle \langle x|\bfrho|x\rangle \langle x|,
\end{align}
whose complement $\Delta^{A\to E}$ has the same form as $\Delta^{A\to B}$. (Here subscripts have been suppressed to avoid cluttering the notation, but it should be clear from the context whether $|x\rangle$ denotes $|x\rangle_A$, $|x\rangle_B$, or $|x\rangle_E$ in the expressions for $\mathcal{N}^{A\to B}$, $\mathcal{N}_c^{A\to E}$, $\Delta^{A\to B}$, and $\Delta^{A\to E}$.) We can easily check that
\begin{align}
\mathcal{N}^{A\to E}_c = \mathcal{N}^{C\to E}_c \circ \Delta^{B \to C} \circ \mathcal{N}^{A\to B};
\end{align} 
therefore $\mathcal{N}_c\circ \Delta$ degrades $\mathcal{N}$ to $\mathcal{N}_c$. Thus $\mathcal{N}$ is degradable and $Q(\mathcal{N}) = Q_1(\mathcal{N})$.

Further examples of degradable channels are discussed in Exercise \ref{ex:degradable}. 

We may also say that a channel $\mathcal{N}^{A\to B}$ is \emph{antidegradable} if we can obtain that channel from its complementary channel by composing $\mathcal{N}^{A\to E}_c$ with a channel mapping $E$ to $B$:
\begin{align}
\mathcal{N}^{A\to B} = \mathcal{T}^{E\to B} \circ \mathcal{N}^{A\to E}_c.
\end{align}
In this sense, Bob receives a noisier output than Eve. It is easy to see that the quantum capacity of an antidegradable channel must vanish. Eve can simulate Bob's output by applying $\mathcal{T}^{E\to B}$ to her output. Therefore, if a quantum state is transmitted through many uses of the channel $\mathcal{N}^{A\to B}$ from Alice to Bob, and Bob can decode the state with high fidelity, than Eve can decode it as well. Hence a nonzero capacity for an antidegradable channel is disallowed by the no-cloning theorem.







\subsection{Capacity of the depolarizing channel}
\label{subsec:capacity-depolarizing}
As we discussed in Chapter 3, the depolarizing channel with error probability $p$ acts on the input density operator $\bfrho$ of a single qubit according to
\begin{equation}
    \mathcal{N}^{A\to B}_p(\bfrho) = (1-p)\bfrho +\frac{p}{3}\left(\bfX\bfrho \bfX +\bfY\bfrho \bfY +\bfZ\bfrho \bfZ \right),
\end{equation}
where $\{\bfX,\bfY,\bfZ\}$ denote the Pauli matrices. Let's compute the one-shot capacity of this channel. 

Because of the channel's symmetry, we can easily guess that the coherent information will be optimized by an input density operator that is maximally mixed. Therefore we consider the action of the channel on the maximally entangled state $|\phi^+\rangle_{RA}=\frac{1}{\sqrt{2}}(\left(|00\rangle+|11\rangle\right)$, where $A$ is the channel input and $R$ is the reference system, obtaining the output
\begin{equation}
\label{eq:rho-RB-p}
    \bfrho_{RB}= (1-p)|\phi^+\rangle\langle \phi^+|+\frac{p}{3}\left(|\psi^+\rangle\langle\psi^+|+ |\psi^-|\langle \psi^-|+|\phi^-\rangle\langle \phi^-|\right).
\end{equation}
The one-shot capacity is the coherent information
\begin{equation}
    Q_1(\mathcal{N}_p) = I_c(R~\rangle B) = H(B)-H(RB)
\end{equation}
if this quantity is nonnegative, and zero otherwise.

Because $\bfrho_{RB}$ is a mixture of Bell states, the marginal density operator $\bfrho_B$ is maximally mixed and hence $H(B)=1$. Since the eigenvalues of $\bfrho_{RB}$ are $\left(1-p,\frac{p}{3},\frac{p}{3},\frac{p}{3}\right)$, we have 
\begin{align}
    &H(RB) = -(1-p)\log_2(1-p)-p\log_2\left(\frac{p}{3}\right)= H_2(p) + p\log_2 3\nonumber\\
    &\implies I_c(R~\rangle B) = 1-H_2(p) - p\log_2 3,
\end{align}
which is positive for $p< p_1\approx .18929 $. Thus the one-shot capacity $Q_1(\mathcal{N}_p)$ is nonzero for error probability $p<p_1$. 

This expression for $Q_1(\mathcal{N}_p)$ has a simple interpretation --- it is the communication rate achieved by random stabilizer codes such that the probability of a decoding error vanishes asymptotically. To see why, suppose that the number $n$ of channel uses is very large, so that the number of transmitted qubits with errors is close to $pn$ with high probability. Approximating the number of ways to choose which qubits have errors by $\binom{n}{np}$, and noticing that for each damaged qubit there are three possible uniformly distributed Pauli errors, we conclude that the number of typical errors is
\begin{equation}
    \approx {\binom{n}{np}}3^{pn}\approx 2^{n\left(H_2(p) + p \log_2 3\right)}.
\end{equation}

Now consider $n$-qubit stabilizer codes with $k$ logical qubits, and imagine averaging uniformly over such codes. These codes have $n-k$ independent commuting stabilizer generators, and measuring all these generators yields an error syndrome with $n-k$ bits. If the number of possible syndromes $2^{n-k}$ is much larger than the number of typical errors, then the observed syndrome is very unlikely to point to any particular typical error merely by accident; instead with high probability the syndrome will point to a unique typical error, the one that actually occurred, which can therefore be decoded correctly.

Concretely, the probability of a decoding error due to a syndrome decoding ambiguity can be estimated as
\begin{align}
\textrm{error prob}\approx\frac{\# {\textrm{ typical errors}}}{\# {\textrm{ syndromes}}} \approx \frac{2^{n\left(H_2(p) + p \log_2 3\right)}}{2^{n-k}}=2^{-n\left(Q_1(\mathcal{N}_p) -\bar R\right)},
\end{align}
where $\bar R = k/n$ is the encoding rate. We conclude that the error probability approaches zero as $n\to\infty$ for any rate $\bar R$ less than $Q_1(\mathcal{N}_p)$.

We used a similar counting argument to obtain a lower bound on the classical capacity $C(p)= 1-H_2(p)$ of the binary symmetric channel in \S\ref{subsec:shannon-noisy}. In that case we could find a matching upper bound by noting that if the communication rate were larger than $C(p)$ then there would not enough distinct syndromes to distinguish among all the typical errors. Curiously, this argument fails to provide a upper bound on the quantum communication rate achieved by stabilizer codes for a very noisy depolarizing channel $\mathcal{N}_p$ where $p$ is close to $p_1$. 

Counting syndromes does not yield an upper bound on the quantum capacity because quantum codes can be highly \emph{degenerate}; that is, there may be many nontrivial errors that act trivially on the code space. We encountered an instructive example of a degenerate stabilizer code in Chapter 7: the toric code, in which $\bfZ$ errors forming closed loops on the two-dimensional qubit lattice (or $\bfX$ errors forming closed loops on the dual lattice) are contained in the code stabilizer.  A more mundane example is the quantum repetition code spaced by codewords $|0\rangle^{\otimes n}$ and $|1\rangle^{\otimes n}$. In this case $\bfZ_i\bfZ_j$ is in the code stabilizer, where $i$ and $j$ label any pair among the $n$ qubits.

It turns out that a positive capacity can be attained for $p$ slightly above $p_1$ by concatenating the length-$m$ quantum repetition code with a random stabilizer code for some values of $m>1$. Correspondingly, one finds that, for some values of the number $m$ of channels uses and of the error probability $p$, the coherent information per channel use is higher than $Q_1(\mathcal{N}_p)$.

Consider the input to $m$ channels uses
\begin{align}
    |\Phi\rangle_{RA_1A_2 \cdots A_m} = \frac{1}{\sqrt{2}}\left(|0\rangle^{\otimes{(m+1)}}+|1\rangle^{\otimes{(m+1)}}\right)_{RA_1A_2 \cdots A_m},
\end{align}
where $R$ is the reference system. After $I\otimes \left(\mathcal{N}_p^{A\to B}\right)^m$ acts on this state, we may compute explicitly $\frac{1}{m}I_c(R \rangle B_1B_2\cdots B_m)$. One finds that for $m=5$ (which turns out to be the optimal case) this quantity, and therefore also $Q(\mathcal{N}_p)$, is positive up to $p< p_5$, where $p_5\approx .19036$ is strictly larger than $p_1$. 

For small $p$, $\frac{1}{m}I_c(R \rangle B_1B_2\cdots B_m)$ is close to $\frac{1}{m}$, hence much smaller for $m>1$ than for $m=1$. But if the channel is sufficiently noisy, the coherent information is superadditive, and one finds a higher value for $m>1$, indicating that the quantum capacity exceeds $Q_1(\mathcal{N}_p)$.

The value of $p_*$, the largest value of $p$ for which the capacity is nonzero, is of considerable interest, but it is still unknown. Exercise \ref{ex:approximate-cloning} shows that the depolarizing channel is antidegradable (and therefore has vanishing capacity) for $p\ge .25$; hence $p_*\le .25$. On the other hand, the best currently known lower bound is $p_*\ge .19088$. Perhaps someday a single-letter formula will be found for the quantum capacity of the depolarizing channel, but for now the superadditivity of coherent information has hindered efforts to obtain a precise estimate of the capacity in the extremely noisy regime. It's humbling. 

\section{Quantum Protocols}


Using the decoupling principle in an i.i.d.~setting, we can prove achievable rates for two fundamental quantum protocols. These are fondly known as the father and mother protocols, so named because each spawns a brood of interesting corollaries. We will formulate these protocols and discuss some of their ``children'' in this section, postponing the proofs until \S\ref{sec:decoupling-inequality}.

\subsection{Father: Entanglement-assisted quantum communication}
\label{subsec:father-protocol}

The father protocol is a scheme for entanglement-assisted quantum communication. Through many uses of a noisy quantum channel $\mathcal{N}^{A\to B}$, this protocol sends quantum information with high fidelity from Alice to Bob, while also consuming some previously prepared quantum entanglement shared by Alice and Bob. The task performed by the protocol is summarized by the {\em father resource inequality}
\begin{align}
\left\langle \mathcal{N}^{A\to B}:\bfrho_A\right\rangle +\frac{1}{2}I(R;E)[qq] \ge  \frac{1}{2}I(R;B)[q\to q],
\label{eq:father-resource-inequality}
\end{align}
where the resources on the left-hand side can be used to achieve the result on the right-hand side, in an asymptotic i.i.d.~setting. That is, the quantum channel $\mathcal{N}$ may be used $n$ times to transmit $\frac{n}{2}I(R;B) - o(n)$ qubits with fidelity $F\ge 1-o(1)$, while consuming $\frac{n}{2}I(R;E) + o(n)$ ebits of entanglement shared between sender and receiver. These entropic quantities are evaluated in a tripartite pure state $\phi_{RBE}$, obtained by applying the Stinespring dilation $\bfU^{A\to BE}$ of $\mathcal{N}^{A\to B}$ to the purification $\psi_{RA}$ of the input density operator $\bfrho_A$. Eq.(\ref{eq:father-resource-inequality}) means that for any input density operator $\bfrho_A$, there exists a coding procedure that achieves the quantum communication at the specified rate by consuming entanglement at the specified rate. 


To remember the father resource inequality, it helps to keep in mind that $I(R;B)$ quantifies something {\em good}, the correlation with the reference system which survives transmission through the channel, while $I(R;E)$ quantifies something {\em bad}, the correlation between the reference system $R$ and the channel's environment $E$, which causes the transmitted information to decohere. The larger the good quantity $I(R;B)$, the higher the rate of quantum communication. The larger the bad quantity $I(R;E)$, the more entanglement we need to consume to overcome the noise in the channel. To remember the factor of $\frac{1}{2}$ in front of $I(R;B)$, consider the case of a noiseless quantum channel, where $\psi_{RA}$ is maximally entangled; in that case there is no environment, 
\begin{align}
\phi_{RB} = \frac{1}{\sqrt{d}} \sum_{x=0}^{d-1} |x\rangle_R\otimes|x\rangle_B,
\end{align}
and $\frac{1}{2}I(R;B) = H(R)=H(B)=\log_2 d$ is just the number of qubits in $A$. To remember the factor of $\frac{1}{2}$ in front of $I(R;E)$, consider the case of a noiseless {\em classical} channel (what we called the completely dephasing channel in \S\ref{subsec:degradable-channels}), where the quantum information completely decoheres in a preferred basis; in that case 
\begin{align}
\phi_{RBE} = \frac{1}{\sqrt{d}} \sum_{x=0}^{d-1} |x\rangle_R\otimes|x\rangle_B\otimes |x\rangle_E,
\end{align}
and $I(R;B) = I(R;E) = H(R)=H(B)=\log_2 d$. Then the father inequality merely expresses the power of quantum teleportation: we can transmit $\frac{n}{2}$ qubits by consuming $\frac{n}{2}$ ebits and sending $n$ bits through the noiseless classical channel. 

Before proving the father resource inequality, we will first discuss a few of its interesting consequences. 

\subsubsection{Entanglement-assisted classical communication.}
Suppose Alice wants to send classical information to Bob, rather than quantum information. Then we can use superdense coding to turn the quantum communication achieved by the father protocol into classical communication, at the cost of consuming some additional entanglement. By invoking the superdense coding resource inequality
\begin{align}
SD:\quad [q\to q] + [qq] \ge 2[c\to c]
\end{align}
$\frac{n}{2}I(R;B)$ times, and combining with the father resource inequality, we obtain $I(R;B)$ bits of classical communication per use of the channel while consuming a number of ebits
\begin{align}
\frac{1}{2} I(R;E) + \frac{1}{2} I(R;B) = H(R)
\end{align}
per channel use. Thus we obtain an achievable rate for entanglement-assisted classical communication through the noisy quantum channel:
\begin{align}
\left\langle \mathcal{N}^{A\to B}:\bfrho_A\right\rangle +H(R)[qq] \ge  I(R;B)[c\to c].
\label{eq:ent-ass-classical}
\end{align}
We may define the {\em entanglement-assisted classical capacity} $C_E(\mathcal{N})$ as the supremum over achievable rates of classical communication per channel use, assuming that an unlimited amount of entanglement is available at no cost. Then the resource inequality eq.(\ref{eq:ent-ass-classical}) implies
\begin{align}
C_E(\mathcal{N})\ge \max_A I(R;B).
\label{eq:ent-ass-class-capacity}
\end{align}
In this case there is a matching upper bound, so eq.(\ref{eq:ent-ass-class-capacity}) is really an equality, and hence a single-letter formula for the entanglement-assisted classical capacity. Furthermore, eq.(\ref{eq:ent-ass-classical}) tells us a rate of entanglement consumption which suffices to achieve the capacity. If we disregard the cost of entanglement, the father protocol shows that a rate can be achieved for entanglement-assisted quantum communication which is half the entanglement-assisted classical capacity $C_E(\mathcal{N})$ of the noisy channel $\mathcal{N}$. That's clearly true, since by consuming entanglement we can use teleportation to convert $n$ bits of classical communication into $n/2$ qubits of quantum communication. 

We also note that for the case where $\mathcal{N}$ is a noisy classical channel, eq.(\ref{eq:ent-ass-class-capacity}) matches Shannon's classical capacity. In that case, eq.(\ref{eq:ent-ass-classical}) asserts that Shannon's classical communication rate is achievable by consuming some entanglement, but it turns out that no consumption of entanglement is actually necessary. 

\subsubsection{Quantum channel capacity.} It may be that Alice wants to send quantum information to Bob, but Alice and Bob are not so fortunate as to have pre-existing entanglement at their disposal. They can still make use of the father protocol, if we are willing to loan them some entanglement, which they are later required to repay. In this case we say that the entanglement {\em catalyzes} the quantum communication. Entanglement is needed to activate the process to begin with, but at the conclusion of the process no net entanglement has been consumed. 

In this catalytic setting, Alice and Bob borrow $\frac{1}{2}I(R;E)$ ebits of entanglement per use of the channel to get started, execute the father protocol, and then sacrifice some of the quantum communication they have generated to replace the borrowed entanglement via the resource inequality
\begin{align}
[q\to q] \ge [qq]. 
\end{align}
After repaying their debt, Alice and Bob retain a number of qubits of quantum communication per channel use
\begin{align}
\frac{1}{2}I(R;B) - \frac{1}{2}I(R;E) = H(B) - H(E) = I_c(R~\rangle B),
\end{align}
the channel's coherent information from $R$ to $B$. We therefore obtain the achievable rate for quantum communication
\begin{align}
\left\langle \mathcal{N}^{A\to B}:\bfrho_A\right\rangle  \ge  I_c(R~\rangle B)[q\to q],
\end{align}
albeit in the catalyzed setting. It can actually be shown that this same rate is achievable without invoking catalysis (see \S\ref{subsec:quantum-capacity-revisited}). As already discussed in \S\ref{subsec:coherent-information}, though, because of the superadditivity of coherent information this resource inequality does not yield a general single-letter formula for the quantum channel capacity $Q(\mathcal{N})$. 

\subsection{Mother: Quantum state transfer}
\label{subsec:mother}
In the mother protocol, Alice, Bob, and Eve initially share a tripartite pure state $\phi_{ABE}$; thus Alice and Bob together hold the purification of Eve's system $E$. Alice wants to send her share of this purification to Bob, using as few qubits of noiseless quantum communication as possible. Therefore, Alice divides her system $A$ into two subsystems $A_1$ and $A_2$, where $A_1$ is as small as possible and $A_2$ is uncorrelated with $E$. She keeps $A_2$ and sends $A_1$ to Bob. After receiving $A_1$, Bob divides $A_1B$ into two subsystems $B_1$ and $B_2$, where $B_1$ purifies $E$ and $B_2$ purifies $A_2$. Thus, at the conclusion of the protocol, Bob holds the purification of $E$ in $B_1$, and in addition Alice and Bob share a bipartite pure state in $A_2B_2$. The protocol is portrayed in the following diagram:

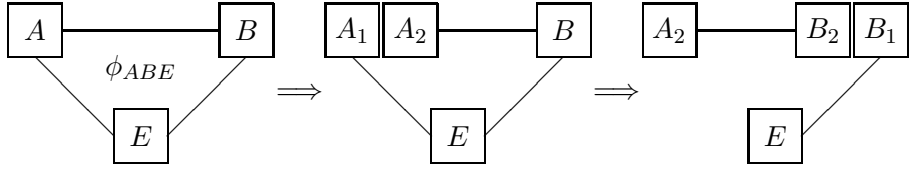
\begin{center}
\begin{picture}(340,60)
\put(0,40){\framebox(20,20){$A$}}
\put(40,0){\framebox(20,20){$E$}}
\put(80,40){\framebox(20,20){$B$}}
\put(20,50){\line(1,0){60}}
\put(10,40){\line(1,-1){30}}
\put(60,10){\line(1,1){30}}
\put(40,30){\makebox(20,12){$\phi_{ABE}$}}
\put(100,20){\makebox(20,12){$\implies$}}
\put(120,40){\framebox(20,20){$A_1$}}
\put(142,40){\framebox(20,20){$A_2$}}
\put(160,0){\framebox(20,20){$E$}}
\put(200,40){\framebox(20,20){$B$}}
\put(162,50){\line(1,0){38}}
\put(130,40){\line(1,-1){30}}
\put(180,10){\line(1,1){30}}
\put(220,20){\makebox(20,12){$\implies$}}
\put(240,40){\framebox(20,20){$A_2$}}
\put(280,0){\framebox(20,20){$E$}}
\put(298,40){\framebox(20,20){$B_2$}}
\put(320,40){\framebox(20,20){$B_1$}}
\put(260,50){\line(1,0){38}}
\put(300,10){\line(1,1){30}}
\end{picture}
\end{center}

In the i.i.d.~version of the mother protocol, the initial state is $\phi_{ABE}^{\otimes n}$, and the task achieved by the protocol is summarized by the mother resource inequality
\begin{equation}
\langle \phi_{ABE} \rangle + \frac{1}{2}I(A;E)[q\to q] \ge \frac{1}{2}I(A;B) [qq] + \langle \phi'_{B_1E} \rangle,
\label{eq:mother-resource-inequality}
\end{equation}
where the resources on the left-hand side can be used to achieve the result on the right-hand side, in an asymptotic i.i.d.~setting, and the entropic quantities are evaluated in the state $\phi_{ABE}$. That is, if $A_1^{(n)}$ denotes the state Alice sends and $A_2^{(n)}$ denotes the state she keeps, then for any positive $\varepsilon$, the state of $A_2^{(n)}E^n$ is $\varepsilon$-close in the $L^1$ norm to a product state, where $\log \left|A_1^{(n)}\right| = \frac{n}{2}I(A;E) + o(n)$, while $A_2^{(n)}B_2^{(n)}$ contains $\frac{n}{2} I(A;B)-o(n)$ shared ebits of entanglement. Eq.(\ref{eq:mother-resource-inequality}) means that for any input pure state $\phi_{ABE}$ there is a way to choose the subsystem $A_2^{(n)}$ of the specified dimension such that $A_2^{(n)}$ and $E^n$ are nearly uncorrelated and the specified amount of entanglement is harvested in $A_2^{(n)}B_2^{(n)}$.

The mother protocol is in a sense {\em dual} to the father protocol. While the father protocol consumes entanglement to achieve quantum communication, the mother protocol consumes quantum communication and harvests entanglement. For the mother, $I(A;B)$ quantifies the correlation between Alice and Bob at the beginning of the protocol (something {\em good}), and $I(A;E)$ quantifies the noise in the initial shared entanglement (something {\em bad}). The mother protocol can also be viewed as a quantum generalization of the Slepian-Wolf distributed compression protocol discussed in \S\ref{subsec:distributed-source}. The mother protocol merges Alice's and Bob's shares of the purification of $E$ by sending Alice's share to Bob, much as distributed source coding merges the classical correlations shared by Alice and Bob by sending Alice's classical information to Bob. For this reason the mother protocol has been called the {\em fully quantum Slepian-Wolf protocol};  the modifier ``fully'' will be clarified below, when we discuss \textit{state merging}, a variant on quantum state transfer in which classical communication is assumed to be freely available. For the mother (or father) protocol, $\frac{1}{2}I(A;E)$ (or $\frac{1}{2}I(R;E)$) quantifies the price we pay to execute the protocol, while $\frac{1}{2}I(A;B)$ (or $\frac{1}{2}I(R;B)$) quantifies the reward we receive.

We may also view the mother protocol as a generalization of the entanglement concentration protocol discussed in \S\ref{sec:concentration-dilution}, extending that discussion in three ways:
\begin{enumerate}
\item The initial entangled state shared by Alice and Bob may be mixed rather than pure.
\item The communication from Alice to Bob is quantum rather than classical.
\item The amount of communication that suffices to execute the protocol is quantified by the resource inequality. 
\end{enumerate}
Also note that if the state of $AE$ is pure (uncorrelated with $B$), then the mother protocol reduces to Schumacher compression. In that case $\frac{1}{2} I(A;E) = H(A)$, and the mother resource inequality states that the purification of $E^n$ can be transferred to Bob with high fidelity using $nH(A) + o(n)$ qubits of quantum communication. 

Before proving the mother resource inequality, we will first discuss a few of its interesting consequences. 

\subsubsection{Hashing inequality.}
Suppose Alice and Bob wish to distill entanglement from many copies of the state $\phi_{ABE}$, using only local operations and classical communication (LOCC). In the catalytic setting, they can borrow some quantum communication, use the mother protocol to distill some shared entanglement, and then use classical communication and their harvested entanglement to repay their debt via quantum teleportation. Using the teleportation resource inequality
\begin{align}
TP:\quad [qq] + 2[c\to c] \ge [q\to q]
\end{align}
$\frac{n}{2} I(A;E)$ times, and combining with the mother resource inequality, we obtain
\begin{equation}
\langle \phi_{ABE} \rangle + I(A;E)[c\to c] \ge I_c(A\rangle B) [qq] + \langle \phi'_{B_1E} \rangle,
\label{eq:hashing-inequality}
\end{equation}
since the net amount of distilled entanglement is $\frac{1}{2}I(A;B)$ per copy of $\phi$ achieved by the mother minus the $\frac{1}{2}I(A;E)$ per copy consumed by teleportation, and
\begin{align}
\frac{1}{2} I(A;B) - \frac{1}{2} I(A;E) = H(B) - H(E) = I_c(A\rangle B).
\end{align}
Eq.(\ref{eq:hashing-inequality}) is the {\em hashing inequality}, which quantifies an achievable rate for distilling ebits of entanglement shared by Alice and Bob from many copies of a mixed state $\bfrho_{AB}$, using one-way classical communication, assuming that $I_c(A\rangle B)=-H(A|B)$ is positive. Furthermore, the hashing inequality tells us how much classical communication suffices for this purpose. 

In the case where the state $\bfrho_{AB}$ is pure, $I_c(A\rangle B) = H(A) - H(AB) = H(A)$ and there is no environment $E$; thus we recover our earlier conclusion about concentration of pure-state bipartite entanglement --- that $H(A)$ Bell pairs can be extracted per copy, with a negligible classical communication cost.

As another example, suppose that Alice prepares $n$ copies of the two-qubit Bell pair $|\phi^+\rangle$, and then sends half of each pair to Bob via the depolarizing channel with error probability $p$, so that after transmission each noisy Bell pair shared by Alice and Bob has fidelity $1-p$ with $|\phi^+\rangle$ as in eq.(\ref{eq:rho-RB-p}). Then the same computation as in \S\ref{subsec:capacity-depolarizing} yields 
\begin{align}
    I_c(A\rangle B) &= H(B)-H(AB) = 1 - H_2(p) - p \log_2 3, \nonumber\\
    I(A,E) &= H(A) - I_c(A\rangle B) = H_2(p) + p \log_2 3 .
\end{align}
The interpretation is similar to our explanation of how to achieve the one-shot capacity of this channel: Alice and Bob can use a random stabilizer code to achieve this distillation rate. Alice measures the code's $n-k$ stabilizer generators on her $n$ qubits, and sends the measurement outcomes to Bob. Bob measures the same stabilizer generators on his $n$ qubits, and notes where his syndrome bits differ from Alice's. Using this information, Bob infers what Pauli operators to apply to his qubits so that the two syndromes now agree; ensuring that Alice and Bob now hold $k$ encoded Bell pairs. If the rate $\bar R = k/n$ is less than $I_c(A\rangle B)$, then for large $n$ Alice and Bob recover $k$ encoded copies of $|\phi^+\rangle$ with high probability. 
\subsubsection{State merging.} 
\label{subsubsec:state-merging}
Suppose Alice and Bob share the purification of Eve's state, and Alice wants to transfer her share of the purification to Bob, where now unlimited {\em classical} communication from Alice to Bob is available at no cost. In contrast to the mother protocol, Alice wants to achieve the transfer with as little one-way quantum communication as possible, even if she needs to send more bits in order to send fewer qubits. 

In the catalytic setting, Alice and Bob can borrow some quantum communication, perform the mother protocol, then use teleportation and the entanglement extracted by the mother protocol to repay some of the borrowed quantum communication. Combining teleportation of $\frac{n}{2}I(A;B)$ qubits with the mother resource inequality, we obtain
\begin{equation}
\langle \phi_{ABE} \rangle + H(A|B)[q\to q] +I(A;B)[c\to c] \ge  \langle \phi'_{B_1E} \rangle,
\label{eq:state-merging}
\end{equation}
using
\begin{align}
\frac{1}{2} I(A;E) - \frac{1}{2}I(A;B) = H(E) - H(B) = H(AB)-H(B) = H(A|B).
\end{align} 
Eq.(\ref{eq:state-merging}) is the {\em state-merging inequality}, expressing how much quantum and classical communication suffices to achieve the state transfer in an i.i.d.~setting, assuming that $H(A|B)$ is nonnegative. 

Like the mother protocol, this state merging protocol can be viewed as a (partially) quantum version of the Slepian-Wolf protocol for merging classical correlations. In the classical setting, $H(X|Y)$ quantifies Bob's remaining ignorance about Alice's information $X$ when Bob knows only $Y$; correspondingly, Alice can reveal $X$ to Bob by sending $H(X|Y)$ bits per letter of $X$. Similarly, state merging provides an operational meaning to the quantum conditional information $H(A|B)$, as the number of qubits per copy of $\phi$ that Alice sends to Bob to convey her share of the purification of $E$, assuming classical communication is free. In this sense we may regard $H(A|B)$ as a measure of Bob's remaining ``ignorance'' about the shared purification of  $E$ when he holds only $B$.

Classically, $H(X|Y)$ is nonnegative, and zero if and only if Bob is already certain about $XY$, but quantumly $H(A|B)$ can be negative. How can Bob have ``negative uncertainty'' about the quantum state of $AB$? If $H(A|B)<0$, or equivalently if $I(A;E) < I(A;B)$, then the mother protocol yields more quantum entanglement than the amount of quantum communication it consumes. Therefore, when $H(A|B)$ is negative ({\em i.e.} $I_c(A\rangle B)$ is positive), the mother resource inequality implies the Hashing inequality, asserting that classical communication from Alice to Bob not only achieves state transfer, but also distills $-H(A|B)$ ebits of entanglement per copy of $\phi$. These distilled ebits can be deposited in the entanglement bank, to be withdrawn as needed in future rounds of state merging, thus reducing the quantum communication cost of those future rounds. Bob's ``negative uncertainty'' today reduces the quantum communication cost of tasks to be performed tomorrow.

\subsection{Operational meaning of strong subadditivity}
\label{subsec:operational-ssa}

The observation that $H(A|B)$ is the quantum communication cost of state merging allows us to formulate a simple operational proof of the strong subadditivity of von Neumann entropy, expressed in the form
\begin{align}
H(A|BC) \le H(A|B), \quad {\rm or} \quad -H(A|B) \le -H(A|BC).
\label{eq:SSA-as-conditional-entropy}
\end{align}
When  $H(A|B)$ is positive, eq.(\ref{eq:SSA-as-conditional-entropy}) is the obvious statement that it is no harder to merge Alice's system with Bob's if Bob holds $C$ as well as $B$. When $H(A|B)$ is negative, eq.(\ref{eq:SSA-as-conditional-entropy}) is the obvious statement that Alice and Bob can distill no less entanglement using one-way classical communication if Bob holds $C$ as well as $B$.

To complete this argument, we need to know that $H(A|B)$ is not only {\em achievable} but also that it is the {\em optimal} quantum communication cost of state merging, and that $-H(A|B)$ ebits is the optimal yield of hashing. The optimality follows from the principle that, for a bipartite pure state, $k$ qubits of quantum communication cannot increase the shared entanglement of $AB$ by more than $k$ ebits. 

If $H(A|B)$ is negative, consider cutting the system $ABE$ into the two parts $AE$ and $B$, as in the following figure:

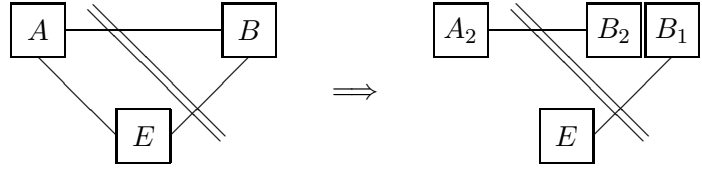
\begin{center}
\begin{picture} (260,60)
\put(0,40){\framebox(20,20){$A$}}
\put(40,0){\framebox(20,20){$E$}}
\put(80,40){\framebox(20,20){$B$}}
\put(20,50){\line(1,0){60}}
\put(10,40){\line(1,-1){30}}
\put(60,10){\line(1,1){30}}
\put(29,58){\line(1,-1){50}}
\put(31,60){\line(1,-1){50}}
\put(120,20){\makebox(20,12){$\implies$}}
\put(160,40){\framebox(20,20){$A_2$}}
\put(200,0){\framebox(20,20){$E$}}
\put(218,40){\framebox(20,20){$B_2$}}
\put(240,40){\framebox(20,20){$B_1$}}
\put(180,50){\line(1,0){38}}
\put(220,10){\line(1,1){30}}
\put(189,58){\line(1,-1){50}}
\put(191,60){\line(1,-1){50}}
\end{picture}
\end{center}
\noindent In the hashing protocol, applied to $n$ copies of $\phi_{ABE}$, the entanglement across this cut at the beginning of the protocol is $nH(B)$. By the end of the protocol $E^n$ has decoupled from $A_2^{(n)}$ and has entanglement $nH(E)$ with $B_1^{(n)}$, ignoring $o(n)$ corrections. If $k$ ebits shared by Alice and Bob are distilled, the final entanglement across the $AE$-$B$ cut is
\begin{align}
nH(E) + k \le nH(B) \implies \frac{k}{n} \le H(B) - H(E) = - H(A|B).
\end{align}
This inequality holds because LOCC cannot increase the entanglement across the cut, and implies that no more than $-H(A|B)$ ebits of entanglement per copy of $\phi_{ABE}$ can be distilled in the hashing protocol, asymptotically. (To be clear, this argument does not exclude entanglement distillation via LOCC at a rate higher than $-H(A|B)$; rather it shows that a protocol achieving distillation at a higher rate could not succeed at transferring the purification of $E$ to $B_1$.)

On the other hand, if $H(A|B)$ is positive, at the conclusion of state merging $B_1^{(n)}$ is entangled with $E^n$, and the entanglement across the $AE$-$B$ cut is at least $nH(E)$. To achieve this increase in entanglement, the number of qubits sent from Alice to Bob must be at least
\begin{align}
k \ge nH(E) - nH(B) \implies \frac{k}{n} \ge H(E) - H(B) = H(A|B).
\end{align}
This inequality holds because the entanglement across the cut cannot increase by more than the quantum communication across the cut, and implies that at least $H(A|B)$ qubits must be sent per copy of $\phi_{ABE}$ to achieve state merging. 

To summarize, we have proven strong subadditivity, not by the traditional route of sophisticated matrix analysis, but via a less direct method. This proof is built on two cornerstones of quantum information theory --- the decoupling principle and the theory of typical subspaces --- which are essential ingredients in the proof of the mother resource inequality. 

\subsection{Negative conditional entropy in thermodynamics}
\label{subsec:negative-entropy}

As a further application of the decoupling mother resource inequality, we now revisit {\em Landauer's Principle}, developing another perspective on the implications of negative quantum conditional entropy. Recall that {\em erasure} of a bit is a process which maps the bit to 0 irrespective of its initial value. This process is {\em irreversible} --- knowing only the final state 0 after erasure, we cannot determine whether the initial state before erasure was 0 or 1. 
Irreversibility implies that erasure incurs an unavoidable thermodynamic cost. According to Landauer's Principle, erasing a bit at temperature $T$ requires work no less than $W = kT \ln 2$. 

A specific erasure procedure is analyzed in Exercise \ref{ex:erasure}. Suppose a two-level quantum system has energy eigenstates $|0\rangle$, $|1\rangle$ with corresponding eigenvalues $E_0=0$ and $E_1=E\ge 0$. Initially the qubit is in an unknown mixture of these two states, and the energy splitting is $E=0$. We erase the bit in three steps. In the first step, we bring the bit into contact with a heat bath at temperature $T> 0$, and wait for the bit to come to thermal equilibrium with the bath. In this step the bit ``forgets'' its initial value, but the bit is not yet erased because it has not been reset. In the second step, with the bit still in contact with the bath, we turn on a control field which slowly increases $E$ to a value much larger than $kT$ while maintaining thermal equilibrium all the while, thus resetting the bit to $|0\rangle$. In the third step, we isolate the bit from the bath and turn off the control field, so the two states of the bit become degenerate again. As shown in Exercise \ref{ex:erasure}, work $W= kT\ln 2$ is required to execute step 2, with the energy dissipated as heat flowing from bit to bath. 

We can also run the last two steps backward, increasing $E$ while the bit is isolated from the bath, then decreasing $E$ with the bit in contact with the bath. This procedure maps the state $|0\rangle$ to the maximally mixed state of the bit, extracting work $W= kT\ln 2$ from the bath in the process. 

Erasure is irreversible because  the agent performing the erasure does not know the information being erased. (If a copy of the information were stored in her memory, survival of that copy would mean that the erasure had not succeeded). From an information-theoretic perspective, the reduction in the thermodynamic entropy of the erased bit, and hence the work required to perform the erasure, arises because erasure reduces the agent's {\em ignorance} about the state of the bit, ignorance which is quantified by the Shannon entropy. But to be more precise, it is the {\em conditional} entropy of the system, given the state of the agent's memory, which captures the agent's ignorance before erasure and therefore also the thermodynamic cost of erasing. Thus the minimal work needed to erase system $A$ should be expressed as 
\begin{align}
W(A|O) = H(A|O) kT \ln 2,
\label{eq:landauer-memory}
\end{align}
where $O$ is the memory of the {\em observer} who performs the erasure, and $H(A|O)$ quantifies that observer's ignorance about the state of $A$. 

But what if $A$ and $O$ are quantum systems? We know that if $A$ and $O$ are entangled, then the conditional entropy $H(A|O)$ can be negative. Does that mean we can erase $A$ while {\em extracting} work rather than doing work? 

Yes, we can! Suppose for example that $A$ and $O$ are qubits and their initial state is maximally entangled. By controlling the contact between $AO$ and the heat bath, the observer can extract work $W=2 kT \log 2$ while transforming $AO$ to a maximally mixed state, using the same work extraction protocol as described above. Then she can do work $W= kT \log 2$ to return $A$ to the state $|0\rangle$. The net effect is to erase $A$ while extracting work $W=kT\log 2$, satisfying the equality eq.(\ref{eq:landauer-memory}). 

To appreciate why this trick works, we should consider the joint state of $AO$ rather than the state of $A$ alone. Although the marginal state of $A$ is mixed at the beginning of the protocol and pure at the end, the state of $AO$ is pure at the beginning and mixed at the end. Positive work is extracted by sacrificing the purity of $AO$. 

To generalize this idea, let's consider $n\gg 1$ copies of the state $\bfrho_{AO}$ of system $A$ and memory $O$. Our goal is to map the $n$ copies of $A$ to the erased state $|000\dots 0\rangle$ while using or extracting the optimal amount of work. In fact, the optimal work per copy is given by eq.(\ref{eq:landauer-memory}) in the $n\to\infty$ limit.

To achieve this asymptotic work per copy, the observer first projects $A^n$ onto its typical subspace, succeeding with probability $1-o(1)$. A unitary transformation then rotates the typical subspace to a subsystem $\bar A$ containing $n(H(A) + o(1))$ qubits, while ``erasing'' the complementary qubits as in eq.(\ref{eq:U-compression}). Now it only remains to erase $\bar A$. 

The mother resource inequality ensures that we may decompose $\bar A$ into subsystems $A_1A_2$ such that $A_2$ contains $\frac{n}{2}\left(I(A;O) - o(1)\right)$ qubits and is nearly maximally entangled with a subsystem of $O^n$.  What is important for the erasure protocol is that we may identify a subsystem of $\bar A O^n$ containing $n\left(I(A;O) - o(1)\right)$ qubits which is only distance $o(1)$ away from a pure state. By controlling the contact between this subsystem and the heat bath, we may extract work $W = n(I(A;O) -o(1))kT \log 2 $ while transforming the subsystem to a maximally mixed state. We then proceed to erase $\bar A$, expending work $kT \log |\bar A|  = n(H(A) + o(1))kT \log 2$. The net work cost of the erasure, per copy of $\bfrho_{AO}$, is therefore 
\begin{align}
W = \left(H(A) - I(A;O) + o(1)\right) kT \log 2 = \left( H(A|O) + o(1)\right) kT \log 2,
\end{align}
and the erasure succeeds with probability $1 - o(1)$. A notable feature of the protocol is that only the subsystem of $O^n$ which is entangled with $A_2$ is affected. Any correlation of the memory $O$ with other systems remains intact, and can be exploited in the future to reduce the cost of erasure of those other systems. 

As does the state merging protocol, this erasure protocol provides an operational interpretation of strong subadditivity. For positive $H(A|O)$, 
$H(A|O) \ge H(A|OO')$
means that it is no harder to erase $A$ if the observer has access to both $O$ and $O'$ than if she has access to $O$ alone. For negative $H(A|O)$, 
${-}H(A|OO') \ge {-}H(A|O)$
means that we can extract at least as much work from $AOO'$ as from its subsystem $AO$.

To carry out this protocol and extract the optimal amount of work while erasing $A$, we need to know which subsystem of $O^n$ provides the purification of $A_2$. The decoupling argument ensures that this subsystem exists, but does not provide a constructive method for finding it, and therefore no concrete protocol for erasing at optimal cost. This quandary is characteristic of Shannon theory; for example, Shannon's noisy channel coding theorem ensures the existence of a code that achieves the channel capacity, but does not provide any explicit code construction.

\section{The Decoupling Inequality}
\label{sec:decoupling-inequality}

Achievable rates for quantum protocols are derived by using random codes, much as in classical Shannon theory. But this similarity between classical and quantum Shannon theory is superficial --- at a deeper conceptual level, quantum protocols differ substantially from classical ones. Indeed, the decoupling principle underlies many of the key findings of quantum Shannon theory, providing a unifying theme that ties together many different results. In particular, the mother and father resource inequalities, and hence all their descendants enumerated above, follow from an inequality that specifies a sufficient condition for decoupling. 

This \textit{decoupling inequality} addresses the following question: Suppose that Alice and Eve share a quantum state $\bfsigma_{AE}$, where $A$ is an $n$-qubit system. This state may be mixed, but in general $A$ and $E$ are correlated; that is, $I(A;E) > 0$. Now Alice starts discarding qubits one at a time, where each qubit is a randomly selected two-dimensional subsystem of what Alice holds. Each time Alice discards a qubit, her correlation with $E$ grows weaker. How many qubits should she discard so that the subsystem she retains has a negligible correlation with Eve's system $E$?

To make the question precise, we need to formalize what it means to discard a random qubit. More generally, suppose that $A$ has dimension $|A|$, and Alice decomposes $A$ into subsystems $A_1$ and $A_2$, then discards $A_1$ and retains $A_2$. We would like to consider many possible ways of choosing the discarded system with specified dimension $|A_1|$. Equivalently, we may consider a fixed decomposition $A=A_1A_2$, where we apply a unitary transformation $\bfU$ to $A$ before discarding $A_1$. Then discarding a random subsystem with dimension $|A_1|$ is the same thing as applying a random unitary $\bfU$ before discarding the fixed subsystem $A_1$:

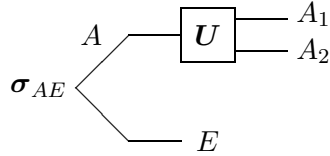
\begin{center}
\begin{picture} (124,60)
\put(0,20){\makebox(20,12){$\bfsigma_{AE}$}}
\put(24,26){\line(1,-1){20}}
\put(44,6){\line(1,0){20}}
\put(64,0){\makebox(20,12){$E$}}
\put(20,40){\makebox(20,12){$A$}}
\put(24,26){\line(1,1){20}}
\put(44,46){\line(1,0){20}}
\put(64,36){\framebox(20,20){$\bfU$}}
\put(84,52){\line(1,0){20}}
\put(104,48){\makebox(20,12){$A_1$}}
\put(84,40){\line(1,0){20}}
\put(104,34){\makebox(20,12){$A_2$}}
\end{picture}
\end{center}

To analyze the consequences of discarding a random subsystem, then, we will need to be able to compute the expectation value of a function $f(\bfU)$ when we average $\bfU$ uniformly over the group of unitary $|A|\times |A|$ matrices. We denote this expectation value as $\mathbb{E}_\bfU\left[f(\bfU)\right]$; to perform computations we will only need to know that $\mathbb{E}_\bfU$ is suitably normalized, and is invariant under left or right multiplication by any constant unitary matrix $\bfV$:
\begin{align}
\mathbb{E}_\bfU\left[1\right]=1,\quad \mathbb{E}_\bfU\left[f(\bfU)\right]= \mathbb{E}_\bfU\left[f(\bfV\bfU)\right]=\mathbb{E}_\bfU\left[f(\bfU\bfV)\right].
\label{eq:expectation-unitary-invariance}
\end{align}
These conditions uniquely define $\mathbb{E}_\bfU\left[f(\bfU)\right]$, which is sometimes described as the integral over the unitary group using the \textit{invariant measure} or \textit{Haar measure} on the group. 

If we apply the unitary transformation $\bfU$ to $A$, and then discard $A_1$, the marginal state of $A_2E$ is
\begin{align}
\bfsigma_{A_2E}(\bfU) := {\rm tr}_{A_1} \left(\left(\bfU_A\otimes \bfI_E\right)\bfsigma_{AE}\left(\bfU_A^\dagger\otimes \bfI_E\right)\right).
\end{align}
The decoupling inequality expresses how close (in the $L^1$ norm) $\bfsigma_{A_2E}$ is to a product state when we average over $\bfU$:
\begin{align}
\left(\mathbb{E}_\bfU\big[~\|\bfsigma_{A_2E}(\bfU) - \bfsigma_{A_2}^{\max}\otimes \bfsigma_E\|_1~\big]\right)^2 \le \frac{|A_2|\cdot|E|}{~|A_1|}~{\rm tr} \left(\bfsigma_{AE}^2\right),
\label{eq:decoupling-inequality}
\end{align}
where 
\begin{align}
\bfsigma_{A_2}^{\max} := \frac{1}{|A_2|} ~\bfI
\end{align}
denotes the maximally mixed state on $A_2$, and $\bfsigma_E$ is the marginal state ${\rm tr}_A\bfsigma_{AE}$. 

This inequality has interesting consequences even in the case where there is no system $E$ at all and $\bfsigma_A$ is pure, where it becomes
\begin{align}
\mathbb{E}_\bfU\big[~\|\bfsigma_{A_2}(\bfU) - \bfsigma_{A_2}^{\max}\|_1~\big] \le \sqrt{\frac{|A_2|}{~|A_1|}~{\rm tr} \left(\bfsigma_{A}^2\right)}= \sqrt{\frac{|A_2|}{|A_1|}}.
\label{eq:random-no-E}
\end{align}
Eq.(\ref{eq:random-no-E}) implies that, for a randomly chosen pure state of the bipartite system $A=A_1A_2$, where $|A_2|/|A_1|  \ll 1$, the density operator on $A_2$ is very nearly maximally mixed with high probability. One can likewise show that the expectation value of the entanglement entropy of $A_1A_2$ is very close to the maximal value: $\mathbb{E}\left[H(A_2)\right] \ge \log_2 |A_2| - |A_2|/\left(2|A_1|\ln 2\right)$. Thus, if for example $A_2$ is 50 qubits and $A_1$ is 100 qubits, the typical entropy deviates from maximal by only about $2^{-50}\approx 10^{-15}$.

\subsection{Proof of the decoupling inequality}
\label{subsec:decoupling-proof}
To prove the decoupling inequality, we will first bound the distance between $\bfsigma_{A_2E}$ and a product state in the $L^2$ norm, and then use the Cauchy-Schwarz inequality to obtain a bound on the $L^1$ distance. Eq.(\ref{eq:decoupling-inequality}) follows from
\begin{align}
\mathbb{E}_\bfU\big[~\|\bfsigma_{A_2E}(\bfU) - \bfsigma_{A_2}^{\max}\otimes \bfsigma_E\|_2^2~\big] \le \frac{1}{|A_1|}~{\rm tr} \left(\bfsigma_{AE}^2\right),
\label{eq:decoupling-L2-norm}
\end{align}
combined with
\begin{align}
\left(\mathbb{E}\left[{f(\bfU)}\right]\right)^2\le \mathbb{E}\left[{f(\bfU)^2}\right]\quad{\rm and}\quad
\|M\|_1^2 \le d \|M\|_2^2
\end{align}
(for nonnegative $f$), which implies 
\begin{align}
\left(\mathbb{E}\left[\|\cdot\|_1\right]\right)^2 \le \mathbb{E}\left[\|\cdot \|_1^2 \right]\le |A_2|\cdot |E| \cdot \mathbb{E}\left[\|\cdot\|_2^2\right].
\end{align}
We also note that
\begin{align}
\|\bfsigma_{A_2E} - \bfsigma_{A_2}^{\max}\otimes \bfsigma_E\|_2^2 &={\rm tr} \left(\bfsigma_{A_2E} - \bfsigma_{A_2}^{\max}\otimes \bfsigma_E\right)^2 \notag\\
&= {\rm tr}\left(\bfsigma_{A_2E}^2\right) - \frac{1}{|A_2|} {\rm tr}\left( \bfsigma_E^2\right),
\end{align}
because 
\begin{align}
{\rm tr}\left( \bfsigma_{A_2}^{\max}\right)^2 = \frac{1}{|A_2|};
\end{align}
therefore, to prove eq.(\ref{eq:decoupling-L2-norm}) it suffices to show
\begin{align}
\mathbb{E}_\bfU\left[{\rm tr}\left(\bfsigma_{A_2E}^2(\bfU)\right)\right]\le \frac{1}{|A_2|} {\rm tr}\left(\bfsigma_E^2\right) +\frac{1}{|A_1|} {\rm tr}\left(\bfsigma_{AE}^2\right).
\label{eq:decoupling-inequality-traces}
\end{align}

We can facilitate the computation of $\mathbb{E}_\bfU\left[{\rm tr}\left(\bfsigma_{A_2E}^2(\bfU)\right)\right]$ using a clever trick. For any bipartite system $BC$, imagine introducing a second copy $B'C'$ of the system. Then (Exercise \ref{ex:decoupling})
\begin{align}
{\rm tr}_C \left(\bfsigma_C^2\right) = {\rm tr}_{BCB'C'}\left( \bfI_{BB'}\otimes\bfS_{CC'}\right)\left(\bfsigma_{BC}\otimes \bfsigma_{B'C'}\right),
\label{eq:swap-in-trace-trick}
\end{align} 
where $\bfS_{CC'}$ denotes the swap operator, which acts as
\begin{equation}
\bfS_{CC'}:|i\rangle_C\otimes |j\rangle_{C'} \mapsto |j\rangle_C\otimes |i\rangle_{C'}.
\end{equation}
In particular, then,
\begin{align}
&{\rm tr}_{A_2E} \left(\bfsigma_{A_2E}^2(\bfU)\right) \notag\\
&= {\rm tr}_{AEA'E'}\left(\bfI_{A_1A_1'}\otimes \bfS_{A_2A_2'}\otimes\bfS_{EE'}\right)\left(\bfsigma_{AE}(\bfU)\otimes \bfsigma_{A'E'}(\bfU)\right)\notag\\
&= {\rm tr}_{AEA'E'}\left(\bfM_{AA'}(\bfU)\otimes\bfS_{EE'}\right)\left(\bfsigma_{AE}\otimes \bfsigma_{A'E'}\right),
\label{eq:decoupling-tensor}
\end{align} 
where
\begin{align}
\bfM_{AA'}(\bfU)= \left(\bfU_A^\dagger\otimes\bfU_{A'}^\dagger\right)\left(\bfI_{A_1A_1'}\otimes \bfS_{A_2A_2'}\right)\left(\bfU_A\otimes\bfU_{A'}\right).
\end{align}
The expectation value of $\bfM_{AA'}(\bfU)$ is evaluated in Exercise \ref{ex:decoupling}; there we find
\begin{align}
\mathbb{E}_{\bfU}\left[ \bfM_{AA'}(\bfU)\right] = c_\bfI \bfI_{AA'} + c_{\bfS} \bfS_{AA'}
\label{eq:exp-value-2-2-poly}
\end{align}
where
\begin{align}
c_\bfI &= \frac{1}{|A_2|} \left(\frac{1 - 1/|A_1|^2}{1-1/|A|^2}\right) \le \frac{1}{|A_2|}, \notag\\
c_{\bfS} &= \frac{1}{|A_1|} \left(\frac{1 - 1/|A_2|^2}{1-1/|A|^2}\right) \le \frac{1}{|A_1|}.
\label{eq:c_bfI-c_bfS}
\end{align}
Plugging into eq.(\ref{eq:decoupling-tensor}), we then obtain
\begin{align}
&\mathbb{E}_\bfU\left[{\rm tr}_{A_2E} \left(\bfsigma_{A_2E}^2(\bfU)\right)\right]\notag\\
&\le 
{\rm tr}_{AEA'E'}\left(\left(\frac{1}{|A_2|} \bfI_{AA'} + \frac{1}{|A_1|} \bfS_{AA'}\right)\otimes\bfS_{EE'}\right)\left(\bfsigma_{AE}\otimes \bfsigma_{A'E'}\right)\notag\\
&= \frac{1}{|A_2|}{\rm tr}\left(\bfsigma_E^2\right) + \frac{1}{|A_1|}\left(\bfsigma_{AE}^2\right),
\end{align}
thus proving eq.(\ref{eq:decoupling-inequality-traces}) as desired.

\subsection{Proof of the mother inequality}

The mother inequality eq.(\ref{eq:mother-resource-inequality}) follows from the decoupling inequality eq.(\ref{eq:decoupling-inequality}) in an i.i.d.~setting. Suppose Alice, Bob, and Eve share the pure state $\phi_{ABE}^{\otimes n}$. Then there are jointly typical subspaces of $A^n$, $B^n$, and $E^n$, which we denote by $\bar A$, $\bar B$, $\bar E$, such that 
\begin{align}
\left| \bar A\right| = 2^{nH(A) + o(n)}, \quad \left| \bar B\right| = 2^{nH(B) + o(n)}, \quad \left| \bar E\right| = 2^{nH(E) + o(n)}.
\end{align}
Furthermore, the normalized pure state $\phi'_{\bar A \bar B \bar E}$ obtained by projecting $\phi_{ABE}^{\otimes n}$ onto $\bar A\otimes \bar B\otimes \bar E$ deviates from $\phi_{ABE}^{\otimes n}$ by distance $o(1)$ in the $L^1$ norm.

In order to transfer the purification of $E^n$ to Bob, Alice first projects $A^n$ onto its typical subspace, succeeding with probability $1 - o(1)$, and compresses the result. She then divides her compressed system $\bar A$ into two parts $\bar A_1\bar A_2$, and applies a random unitary to $\bar A$ before sending $\bar A_1$ to Bob. Quantum state transfer is achieved if $\bar A_2$ decouples from $\bar E$. 

Because $\phi'_{\bar A \bar B \bar E}$ is close to $\phi_{ABE}^{\otimes n}$, we can analyze whether the protocol is successful by supposing the initial state is $\phi'_{\bar A\bar B \bar E}$ rather than $\phi_{ABE}^{\otimes n}$. According to the decoupling inequality
\begin{align}
&\left(\mathbb{E}_\bfU\big[~\|\bfsigma_{\bar A_2\bar E}(\bfU) - \bfsigma_{\bar A_2}^{\max}\otimes \bfsigma_{\bar E}\|_1~\big]\right)^2 \le \frac{|\bar A|\cdot |\bar E|}{~|\bar A_1|^2}~{\rm tr} \left(\bfsigma_{\bar A\bar E }^2\right)\notag\\
&= \frac{1}{~|\bar A_1|^2} 2^{n(H(A) +H(E)+o(1))}~{\rm tr} \left(\bfsigma_{\bar A\bar E }^2\right)= \frac{1}{~|\bar A_1|^2} 2^{n(H(A) +H(E)-H(B)+o(1))};
\label{eq:decoupling-inequality-typical}
\end{align}
here we have used properties of typical subspaces in the second line, as well as the property that $\bfsigma_{\bar A \bar E}$ and $\bfsigma_{\bar B}$ have the same nonzero eigenvalues, because $\phi'_{\bar A\bar B \bar E}$ is pure.

Eq.(\ref{eq:decoupling-inequality-typical}) bounds the $L^1$ distance of $\bfsigma_{\bar A_2\bar E}(\bfU)$ from a product state when averaged over all unitaries, and therefore suffices to ensure the existence of at least one unitary transformation $\bfU$ such that the $L^1$ distance is bounded above by the right-hand side. Therefore, by choosing this $\bfU$, Alice can decouple $\bar A_2$ from $E^n$ to $o(1)$ accuracy in the $L^1$ norm by sending to Bob
\begin{align}
\log_2 |\bar A_1| = \frac{n}{2}\left(H(A)+H(E) - H(B) + o(1)\right) = \frac{n}{2} \left(I(A;E) + o(1)\right)
\label{eq:log1-decouple}
\end{align}
qubits, suitably chosen from the (compressed) typical subspace of $A^n$. Alice retains $|\bar A_2| = nH(A) - \frac{n}{2}I(A;E) -o(n)$ qubits of her compressed system, which are nearly maximally mixed and uncorrelated with $E^n$; hence at the end of the protocol she shares with Bob this many qubit pairs, which have high fidelity with a maximally entangled state. Since $\phi_{ABE}$ is pure, and therefore $H(A) = \frac{1}{2}\left(I(A;E) + I(A;B)\right)$, we conclude that Alice and Bob distill $\frac{n}{2}I(A;B)-o(n)$ ebits of entanglement, thus proving the mother resource inequality.

We can check that this conclusion is plausible using a crude counting argument. Disregarding the $o(n)$ corrections in the exponent, the state $\phi_{ABE}^{\otimes n}$ is nearly maximally mixed on a typical subspace of $A^nE^n$ with dimension $2^{nH(AE)}$, {\it i.e.} the marginal state on $\bar A \bar E$ can be realized as a nearly uniform ensemble of this many mutually orthogonal states. If $\bar A_1$ is randomly chosen and sufficiently small, we expect that, for each state in this ensemble, $\bar A_1$ is nearly maximally entangled with a subsystem of the much larger system $\bar A_2 \bar E$, and that the marginal states on $\bar A_2\bar E$ arising from different states in the $\bar A\bar E$ ensemble have a small overlap. Therefore, we anticipate that tracing out $\bar A_1$ yields a state on $\bar A_2 \bar E$ which is nearly maximally mixed on a subspace with dimension $|\bar A_1| 2^{nH(AE)}$. Approximate decoupling occurs when this state attains full rank on $\bar A_2 \bar E$, since in that case it is close to maximally mixed on $\bar A_2 \bar E$ and therefore close to a product state on its support. The state transfer succeeds, therefore, provided
\begin{align}
&|\bar A_1| 2^{nH(AE)}\approx |\bar A_2|\cdot |\bar E| =\frac{|\bar A|\cdot |\bar E|}{|\bar A_1|}\approx \frac{2^{n(H(A)+H(E))}}{|\bar A_1|} \notag\\
&\implies |\bar A_1|^2 \approx 2^{n I(A;E)},
\end{align}
as in eq.(\ref{eq:log1-decouple}).

Our derivation of the mother resource inequality, based on random coding, does not exhibit any concrete protocol that achieves the claimed rate, nor does it guarantee the existence of any protocol in which the required quantum processing can be executed efficiently. Concerning the latter point, it is notable that our derivation of the decoupling inequality applies not just to the expectation value averaged uniformly over the unitary group, but also to any average over unitary transformations which satisfies eq.(\ref{eq:exp-value-2-2-poly}). In fact, this identity is satisfied by a uniform average over the Clifford group, which means that there is some Clifford transformation on $\bar A$ which achieves the rates specified in the mother resource inequality. Any Clifford transformation on $n$ qubits can be reached by a circuit with $O(n^2)$ gates. Since it is also known that Schumacher compression can be achieved by a polynomial-time quantum computation, Alice's encoding operation can be carried out efficiently.

In fact, after compressing, Alice encodes the quantum information she sends to Bob using a stabilizer code (with Clifford encoder $\bfU$), and Bob's task, after receiving $\bar A_1$ is to correct the erasure of $\bar A_2$. Bob can replace each erased qubit by the standard state $|0\rangle$, and then measure the code's check
operators. With high probability, there is a unique Pauli operator acting on the erased
qubits that restores Bob's state to the code space, and the recovery operation can be
efficiently computed using linear algebra. Hence, Bob's part of the mother protocol, like Alice's, can be executed efficiently. (Projections onto typical subspaces are used in the proof of the mother inequality, but not in the protocol itself, so we don't need to worry about whether these projections can be performed efficiently.)

\subsection{Proof of the father inequality}
\label{subsec:proof-father}

\subsubsection{One-shot version.}
In the one-shot version of the father protocol, Alice and Bob share a pair of maximally entangled systems $A_1B_1$, and in addition Alice holds input state $\bfrho_{A_2}$ of system $A_2$ which she wants to convey to Bob. Alice encodes $\bfrho_{A_2}$ by applying a unitary transformation $\bfV$ to $A=A_1A_2$, then sends $A$ to Bob via the noisy quantum channel $\mathcal{N}^{A\to B_2}$. Bob applies a decoding map $\mathcal{D}^{B_1B_2\to \tilde A_2}$ jointly to the channel output and his half of the entangled state he shares with Alice, hoping to recover Alice's input state with high fidelity:

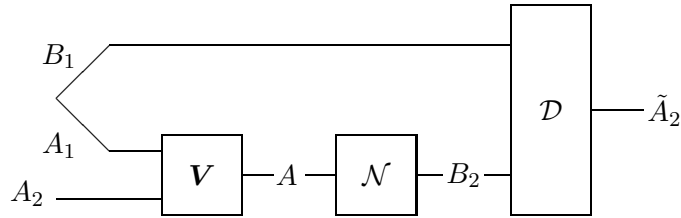
\begin{center}
\begin{picture} (256,80)
\put(16,20){\makebox(10,12){$A_1$}}
\put(16,54){\makebox(10,12){$B_1$}}
\put(4,2){\makebox(10,12){$A_2$}}
\put(20,44){\line(1,-1){20}}
\put(20,44){\line(1,1){20}}
\put(40,64){\line(1,0){152}}
\put(40,24){\line(1,0){20}}
\put(20,6){\line(1,0){40}}
\put(60,0){\framebox(30,30){$\bfV$}}
\put(90,15){\line(1,0){12}}
\put(102,9){\makebox(10,12){$A$}}
\put(114,15){\line(1,0){12}}
\put(126,0){\framebox(30,30){$\mathcal{N}$}}
\put(156,15){\line(1,0){10}}
\put(168,9){\makebox(12,12){$B_2$}}
\put(182,15){\line(1,0){10}}
\put(192,0){\framebox(30,79){$\mathcal{D}$}}
\put(182,15){\line(1,0){10}}
\put(222,39.5){\line(1,0){20}}
\put(244,33.5){\makebox(12,12){$\tilde A_2$}}
\end{picture}
\end{center}
\noindent We would like to know how much shared entanglement suffices for Alice and Bob to succeed. 

This question can be answered using the decoupling inequality. First we introduce a reference system $R'$ which is maximally entangled with $A_2$; then Bob succeeds if his decoder can extract the purification of $R'$. Because the system $R'B_1$ is maximally entangled with $A_1A_2$, the encoding unitary $\bfV$ acting on $A_1A_2$ can be replaced by its transpose $\bfV^T$ acting on $R' B_1$. We may also replace $\mathcal{N}$ by its Stinespring dilation $\bfU^{A_1A_2\to B_2E}$, so that the extended output state $\phi$ of $R'B_1B_2 E$ is pure: 

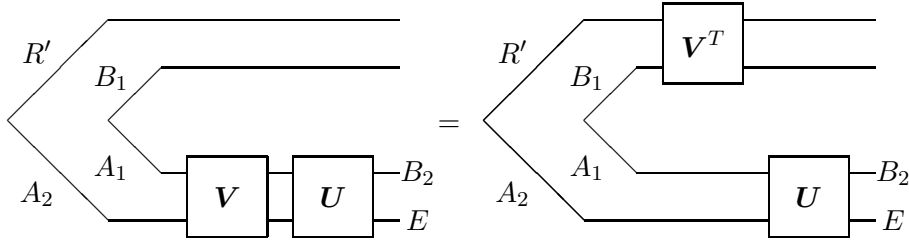
\begin{center}
\begin{picture} (344,90)
\put(38,20){\makebox(10,12){$A_1$}}
\put(38,54){\makebox(10,12){$B_1$}}
\put(10,10){\makebox(10,12){$A_2$}}
\put(10,64){\makebox(10,12){$R'$}}
\put(42,44){\line(1,-1){20}}
\put(42,44){\line(1,1){20}}
\put(62,64){\line(1,0){90}}
\put(62,24){\line(1,0){10}}
\put(4,44){\line(1,-1){38}}
\put(4,44){\line(1,1){38}}
\put(42,82){\line(1,0){110}}
\put(42,6){\line(1,0){30}}
\put(72,0){\framebox(30,30){$\bfV$}}
\put(102,6){\line(1,0){10}}
\put(102,24){\line(1,0){10}}
\put(112,0){\framebox(30,30){$\bfU$}}
\put(142,6){\line(1,0){10}}
\put(142,24){\line(1,0){10}}
\put(154,18){\makebox(10,12){$B_2$}}
\put(154,0){\makebox(10,12){$E$}}
\put(166,36){\makebox(10,12){$=$}}
\put(184,44){\line(1,-1){38}}
\put(184,44){\line(1,1){38}}
\put(218,20){\makebox(10,12){$A_1$}}
\put(218,54){\makebox(10,12){$B_1$}}
\put(190,10){\makebox(10,12){$A_2$}}
\put(190,64){\makebox(10,12){$R'$}}
\put(222,44){\line(1,-1){20}}
\put(222,44){\line(1,1){20}}
\put(242,64){\line(1,0){10}}
\put(282,64){\line(1,0){50}}
\put(242,24){\line(1,0){50}}
\put(184,44){\line(1,-1){38}}
\put(184,44){\line(1,1){38}}
\put(282,82){\line(1,0){50}}
\put(222,82){\line(1,0){30}}
\put(222,6){\line(1,0){70}}
\put(252,58){\framebox(30,30){$\bfV^T$}}
\put(292,0){\framebox(30,30){$\bfU$}}
\put(322,6){\line(1,0){10}}
\put(322,24){\line(1,0){10}}
\put(334,18){\makebox(10,12){$B_2$}}
\put(334,0){\makebox(10,12){$E$}}
\end{picture}
\end{center}

\noindent Finally we invoke the decoupling principle --- if $R'$ and $E$ decouple, then $R'$ is purified by a subsystem of $B_1 B_2$, which means that Bob can recover $\bfrho_{A_2}$ with a suitable decoding map.

If we consider $\bfV$, and hence also $\bfV^T$, to be a random unitary, then we may describe the situation this way: We have a tripartite pure state $\phi_{RB_2 E}$, where $R = R'B_1$, and we would like to know whether the marginal state of $R' E$ is close to a product state when the random subsystem $B_1$ is discarded from $R$. This is exactly the question addressed by the decoupling inequality, which in this case may be expressed as 
\begin{align}
\left(\mathbb{E}_\bfV\big[~\|\bfsigma_{R'E}(\bfV) - \bfsigma_{R'}^{\max}\otimes \bfsigma_E\|_1~\big]\right)^2 \le \frac{|R|\cdot|E|}{~|B_1|^2}~{\rm tr} \left(\bfsigma_{RE}^2\right),
\label{eq:decoupling-for-father}
\end{align}
Eq.(\ref{eq:decoupling-for-father}) asserts that the $L^1$ distance from a product state is bounded above when averaged uniformly over all unitary $\bfV$'s; therefore there must be some particular encoding unitary $\bfV$ that satisfies the same bound. We conclude that near-perfect decoupling of $R'E$, and therefore high-fidelity decoding of $B_2$, is achievable provided that 
\begin{align}
|A_1| = |B_1| \gg |R'|\cdot|E| ~{\rm tr} \left(\bfsigma_{RE}^2\right) = |A_2| \cdot |E| 
~ {\rm tr} \left(\bfsigma_{B_2}^2\right),
\end{align}
where to obtain the second equality we use the purity of $\phi_{RB_2 E}$ and recall that the reference system $R'$ is maximally entangled with $A_2$, and therefore has the same dimension. 

\subsubsection{i.i.d.~version.} In the i.i.d.~version of the father protocol, Alice and Bob achieve high fidelity entanglement-assisted quantum communication through $n$ uses of the quantum channel $\mathcal{N}^{A\to B}$. The code they use for this purpose can be described in the following way:  Consider an input density operator $\bfrho_A$ of system $A$, which is purified by a reference system $R$. Sending the purified input state $\psi_{RA }$ through $\bfU^{A\to BE}$, the isometric dilation of $\mathcal{N}^{A\to B}$, generates the tripartite pure state $\phi_{RBE}$. Evidently applying $\left(\bfU^{A\to BE}\right)^{\otimes n}$ to $\psi_{RA}^{\otimes n}$ produces $\phi_{RBE}^{\otimes n}$. 

But now suppose that before transmitting the state to Bob, Alice projects $A^n$ onto its typical subspace $\bar A$, succeeding with probability $1 - o(1)$ in preparing a state of $\bar A \bar R$ that is nearly maximally entangled, where $\bar R$ is the typical subspace of $R^n$. Imagine dividing $\bar R$ into a randomly chosen subsystem $B_1$ and its complementary subsystem $R'$; then there is a corresponding decomposition of $\bar A=A_1A_2$ such that $A_1$ is very nearly maximally entangled with $B_1$ and $A_2$ is very nearly maximally entangled with $R'$.  

If we interpret $B_1$ as Bob's half of an entangled state of $A_1B_1$ shared with Alice, this becomes the setting where the one-shot father protocol applies, if we ignore the small deviation from maximal entanglement in $A_1B_1$ and $R'A_2$. As for our analysis of the i.i.d.~mother protocol, we apply the one-shot father inequality not to $\phi_{RBE}^{\otimes n}$, but rather to the nearby state $\phi'_{\bar R\bar B\bar E}$, where $\bar B$ and $\bar E$ are the typical subspaces of $B^n$ and $E^n$ respectively. Applying eq.(\ref{eq:decoupling-for-father}), and using properties of typical subspaces, we can bound the square of the $L^1$ deviation of $R'\bar E$ from a product state, averaged over the choice of $B_1$, by 
\begin{align}
\frac{ |\bar R|\cdot|\bar E|}{~|B_1|^2}~{\rm tr} \left(\bfsigma_{\bar B}^2\right) = \frac{2^{n(H(R) + H(E)- H(B)+o(1))}}{|B_1|^2} = \frac{2^{n(I(R;E)+o(1))}}{|B_1|^2};
\end{align}
hence the bound also applies for some particular way of choosing $B_1$. This choice defines the code used by Alice and Bob in a protocol which consumes 
\begin{align}
\log_2 |B_1| = \frac{n}{2}I(R;E) + o(n)
\end{align}
ebits of entanglement, and conveys from Alice to Bob
\begin{align}
nH(R) -\frac{n}{2} I(R;E) - o(n) = \frac{n}{2}I(R;B) - o(n)
\end{align}
high-fidelity qubits. This proves the father resource inequality. 


\subsection{Quantum channel capacity revisited}
\label{subsec:quantum-capacity-revisited}

In \S\ref{subsec:father-protocol} we showed that the coherent information is an achievable rate for quantum communication over a noisy quantum channel. That derivation, a corollary of the father resource inequality, applied to a catalytic setting, in which shared entanglement between sender and receiver can be borrowed and later repaid. It is useful to see that the same rate is achievable without catalysis, a result we can derive from an alternative version of the decoupling inequality. 

This version applies to the setting depicted here:

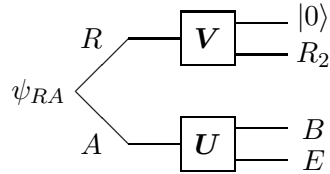
\begin{center}
\begin{picture} (124,60)
\put(0,20){\makebox(20,12){$\psi_{RA}$}}
\put(24,26){\line(1,-1){20}}
\put(44,6){\line(1,0){20}}
\put(20,40){\makebox(20,12){$R$}}
\put(20,0){\makebox(20,12){$A$}}
\put(24,26){\line(1,1){20}}
\put(44,46){\line(1,0){20}}
\put(64,36){\framebox(20,20){$\bfV$}}
\put(64,-4){\framebox(20,20){$\bfU$}}
\put(84,52){\line(1,0){20}}
\put(104,48){\makebox(20,12){$|0\rangle$}}
\put(84,40){\line(1,0){20}}
\put(104,34){\makebox(20,12){$R_2$}}
\put(84,0){\line(1,0){20}}
\put(84,12){\line(1,0){20}}
\put(104,6){\makebox(20,12){$B$}}
\put(104,-6){\makebox(20,12){$E$}}
\end{picture}
\end{center}

\noindent A density operator $\bfrho_A$ for system $A$, with purification $\psi_{RA}$, is transmitted through a channel $\mathcal{N}^{A\to B}$ which has the isometric dilation $\bfU^{A\to BE}$. The reference system $R$ has a decomposition into subsystems $R_1 R_2$. We apply a random unitary transformation $\bfV$ to $R$, then project $R_1$ onto a fixed vector $|0\rangle_{R_1}$, and renormalize the resulting state. In effect, then we are projecting $R$ onto a subspace with dimension $|R_2|$, which purifies a corresponding code subspace of $A$. This procedure prepares a normalized pure state $\phi_{R_2 BE}$, and a corresponding normalized marginal state $\sigma_{R_2 E}$ of $R_2 E$.

If $R_2$ decouples from $E$, then $R_2$ is purified by a subsystem of $B$, which means that the code subspace of $A$ can be recovered by a decoder applied to $B$. A sufficient condition for approximate decoupling can be derived from the inequality
\begin{align}
\left(\mathbb{E}_\bfV\big[~\|\bfsigma_{R_2E}(\bfV) - \bfsigma_{R_2}^{\max}\otimes \bfsigma_E\|_1~\big]\right)^2 \le |R_2|\cdot |E|~{\rm tr} \left(\bfsigma_{RE}^2\right).
\label{eq:decoupling-inequality-projected}
\end{align}
Eq.(\ref{eq:decoupling-inequality-projected}) resembles eq.(\ref{eq:decoupling-inequality}) and can be derived by a similar method. Note that the right-hand side of eq.(\ref{eq:decoupling-inequality-projected}) is enhanced by a factor of $|R_1|$ relative to the right-hand side of eq.(\ref{eq:decoupling-inequality}). This factor arises because after projecting $R_1$ onto the fixed state $|0\rangle$ we need to renormalize the state by multiplying by $|R_1|$, while on the other hand the projection suppresses the expected distance squared from a product state by a factor $|R_1|$.

In the i.i.d.~setting where the noisy channel is used $n$ times, we consider $\phi_{RBE}^{\otimes n}$, and project onto the jointly typical subspaces $\bar R$, $\bar B$, $\bar E$ of $R^n$, $B^n$, $E^n$ respectively, succeeding with high probability. We choose a code by projecting $\bar R$ onto a random subspace with dimension $|R_2|$. Then, the right-hand side of eq.(\ref{eq:decoupling-inequality-projected}) becomes
\begin{align}
|R_2| \cdot 2^{n(H(E) - H(B)+ o(1))},
\end{align}
and since the inequality holds when we average uniformly over $\bfV$, it surely holds for some particular $\bfV$. That unitary defines a code which achieves decoupling and has the rate
\begin{align}
\frac{1}{n}\log_2 |R_2| = H(E) - H(B) - o(1) = I_c(R~\rangle B) - o(1).
\end{align}
Hence the coherent information is an achievable rate for high-fidelity quantum communication over the noisy channel. 

\subsection{Black holes as mirrors}
As our final application of the decoupling inequality, we consider a highly idealized model of black hole dynamics. Suppose that Alice holds a $k$-qubit system $A$ which she wants to conceal from Bob. To be safe, she discards her qubits by tossing them into a large black hole, where she knows Bob will not dare to follow. The black hole $B$ is an $(n{-}k)$-qubit system, which grows to $n$ qubits after merging with $A$, where $n$ is much larger than $k$.

Black holes are not really completely black --- they emit Hawking radiation. But qubits leak out of an evaporating black hole very slowly, at a rate per unit time which scales like $n^{-1/2}$. Correspondingly, it takes time $\Theta(n^{3/2})$ for the black hole to radiate away a significant fraction of its qubits. Because the black hole Hilbert space is so enormous, this is a very long time, about $10^{67}$ years for a solar mass black hole, for which $n\approx 10^{78}$. Though Alice's qubits might not remain secret forever, she is content knowing that they will be safe from Bob for $10^{67}$ years. 

But in her haste, Alice fails to notice that her black hole is very, very old. It has been evaporating for so long that it has already radiated away more than half of its qubits. Let's assume that the joint state of the black hole and its emitted radiation is pure, and furthermore that the radiation is a Haar-random subsystem of the full system. 

Because the black hole $B$ is so old, $|B|$ is much smaller than the dimension of the radiation subsystem; therefore, as in eq.(\ref{eq:random-no-E}), we expect the state of $B$ to be very nearly maximally mixed with high probability. We denote by $R_B$ the subsystem of the emitted radiation which purifies $B$; thus the state of $BR_B$ is very nearly maximally entangled. We assume that $R_B$ has been collected by Bob and is under his control. 

To keep track of what happens to Alice's $k$ qubits, we suppose that her $k$-qubit system $A$ is maximally entangled with a reference system $R_A$. After $A$ enters the black hole, Bob waits for a while, until the $k'$-qubit system $A'$ is emitted in the black hole's Hawking radiation. After retrieving $A'$, Bob hopes to recover the purification of $R_A$ by applying a suitable decoding map to $A' R_B$. Can he succeed?

We've learned that Bob can succeed with high fidelity if the remaining black hole system $B'$ decouples from Alice's reference system $R_A$. Let's suppose that the qubits emitted in the Hawking radiation are chosen randomly; that is, $A'$ is a Haar-random $k'$-qubit subsystem of the $n$-qubit system $AB$, as depicted here:

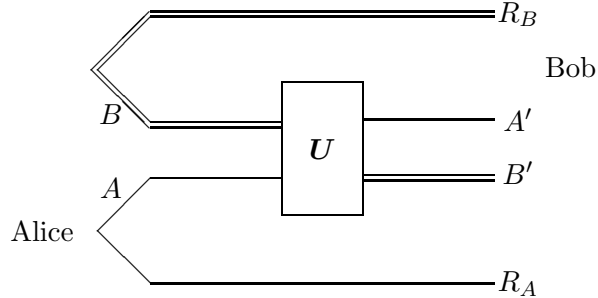
\begin{center}
\begin{picture} (234,112)
\put(42,26){\line(1,-1){20}}
\put(42,26){\line(1,1){20}}
\put(62,46){\line(1,0){50}}
\put(62,6){\line(1,0){130}}
\put(196,0){\makebox(10,12){$R_A$}}
\put(196,42){\makebox(10,12){$B'$}}
\put(42,36){\makebox(10,12){$A$}}
\put(16,20){\makebox(10,12){Alice}}
\put(112,32){\framebox(30,50){$\bfU$}}
\put(142,47){\line(1,0){50}}
\put(142,45){\line(1,0){50}}
\put(42,87){\line(1,-1){20}}
\put(42,87){\line(1,1){20}}
\put(39.5,87){\line(1,-1){22}}
\put(39.5,87){\line(1,1){22}}
\put(62,67){\line(1,0){50}}
\put(62,65){\line(1,0){50}}
\put(42,64){\makebox(10,12){$B$}}
\put(142,68){\line(1,0){50}}
\put(62,107){\line(1,0){130}}
\put(62,109){\line(1,0){130}}
\put(196,102){\makebox(10,12){$R_B$}}
\put(196,62){\makebox(10,12){$A'$}}
\put(216,82){\makebox(10,12){Bob}}
\end{picture}
\end{center}

\noindent The double lines indicate the very large systems $B$ and $B'$, and single lines the smaller systems $A$ and $A'$. Because the radiated qubits are random, we can determine whether $R_AB'$ decouples using the decoupling inequality, which for this case becomes
\begin{align}
\mathbb{E}_\bfU\big[~\|\bfsigma_{B'R_A}(\bfU) - \bfsigma_{B'}^{\max}\otimes \bfsigma_{R_A}\|_1~\big] \le \sqrt{\frac{|AB R_A|}{~|A'|^2}~{\rm tr} \left(\bfsigma_{ABR_A}^2\right)}.
\label{eq:decoupling-inequality-blackhole}
\end{align}
Because the state of $AR_A$ is pure, and $B$ is maximally entangled with $R_B$, we have ${\rm tr} \left(\bfsigma_{ABR_A}^2\right) = 1 / |B|$, and therefore the Haar-averaged $L^1$ distance of $\bfsigma_{B'R_A}$ from a product state is bounded above by
\begin{align}
\sqrt{\frac{|AR_A|}{|A'|^2}} = \frac{|A|}{|A'|}.
\end{align}
Thus, if Bob waits for only $k' = k+c$ qubits of Hawking radiation to be emitted after Alice tosses in her $k$ qubits, Bob can decode her qubits with excellent fidelity $F \ge 1 - 2^{-c}$. 

Alice made a serious mistake. Rather than waiting for $\Omega(n)$ qubits to emerge from the black hole, Bob can already decode Alice's secret quite well when he has collected just a few more than $k$ qubits. And Bob is an excellent physicist, who knows enough about black hole dynamics to infer the encoding unitary transformation $\bfU$, information he uses to find the right decoding map. 

We could describe the conclusion, more prosaically, by saying that the random unitary $\bfU$ applied to $AB$ encodes a good quantum error-correcting code, which achieves high-fidelity entanglement-assisted transmission of quantum information though an erasure channel with a high erasure probability. Of the $n$ input qubits, only $k'$ randomly selected qubits are received by Bob; the rest remain inside the black hole and hence are inaccessible. The input qubits, then, are erased with probability $p = (n -k')/n$, while nearly error-free qubits are recovered from the input qubits at a rate 
\begin{align}
R= \frac{k}{n} = 1 - p -\frac{k'-k}{n};
\end{align}
in the limit $n\to\infty$ with $c=k'-k$ fixed, this rate approaches $1-p$, the entanglement-assisted quantum capacity of the erasure channel. 

So far, we've assumed that the emitted system $A'$ is a randomly selected subsystem of $AB$. That won't be true for a real black hole. However, it is believed that the internal dynamics of actual black holes mixes quantum information quite rapidly (the {\em fast scrambling conjecture}). For a black hole with temperature $T$, it takes time of order $\hbar / kT$ for each qubit to be emitted in the Hawking radiation, and a time longer by only a factor of $\log n$ for the dynamics to mix the black hole degrees of freedom sufficiently for our decoupling estimate to hold with reasonable accuracy. For a solar mass black hole, Alice's qubits are revealed just a few milliseconds after she deposits them, much faster than the $10^{67}$ years she had hoped for! Because Bob holds the system $R_B$ which purifies $B$, and because he knows the right decoding map to apply to $A'R_B$, the black hole behaves like an information mirror --- Alice's qubits bounce right back!

If Alice is more careful, she will dump her qubits into a young black hole instead. If we assume that the initial black hole $B$ is in a pure state, then $\sigma_{ABR_A}$ is also pure, and the Haar-averaged $L^1$ distance of $\bfsigma_{B'R_A}$ from a product state is bounded above by
\begin{align}
\sqrt{\frac{|ABR_A|}{|A'|^2}} = \sqrt{\frac{2^{n+k}}{2^{2k'}}}= \frac{1}{2^c}
\end{align} 
after
\begin{align}
k' = \frac{1}{2} (n+k)+c
\end{align}
qubits are emitted. In this case, Bob needs to wait a long time, until more than half of the qubits in $AB$ are radiated away. Once Bob has acquired $k+2c$ more qubits than the number still residing in the black hole, he is empowered to decode Alice's $k$ qubits with fidelity $F \ge 1 - 2^{-c}$. In fact, there is nothing special about Alice's subsystem $A$; by adjusting his decoding map appropriately, Bob can decode any $k$ qubits he chooses from among the $n$ qubits in the initial black hole $AB$. 

There is far more to learn about quantum information processing by black holes, an active topic of current research, but we will not delve further into this fascinating topic here. We can be confident, though, that the tools and concepts of quantum information theory discussed in this book will be helpful for addressing the many unresolved mysteries of quantum gravity, as well as many other open questions in the physical sciences. 




\section{Summary}

{\bf Shannon entropy and classical data compression.}  The {\it Shannon
entropy} of an ensemble $X = \{x,p(x)\}$ is $H(X) := \mathbb{E}_X[-\log
p(x)]$; it quantifies the compressibility of classical information.  A
message $n$ letters long, where each letter is drawn independently from $X$,
can be compressed to $H(X)$ bits per letter (and no further), yet can still be
decoded with arbitrarily good accuracy as $n \rightarrow \infty$.

{\bf Conditional entropy and information merging.} The {\it conditional entropy} $H(X|Y)= H(XY) - H(Y)$ quantifies how much the information source $X$ can be compressed when $Y$ is known. If $n$ letters are drawn from $XY$, where Alice holds $X$ and Bob holds $Y$, Alice can convey $X$ to Bob by sending $H(X|Y)$ bits per letter, asymptotically as $n \rightarrow \infty$.

{\bf Mutual information and classical channel capacity.}  The {\it mutual
information}  $I(X;Y) = H(X) + H(Y) - H(XY)$ quantifies how information sources $X$ and $Y$ are correlated; when we learn the value of $y$ we acquire (on the average)
$I(X;Y)$ bits of information about $x$, and vice versa.  The capacity of a memoryless noisy classical communication channel is $C = \max_X I(X;Y)$.  This is
the highest number of bits per letter that can be transmitted through $n$ uses of the
channel, using the best possible code, with negligible error probability as $n
\rightarrow \infty$.

{\bf Von Neumann entropy and quantum data compression.}
The {\it von Neumann entropy} of a density operator $\bfrho$ is 
\begin{equation}
H(\bfrho) = - {\rm tr} \bfrho\log\bfrho;
\end{equation}
it quantifies the compressibility of an ensemble of pure
quantum states.  A message $n$ letters long, where each letter is drawn
independently from the ensemble $\{|\varphi(x)\rangle, p(x)\}$, can be compressed
to $H(\bfrho)$ qubits per letter (and no further) where $\bfrho = \sum_X p(x)|\varphi(x)\rangle\langle\varphi(x)|$, yet can still be decoded
with arbitrarily good fidelity as $n \rightarrow \infty$.  

{\bf Entanglement concentration and dilution}.  The {\it entanglement} $E$ of a bipartite pure
state $|\psi\rangle_{AB}$ is $E = H(\bfrho_A)$ where $\bfrho_A = {\rm tr}_B
(|\psi\rangle\langle\psi|)$. With local operations and classical communication, 
we can prepare $n$ copies of
$|\psi\rangle_{AB}$ from $nE$ Bell pairs (but not from fewer), and we can distill
$nE$ Bell pairs  (but not more) from $n$ copies of $|\psi\rangle_{AB}$, asymptotically as $n
\rightarrow \infty$.

{\bf Accessible information}. The {\it Holevo chi} of an ensemble $\mathcal{E} = \{\bfrho(x), p(x)\}$ of quantum
states is
\begin{equation}
\chi(\cale) = H\left(\sum_x p(x) \bfrho(x)\right) - \sum_x p(x) H(\bfrho(x)).
\end{equation} The {\it accessible information} of an ensemble $\cale$
of quantum states is the maximal number of bits of information that can be acquired about
the preparation of the state (on the average) with the best possible
measurement.  The accessible information cannot exceed the Holevo chi
of the ensemble.  
The product-state capacity of a quantum
channel $\mathcal{N}$ is
\begin{equation}
C_1(\mathcal{N}) = \max_\mathcal{E} \chi (\mathcal{N}(\mathcal{E})).
\end{equation}
This is the highest number of classical bits per letter that can be transmitted
through $n$ uses of the quantum channel, with negligible error probability as $n
\rightarrow \infty$, assuming that each codeword is a product state.

{\bf Decoupling and quantum communication}. In a tripartite pure state $\phi_{RBE}$, we say that systems $R$ and $E$ {\em decouple} if the marginal density operator of $RE$ is a product state, in which case $R$ is purified by a subsystem of $B$. A quantum state transmitted through a noisy quantum channel $\mathcal{N}^{A\to B}$ (with isometric dilation $\bfU^{A\to BE}$) can be accurately decoded if a reference system $R$ which purifies channel's input $A$ nearly {\em decouples} from the channel's environment $E$. 

{\bf Father and mother protocols}. The {\it father and mother resource inequalities} specify achievable rates for entanglement-assisted quantum communication and quantum state transfer, respectively. Both follow from the {\it decoupling inequality}, which establishes a sufficient condition for approximate decoupling in a tripartite mixed state. By combining the father and mother protocols with superdense coding and teleportation, we can derive achievable rates for other protocols, including entanglement-assisted classical communication, quantum communication, entanglement distillation, and quantum state merging.

\medskip

{\bf Homage to Ben Schumacher}:

\begin{quote}
    Ben.\\
    He rocks.\\
    I remember\\
    When\\
    He showed me how to fit\\
    A qubit\\
    In a small box.\\
\\
    I wonder how it feels\\
    To be compressed.\\
    And then to pass\\
    A fidelity test.\\
\\
    Or does it feel\\
    At all, and if it does\\
    Would I squeal\\
    Or be just as I was?\\
\\
    If not undone\\
    I'd become as I'd begun\\
    And write a memorandum\\
    On being random.\\
    Had it felt like a belt\\
    Of rum?\\
\\
    And might it be predicted\\
    That I'd become addicted,\\
    Longing for my session\\
    Of compression?\\
\\
    I'd crawl\\
    To Ben again.\\
    And call,\\
    Put down your pen!\\
    Don't stall!\\
    Make me small!
\end{quote}

\section{Bibliographical Notes}

Cover and Thomas \cite{cover-thomas} is an excellent textbook on classical information theory. Shannon's original paper \cite{shannon} is still very much worth reading.

Nielsen and Chuang \cite{nielsen-chuang} provide a clear introduction to some aspects of quantum Shannon theory. Wilde \cite{wilde} is a more up-to-date and very thorough account. 

Properties of entropy are reviewed in \cite{wehrl}. Strong subadditivity of von Neumann entropy was proven by Lieb and Ruskai \cite{lieb-ruskai}, and the condition for equality was derived by Hayden {\em et al.} \cite{hayden-ssa}. The connection between separability and majorization was pointed out by Nielsen and Kempe \cite{nielsen-kempe}.

Bekenstein's entropy bound was formulated in \cite{bekenstein} and derived by Casini \cite{casini}. Entropic uncertainty relations are reviewed in \cite{berta-uncertainty}, and I follow their derivation. The original derivation, by Maassen and Uffink \cite{maassen} uses different methods. 

Schumacher compression was first discussed in \cite{schumacher-compression,schumacher-jozsa}, and Bennett {\em et al.}~\cite{bennett-concentration} devised protocols for  entanglement concentration and dilution. Measures of mixed-state entanglement are reviewed in \cite{horodecki-review}. The reversible theory of mixed-state entanglement was formulated by Brand\~ao and Plenio \cite{brandao-locc} (a flaw in their analysis was addressed in \cite{hayashi-stein,lami-stein}). Squashed entanglement was introduced by Christandl and Winter \cite{christandl-winter-squashed}, and its monogamy discussed by Koashi and Winter \cite{koashi-winter-squashed}. Brand\~ao, Christandl, and Yard \cite{brandao-squashed} showed that squashed entanglement is positive for any nonseparable bipartite state. Doherty, Parrilo, and Spedalieri \cite{doherty} showed that every nonseparable bipartite state fails to be $k$-extendable for some finite $k$. 

The Holevo bound was derived in \cite{holevo-bound}. Peres-Wootters coding was discussed in \cite{peres-wootters}. The product-state capacity formula was derived by Holevo \cite{holevo-capacity} and by Schumacher and Westmoreland \cite{schumacher-westmoreland-capacity}. Hastings \cite{hastings-superadditive} showed that Holevo chi can be superadditive. Horodecki, Shor, and Ruskai \cite{horodecki-entanglement-breaking} introduced entanglement-breaking channels, and additivity of Holevo chi for these channels was shown by Shor \cite{shor-entanglement-breaking}.

Necessary and sufficient conditions for quantum error correction were formulated in terms of the decoupling principle by Schumacher and Nielsen \cite{schumacher-nielsen}; that (regularized) coherent information is an upper bound on quantum capacity was shown by Schumacher \cite{schumacher-coherent}, Schumacher and Nielsen \cite{schumacher-nielsen}, and Barnum {\em et al.} \cite{barnum-coherent}. That coherent information is an achievable rate for quantum communication was conjectured by Lloyd \cite{lloyd-coherent} and by Schumacher \cite{schumacher-coherent}, then proven by Shor \cite{shor-quantum-capacity} and by Devetak \cite{devetak-quantum-capacity}. Devetak and Winter \cite{devetak-winter-hashing} showed it is also an achievable rate for entanglement distillation. The quantum Fano inequality was derived by Schumacher \cite{schumacher-coherent}.

Approximate decoupling was analyzed by Schumacher and Westmoreland \cite{schumacher-westmoreland-approximate}, and used to prove capacity theorems by Devetak \cite{devetak-quantum-capacity}, by Horodecki {\em et al.} \cite{horodecki-merging},  by Hayden {\em et al.} \cite{hayden-decoupling}, and by Abeyesinghe {\em et al.} \cite{abeyesinghe-decoupling}. The entropy of Haar-random subsystems had been discussed earlier, by Lubkin \cite{lubkin}, Lloyd and Pagels \cite{lloyd-entropy}, and Page \cite{page}. Devetak, Harrow, and Winter \cite{devetak-mother-father,devetak-mother-father-long} introduced the mother and father protocols and their descendants. Devatak and Shor \cite{devetak-shor-degradable} introduced degradable quantum channels and proved that coherent information is additive for these channels. Bennett {\em et al.} \cite{bennett-entanglement-assisted,bennett-entanglement-assisted-more} found the single-letter formula for entanglement-assisted classical capacity. Superadditivity of coherent information was discovered by Shor and Smolin \cite{shor-smolin} and by DiVincenzo {\em et al.} \cite{divincenzo-superadditive}. Smith and Yard \cite{smith-yard} found extreme examples of superadditivity, in which two zero-capacity channels have nonzero capacity when used jointly. The achievable rate for state merging was derived by Horodecki {\em et al.} \cite{horodecki-merging}, and used by them to prove strong subadditivity of von Neumann entropy. 

Decoupling was applied to Landuaer's principle by Renner {\em et al.} \cite{renner-landauer}, and to black holes by Hayden and Preskill \cite{hayden-preskill}. The fast scrambling conjecture was proposed by Sekino and Susskind \cite{sekino-susskind}.


\begin{exercises}


\item{\bf Positivity of quantum relative entropy}\label{ex:rel-entropy}
\begin{description}
\item[$a$)] Show that $\ln x \le x-1$ for all positive real x, with equality iff $x=1$.

\item[$b$)] The (classical) relative entropy of a probability distribution $\{p(x)\}$ relative to $\{q(x)\}$ is defined as
\begin{equation}
D(p\| q)\equiv \sum_x p(x)\left(\log p(x) - \log q(x)\right)~.
\end{equation}
Show that 
\begin{equation}
D(p\| q)\ge 0~,
\end{equation}
with equality iff the probability distributions are identical. {\bf Hint}: Apply the inequality from $(a)$ to $\ln \left(q(x)/p(x)\right)$.

\item[$c$)] The quantum relative entropy of the density operator $\bfrho$ with respect to $\bfsigma$ is defined as
\begin{equation}
D(\bfrho\|\bfsigma)= {\rm tr}~\bfrho\left(\log \bfrho - \log \bfsigma\right)~.
\end{equation}
Let $\{p_i\}$ denote the eigenvalues of $\bfrho$ and $\{q_a\}$ denote the eigenvalues of $\bfsigma$. Show that 
\begin{equation}
D(\bfrho\|\bfsigma)= \sum_i p_i\left(\log p_i - \sum_a D_{ia}\log q_a\right)~,
\end{equation}
where $D_{ia}$ is a doubly stochastic matrix. Express $D_{ia}$ in terms of the eigenstates of $\bfrho$ and $\bfsigma$. (A matrix is doubly stochastic if its entries are nonnegative real numbers, where each row and each column sums to one.)

\item[$d$)] Show that if $D_{ia}$ is doubly stochastic, then (for each $i$)
\begin{equation}
\log\left(\sum_a D_{ia}q_a\right)\ge \sum_a D_{ia}\log q_a~,
\end{equation}
with equality only if $D_{ia}=1$ for some $a$.

\item[$e$)] Show that 
\begin{equation}
D(\bfrho\|\bfsigma)\ge D(p\| r)~,
\end{equation}
where $r_i=\sum_a D_{ia}q_a$.

\item[$f$)] Show that $D(\bfrho\|\bfsigma)\ge 0$, with equality iff $\bfrho=\bfsigma$.
\end{description}

\item {\bf Properties of von Neumann entropy} \label{ex:von-neumann-entropy}
\begin{description}
\item[$a$)] Use nonnegativity of quantum relative entropy to prove the {\it subadditivity}
of von Neumann entropy
\begin{equation}
H(\bfrho_{AB}) \leq H(\bfrho_A) + H(\bfrho_B),
\end{equation}
with equality iff $\bfrho_{AB}=\bfrho_A\otimes\bfrho_B$.
{\bf Hint}:  Consider the relative entropy of $\bfrho_{AB}$ and $\bfrho_A \otimes \bfrho_B$.

\item[$b$)] Use subadditivity to prove the concavity of the von Neumann entropy:
\begin{equation}
H(\sum_x p_x\bfrho_x)\ge \sum_x p_x H(\bfrho_x)~.
\end{equation}
{\bf Hint}: Consider
\begin{equation}
\bfrho_{AB}=\sum_x p_x \left(\bfrho_x\right)_A\otimes \left(|x\rangle\langle x|\right)_B~,
\end{equation}
where the states $\{|x\rangle_B\}$ are mutually orthogonal. 

\item[$c$)] Use the condition
\begin{equation}
H(\bfrho_{AB})= H(\bfrho_A)+ H(\bfrho_B)\quad {\rm iff} \quad \bfrho_{AB}=\bfrho_A\otimes \bfrho_B
\end{equation}
to show that, if all $p_x$'s are nonzero,
\begin{equation}
H\left(\sum_x p_x\bfrho_x\right)= \sum_x p_x H(\bfrho_x)~
\end{equation}
iff all the $\bfrho_x$'s are identical.

\end{description}

\item  {\bf Monotonicity of quantum relative entropy} \label{ex:rel-entropy-monotonicity}

Quantum relative entropy has a property called {\it monotonicity}:
\begin{equation}
D(\bfrho_A\|\bfsigma_A) \leq D(\bfrho_{AB} \| \bfsigma_{AB});
\end{equation}
The relative entropy of two density operators on a system $AB$ cannot be less than
the induced relative entropy on the subsystem $A$.
\begin{description}
\item[$a$)] Use monotonicity of quantum relative entropy to prove the strong
subadditivity property of von Neumann entropy.  {\bf Hint}: On a tripartite
system $ABC$, consider the relative entropy of $\bfrho_{ABC}$ and $\bfrho_A \otimes
\bfrho_{BC}$.

\item[$b$)] Use monotonicity of quantum relative entropy to show that the action of a quantum
channel $\mathcal{N}$ cannot increase relative entropy:
\begin{equation}
D(\mathcal{N}(\bfrho) \| \mathcal{N}(\bfsigma)) \leq D(\bfrho\|\bfsigma),
\end{equation}
{\bf Hint}:  Recall that
any quantum channel has an isometric dilation.

\end{description}

\item {\bf Hypothesis testing}

The (classical) relative entropy $D(p\| q)$ charactizes the difference between the distributions $p(x)$ and $q(x)$ in a sense that is useful for hypothesis testing. Suppose that an unknown distribution is sampled $n\ggg 1$ times, and we find that outcome $x$ occurs $np(x)$ times. We would like to assess whether these outcomes are compatible with sampling from a hypothetical distribution $q(x)$. As discussed in \S\ref{subsec:relative-hypothesis}, the probability of obtaining this result if we are actually sampling from $q(x)$ may be estimated as
\begin{equation}
\text{Probability} = 2^{-nD(p\| q)}.
\end{equation}

\begin{description}
\item[$a$)] Suppose
\begin{equation}
q(x) = p(x) + \varepsilon(x), \quad \sum_x \varepsilon(x) = 0.
\end{equation}
Write $D(p\| q)$ as a function of $p(x)$ and $\varepsilon(x)$, expanded to quadratic order in $\varepsilon(x)$. For this purpose, suppose that the logarithms in the definition of relative entropy are natural logs rather than logs to the base 2. 

\item[$b$)] Suppose that 
\begin{equation}
\| \varepsilon \|_1=\sum_x |\varepsilon(x)| = \delta.
\end{equation}
Using the quadratic approximation found in $(a)$, find the minimal value of $D(p\|q)$ and the hypothetical distribution $q(x)=p(x)+\varepsilon(x)$ that minimizes it.

\item[$c$)] Suppose we toss a coin $n$ times, and the outcome ``heads'' is observed $n\left(\frac{1}{2} + \frac{\delta}{2}\right)$ times. Hence $p(\text{heads})= \frac{1}{2} + \frac{\delta}{2}$ and $p(\text{tails})= \frac{1}{2} - \frac{\delta}{2}$. Consider the hypothesis that the coin is unbiased: $q(\text{heads})=q(\text{tails})=\frac{1}{2}$. Compute the relative entropy $D(p\|q)$ to quadratic order in $\delta$.

\item[$d$)] A coin is tossed 1 million times. Use the result of $(c)$  to estimate the probability of the coin coming up heads 505,000 times, assuming that the coin is unbiased. 
\end{description}

\item {\bf Separability and majorization}

The hallmark of entanglement is that in an entangled state the whole is less random than its parts. But in a separable state the correlations are essentially classical and so are expected to adhere to the classical principle that the parts are less disordered than the whole. The objective of this problem is to make this expectation precise by showing that if the bipartite (mixed) state $\bfrho_{AB}$ is separable, then
\begin{equation}
\lambda(\bfrho_{AB})\prec \lambda(\bfrho_A)~, \quad 
\lambda(\bfrho_{AB})\prec \lambda(\bfrho_B)~.
\end{equation}
Here $\lambda(\bfrho)$ denotes the vector of eigenvalues of $\bfrho$, and $\prec$ denotes majorization.

A separable state can be realized as an ensemble of pure product states, so that if $\bfrho_{AB}$ is separable, it may be expressed as
\begin{equation}
\bfrho_{AB}= \sum_a p_a ~|\psi_a\rangle\langle \psi_a|\otimes |\varphi_a\rangle\langle \varphi_a|~.
\end{equation}
We can also diagonalize $\bfrho_{AB}$, expressing it as
\begin{equation}
\bfrho_{AB}=\sum_j r_j |e_j\rangle\langle e_j|~,
\end{equation}
where $\{|e_j\rangle\}$ denotes an orthonormal basis for $AB$; then by the HJW theorem, there is a unitary matrix $V$ such that
\begin{equation}
\sqrt{r_j}|e_j\rangle = \sum_a V_{ja} \sqrt{p_a}|\psi_a\rangle \otimes |\varphi_a\rangle ~.
\end{equation}
Also note that $\bfrho_A$ can be diagonalized, so that 
\begin{equation}
\bfrho_A= \sum_a  p_a|\psi_a\rangle\langle \psi_a| = \sum_\mu s_\mu |f_\mu\rangle \langle f_\mu|~;
\end{equation}
here $\{|f_\mu\rangle\}$ denotes an orthonormal basis for $A$, and by the HJW theorem, there is a unitary matrix $U$ such that 
\begin{equation}
\sqrt{p_a} |\psi_a\rangle = \sum_\mu U_{a\mu}\sqrt{s_\mu}|f_\mu\rangle~.
\end{equation}
Now show that there is a doubly stochastic matrix $D$ such that
\begin{equation}
r_j=\sum_\mu D_{j\mu} s_\mu~.
\end{equation}
That is, you must check that the entries of $D_{j\mu}$ are real and nonnegative, and that $\sum_j D_{j\mu}=1=\sum_\mu D_{j\mu}$. Thus we conclude that $\lambda(\bfrho_{AB}) \prec \lambda (\bfrho_A)$. Just by interchanging $A$ and $B$, the same argument also shows that $\lambda(\bfrho_{AB}) \prec \lambda (\bfrho_B)$.

{\bf Remark}: Note that it follows from the Schur concavity of Shannon entropy that, if $\bfrho_{AB}$ is separable, then
the von Neumann entropy has the properties $H(AB) \ge H(A)$ and $H(AB) \ge H(B)$. Thus, for separable states, conditional entropy is nonnegative: $H(A|B) = H(AB)-H(B) \ge 0$ and $H(B|A) = H(AB)-H(A) \ge 0$. In contrast, if $H(A|B)$ is negative, then according to the hashing inequality the state of $AB$ has positive distillable entanglement $-H(A|B)$, and therefore is surely not separable.

\item {\bf Additivity of squashed entanglement}\label{ex:squashed-additivity}

Suppose that Alice holds systems $A$, $A'$ and Bob holds systems $B$, $B'$. How is the entanglement of $AA'$ with $BB'$ related to the entanglement of $A$ with $B$ and $A'$ with $B'$? In this problem we will show that the squashed entanglement is superadditive,
\begin{align}
E_{\rm sq}(\bfrho_{ABA'B'}) \ge E_{\rm sq}(\bfrho_{AB}) + E_{\rm sq}(\bfrho_{A'B'})
\label{eq:ex-squashed-superadditive}
\end{align}
and is strictly additive for a tensor product,
\begin{align}
E_{\rm sq}(\bfrho_{AB}\otimes \bfrho_{A'B'}) = E_{\rm sq}(\bfrho_{AB}) + E_{\rm sq}(\bfrho_{A'B'}).
\label{eq:ex-squashed-additive}
\end{align}

\begin{description}
\item[$a$)] Use the chain rule for mutual information eq.(\ref{eq:chain-rule-mutual}) and eq.(\ref{eq:chain-rule-mutual-conditioned}) and the nonnegativity of quantum conditional mutual information to show that
\begin{align}
I(AA';BB'|C) \ge I(A;B|C) + I(A';B'|AC),
\label{eq:superadditive-chain-rule}
\end{align}
and show that eq.(\ref{eq:ex-squashed-superadditive}) follows.

\item[$b$)] Show that for any extension $\bfrho_{ABC}\otimes \bfrho_{A'B'C'}$ of the product state $\bfrho_{AB}\otimes \bfrho_{A'B'}$, we have
\begin{align}
I(AA';BB'|CC') \le I(A;B|C) + I(A';B'|C').
\label{eq:subadditive-chain-rule}
\end{align}
Conclude that 
\begin{align}
E_{\rm sq}(\bfrho_{AB}\otimes \bfrho_{A'B'}) \le E_{\rm sq}(\bfrho_{AB}) + E_{\rm sq}(\bfrho_{A'B'}),
\end{align}
which, when combined with eq.(\ref{eq:ex-squashed-superadditive}), implies eq.(\ref{eq:ex-squashed-additive}).
\end{description}

\item {\bf The first law of von Neumann entropy}

Writing the density operator in terms of its {\em modular Hamiltonian} $\bfK$ as in \S\ref{subsec:bekenstein},
\begin{align}
\bfrho = \frac{e^{-\bfK}}{{\rm tr}\left(e^{-\bfK} \right)},
\end{align}
consider how the entropy $S(\bfrho) = - {\rm tr}\left(\bfrho\ln\bfrho\right)$ changes when the density operator is perturbed slightly:
\begin{align}
\bfrho \to \bfrho' = \bfrho + \delta\bfrho.
\end{align}
Since $\bfrho$ and $\bfrho'$ are both normalized density operators, we have ${\rm tr}\left(\delta\bfrho\right) = 0$. Show that
\begin{align}
S(\bfrho') - S(\bfrho) = {\rm tr}\left(\bfrho' \bfK\right) - {\rm tr}\left(\bfrho \bfK\right)+ O\left(\left(\delta\bfrho\right)^2\right);
\end{align}
that is,
\begin{align}
\delta S = \delta \langle \bfK \rangle
\label{eq:first-law}
\end{align}
to first order in the small change in $\bfrho$. This statement generalizes the first law of thermodynamics; for the case of a thermal density operator with $\bfK = \bfH/T$ (where $\bfH$ is the Hamiltonian and $T$ is the temperature), it becomes the more familiar statement
\begin{align}
\delta E = \delta\langle\bfH\rangle = T \delta S.
\end{align}

\item {\bf Distinguishing two nonorthogonal pure states}
\label{ex:distinghishing-two-states}

\begin{description}
\item[$a$)] Consider an ensemble in which the two pure states of a single qubit
\begin{equation}
|\varphi_0\rangle =\begin{pmatrix}
1 \\ 0
\end{pmatrix},
\quad
|\varphi_1\rangle=
\begin{pmatrix}
 \cos(\theta/2) \\ \sin(\theta/2)
\end{pmatrix}
\end{equation}
occur equiprobably, where $\theta\in [0,2\pi)$.  The two states both lie in the $xz$ plane of the Bloch sphere, and the angle between them is $\theta$. Suppose that a state is drawn from this ensemble and then the Pauli observable
\begin{equation}
\bfZ = |0\rangle\langle 0| - |1\rangle\langle 1| 
\end{equation}
is measured. Compute the information gain achieved by this measurement.

\item[$b$)]  Now suppose that, instead of $\bfZ$, the rotated observable 
\begin{equation}
\bfZ' = |\psi\rangle\langle \psi| - |\psi^\perp\rangle\langle \psi^\perp|
\end{equation}
is measured, where 
\begin{equation} |\psi\rangle =
\begin{pmatrix}
 \cos(\theta/4 -\pi/4)\\ \sin(\theta/4-\pi/4)
\end{pmatrix},
\quad |\psi^\perp\rangle =
\begin{pmatrix}
 \cos(\theta/4 +\pi/4)\\ \sin(\theta/4+\pi/4)
\end{pmatrix}.
\end{equation}
Compute the information gain in this case. Make a plot showing the two functions of $\theta$ computed in parts ($a$) and ($b$). Which has the higher information gain?

\item[$c$)] For $\theta = \pi/2$ and for $\theta = 2\pi/3$, find the numerical value of the information gain, in bits, achieved by measuring $\bfZ$ and $\bfZ'$.
\end{description}

\item {\bf The Peres--Wootters POVM.}\label{ex:peres-wootters}

Consider the Peres--Wootters information source described in \S\ref{subsec:peres-wootters}.  It prepares one of the three states
\begin{equation}
|\Phi_a\rangle = |\varphi_a\rangle\otimes|\varphi_a\rangle, \quad a = 1,2,3,
\end{equation}
each occurring with {\it a priori} probability ${1\over 3}$, where the
$|\varphi_a\rangle$'s are defined in eq.(\ref{eq:qubit-trine}).

\begin{description}
\item[$a$)]  Express the density matrix
\begin{equation}
\bfrho = {1\over 3} \left(\sum_a |\Phi_a\rangle\langle\Phi_a|\right),
\end{equation}
in terms of the Bell basis of maximally entangled states $\{|\phi^\pm\rangle,
|\psi^\pm\rangle\}$, and compute $H(\bfrho)$.

\item[$b$)]  For the three vectors $|\Phi_a\rangle, a = 1,2,3$, construct
the ``pretty good measurement'' defined in eq.(\ref{eq:PGM-POVM}).  (Again, expand the
$|\Phi_a\rangle$'s in the Bell basis.)  In this case, the PGM is an orthogonal
measurement.  Express the elements of the PGM basis in terms of the Bell basis.

\item[$c$)]  Compute the mutual information of the PGM outcome and the
preparation.

\end{description}

\item  {\bf An adaptive protocol}

Consider again the Peres-Wootters ensemble in which the three two-qubit states 
\begin{equation}
|\Phi_a\rangle= |\varphi_a\rangle\otimes |\varphi_a\rangle,\quad  a = 1, 2, 3
\end{equation}
occur equiprobably. A state is drawn from this ensemble, and we are to measure the two-qubit state with the goal of gaining information about the value of $a$. Instead of doing the collective measurement described in \S\ref{subsec:peres-wootters}, consider an adaptive strategy in which the two qubits are measured separately. On the first qubit we perform the POVM with the three outcomes
\begin{equation}
\bfE_a = \frac{2}{3} \left( \bfI - |\varphi_a\rangle \langle \varphi_a|\right), \quad a = 1, 2, 3;
\end{equation}
this measurement excludes one of the three states, but provides no information to distinguish the other two states. Then we perform a measurement on the second qubit that is conditioned on the outcome of the first measurement. Guided by the results of Exercise \ref{ex:distinghishing-two-states}, choose the most informative measurement to perform on the second qubit, and then compute the information gain achieved by this adaptive procedure. Compare with the information gain achieved by the collective measurement described in \S\ref{subsec:peres-wootters}.

\item {\bf An ensemble of four pure states}

\begin{description}
\item[$a$)] Consider an ensemble in which the four pure states of a single qubit
\begin{equation}
|\varphi_0\rangle =\begin{pmatrix}
1 \\ 0
\end{pmatrix},
\quad
|\varphi_1\rangle=
\begin{pmatrix}
0 \\ 1
\end{pmatrix},\quad
|\varphi_2\rangle=
\begin{pmatrix}
\frac{1}{\sqrt 2} \\ \frac{1}{\sqrt 2}
\end{pmatrix},\quad
|\varphi_3\rangle=
\begin{pmatrix}
\frac{1}{\sqrt 2} \\ -\frac{1}{\sqrt 2}
\end{pmatrix},
\end{equation}
occur equiprobably. Compute the information gain achieved by measuring $\bfZ$ for a state drawn from this ensemble.

\item[$b$)] Now compute the information gain achieved by measuring $\bfZ'$ for $\theta = \pi/2$. Compare with the result of $(a)$.
\end{description}

\item {\bf An infinite ensemble}
\begin{description}

\item[$a$)] We now consider an ensemble with an infinite number of possible pure states, namely the uniform distribution on $\theta \in [0,2\pi)$ where
\begin{equation}
|\varphi(\theta)\rangle=
\begin{pmatrix}
 \cos(\theta/2) \\ \sin(\theta/2)
\end{pmatrix}.
\end{equation}
That is, when a state is sampled from this ensemble, a value in the interval $[\theta, \theta + d\theta]$ occurs with probability $d\theta/2\pi$. How much information do we gain about the value of $\theta$ by measuring $\bfZ$? To avoid having to take a difference of infinite quantities, answer this question by computing the conditional entropy $H(Y|\theta)$ and using 
\begin{equation}
I(\theta;Y) = H(Y) - H(Y|\theta),
\end{equation}
where $Y$ denotes the probability distribution of measurement outcomes. 

\item[$b$)] Hoping to gain more information, let's try a POVM with many outcomes rather than a two-outcome orthogonal measurement. Choose the POVM elements to be
\begin{equation}
\bfE_k = \frac{2}{n} |\varphi(\theta_k)\rangle\langle \varphi(\theta_k)|, \quad k= 0, 1, 2, \dots ,n{-}1,
\end{equation}
where 
\begin{equation}
\theta_k = \frac{2\pi k}{n}.
\end{equation}
Check the normalization condition
\begin{equation}
\sum_k \bfE_k= \bfI,
\end{equation}
and compute the information gain achieved by this measurement. Compare with the result of $(a)$.

\end{description}

\item {\bf Approximate cloning and the depolarizing channel}
\label{ex:approximate-cloning}

Consider a qubit channel $\mathcal{N}_\theta^{A\to B}$ with the isometric dilation $U_\theta^{A\to BEF}$ which acts on an orthonormal basis as
\begin{align}
|0\rangle_A &\mapsto |\phi_0\rangle_{BEF}= \cos\theta |00\rangle_{BE}|0\rangle_F + \sin\theta |\psi^+\rangle_{BE}|1\rangle_F,\\
|1\rangle_A &\mapsto |\phi_1\rangle_{BEF}= \cos\theta |11\rangle_{BE}|1\rangle_F + \sin\theta |\psi^+\rangle_{BE}|0\rangle_F,
\end{align}
where
\begin{align}
|\psi^+\rangle=\frac{1}{\sqrt{2}}\left(|01\rangle + |10\rangle\right),
\end{align}
and $0\le\theta\le\pi/2$. Here we have split the channel's environment into two parts labeled $E$ and $F$, and we have constructed the isometry to be symmetric under the interchange of $B$ and $E$. Hence the channel $\mathcal{N}_\theta^{A\to B}$ obtained by tracing out $EF$ is identical to the channel $\mathcal{N}_\theta^{A\to E}$ obtained by tracing out $BF$. 
It follows that $\mathcal{N}_\theta^{A\to B}$ has zero capacity. To reach this conclusion, note that if Alice sends a quantum state to Bob via many uses of the isometry $U^{A\to BEF}$, and Bob (who receives the output system $B$) is able to decode the state with high fidelity, then Eve (who receives the output system $E$) receives the same output as Bob and therefore she can decode the state as well. Thus the no-cloning theorem ensures that high-fidelity decoding of the output is not possible. 

\begin{description}
\item[$a$)] The channel $\mathcal{N}_\theta^{A\to B}$ is actually a Pauli channel, which can be expressed as
\begin{align}
\mathcal{N}_\theta(\rho) = f(\theta) \rho +g(\theta) Z\rho Z + h(\theta)\frac{I}{2}.
\end{align}
Find the functions $f$, $g$, and $h$.

\item[$b$)] We may choose a value $\theta=\theta_0$ such that $g(\theta_0)=0$, in which case $\mathcal{N}$ becomes a depolarizing channel with error probability $p$:
\begin{align}
\mathcal{N}_{\theta_0}(\rho)= \left(1-\frac{4p}{3}\right)\rho + \left(\frac{4p}{3}\right)\frac{I}{2}.
\end{align}
Find this value $\theta_0$ and the corresponding value of $p$. Our no-cloning argument shows that the depolarizing channel with this error probability $p$ has zero capacity.

\item[$c$)] Tracing out $F$ from the isometry $U^{A\to BEF}$, we obtain a channel $N^{A\to BE}$, which may be regarded as an approximate cloning machine. If Alice sends a pure state $|\psi\rangle$ through the channel, the marginal density operators $\rho_B$ and $\rho_E$ received by Bob and Eve are identical, each approximating the input state with fidelity
\begin{align}
F = \langle \psi|\rho_B|\psi\rangle = \langle \psi|\rho_E|\psi\rangle.
\end{align}
For the value $\theta=\theta_0$ found in ($b$), compute the fidelity achieved by this approximate cloner.

\item[$d$)] For $\theta\ne \theta_0$, we have $g(\theta)\ne 0$, and therefore the fidelity achieved by the approximate cloner depends on the pure state input $|\psi\rangle$. In that case we may consider the worst-case fidelity $F_\text{min}(\theta)$, the lowest value of $F$ achieved for any pure state input to $\mathcal{N}^{A\to BE}$. Find the value $\theta=\theta_1$ such that this worst-case fidelity is as large as possible and compute $F_\text{min}(\theta_1)$. (Be sure to recall that $g(\theta)$ changes sign at $\theta=\theta_0$.)

\end{description}

\item{\bf Information gain for a quantum state drawn from the uniform ensemble}

Suppose Alice prepares a quantum state drawn from the ensemble $\{\bfrho(x), p(x)\}$ and Bob performs a measurement $\{\bfE(y)\}$ yielding outcome $y$ with probability $p(y|x) = {\rm tr}\left(\bfE(y)\bfrho(x)\right)$. As noted in \S\ref{subsec:how-much-measurement}, Bob's information gain about Alice's preparation is the mutual information $I(X;Y) = H(X) - H(X|Y)$. If $x$ is a continuous variable, while $y$ is discrete, it is more convenient to use the symmetry of mutual information to write $I(X;Y) = H(Y) - H(Y|X)$, where
\begin{align}\label{eq:hw-conditional-continuous}
H(Y|X) = \sum_y \int dx \cdot p(x) \cdot p(y|x)\cdot  \log p(y|x);
\end{align}
here $p(x)$ is a probability {\em density} (that is, $p(x)dx$ is the probability for $x$ to lie in the interval $[x,x+dx]$).

For example, suppose that Alice prepares an arbitrary pure state $|\varphi\rangle$ chosen from the uniform ensemble in a $d$-dimensional Hilbert space, and Bob performs an orthogonal measurement projecting onto the basis $\{|e_y\rangle\}$, hoping to learn something about what Alice prepared. Then Bob obtains outcome $y$ with probability
\begin{align}
p(y|\theta) =  |\langle e_y|\varphi\rangle|^2\equiv \cos^2\theta
\end{align}
where $\theta$ is the angle between $|\varphi\rangle$ and $|e_y\rangle$. Because Alice's ensemble is uniform, Bob's outcomes are also uniformly distributed; hence $H(Y) = \log d$. Furthermore, the measurement outcome $y$ reveals only information about $\theta$; Bob learns nothing else about $|\varphi\rangle$. Therefore, eq.(\ref{eq:hw-conditional-continuous}) implies that the information gain may be expressed as 
\begin{align}\label{I_from_theta}
I(X;Y) = \log d - d\int d\theta \cdot p(\theta) \cdot \cos^2\theta \cdot \log \cos^2 \theta.
\end{align}
Here $p(\theta)d\theta$ is the probability density for the vector $|\varphi\rangle$ to point in a direction making angle $\theta$ with the axis $|e_y\rangle$, where $0\le \theta \le \pi/2$.

\begin{description}
\item[$a$)] Show that
\begin{eqnarray}
p(\theta)\cdot d\theta=-(d-1)\left[1-\cos^2\theta\right]^{d-2} \cdot d\cos^2\theta.
\end{eqnarray}
{\bf Hint}: Choose a basis in which the fixed axis $|e_y\rangle$ is
\begin{equation}
|e_y\rangle= (1,\vec 0)
\end{equation}
and write
\begin{equation}
|\varphi\rangle=(e^{i\phi}\cos\theta,\psi^\perp),
\end{equation}
where $\theta \in [0,\pi/2]$, and $|\psi^\perp\rangle$ denotes a complex $(d{-}1)$-component vector with length $\sin\theta$. Now note that the phase $\phi$ resides on a circle of radius $\cos\theta$ (and hence circumference $2\pi\cos\theta$), while $|\psi^\perp\rangle$ lies on a sphere of radius $\sin\theta$ (thus the volume of the sphere, up to a multiplicative numerical constant, is $\sin^{2d-3}\theta$).

\item[$b$)] Now evaluate the integral eq.~(\ref{I_from_theta}) to show that the information gain from the measurement, in {\em nats}, is 
\begin{equation}
I(X;Y)=\ln d-\left({1\over 2} + {1\over 3} + \cdots +{1\over d}\right)~.
\end{equation}
(Information is expressed in nats if logarithms are natural logarithms; $I$ in nats is related to $I$ in bits by $I_{\rm bits}= I_{\rm nats}/\ln 2$.) {\bf Hint}: To evaluate the integral
\begin{equation}
\int_0^1dx\, (1-x)^px\ln x~,
\end{equation}
observe that 
\begin{equation}
x\ln x= {d\over ds}x^s\Big|_{s=1}~,
\end{equation}
and then calculate $\int_0^1dx\, (1-x)^px^{s}$ by integrating by parts repeatedly.

\item[$c$)]
Show that in the limit of large $d$, the information gain, in {\em bits}, approaches
\begin{equation}
I_{d=\infty}= {1-\gamma\over \ln 2}= .60995\dots,
\end{equation}
where $\gamma=.57721\dots$ is Euler's constant.
\end{description}

Our computed value of $H(Y|X)$ may be interpreted in another way: Suppose we fix an orthogonal measurement, choose a typical state, and perform the measurement repeatedly on that chosen state. Then the measurement outcomes will not be uniformly distributed. Instead the entropy of the outcomes will fall short of maximal by .60995 bits, in the limit of large Hilbert space dimension.

\item {\bf Fano's inequality}\label{ex:fano}
 
Suppose $X = \{x,p(x)\}$ is a probability distribution for a letter $x$ drawn from an alphabet of $d$ possible letters, and that $XY$ is the joint distribution for $x$ and another random variable $y$ which is correlated with $x$. Upon receiving $y$ we estimate the value of $x$ by evaluating a function $\hat x(y)$. We may anticipate that if our estimate is usually correct, then the conditional entropy $H(X|Y)$ must be small. In this problem we will confirm that expectation. 

Let $e\in\{0,1\}$ denote a binary random variable which takes the value $e=0$ if $x = \hat x(y)$ and takes the value $e=1$ if $x\ne \hat x(y)$, and let $XYE$ denote the joint distribution for $x, y, e$. The {\em error probability} $P_e$ is the probability that $e=1$, averaged over this distribution. Our goal is to derive an upper bound on $H(X|Y)$ depending on $P_e$. 

\begin{description}
  \item[$a$)] Show that
\begin{align}
H(X|Y) = H(X|YE) + H(E|Y) - H(E|XY).
\end{align}

\end{description}

Note that $H(E|XY) = 0$ because $e$ is determined when $x$ and $y$ are know, and that $H(E|Y) \le H(E)$ because mutual information is nonnegative. Therefore,
\begin{align}
H(X|Y) \le H(X|YE) + H(E).
\end{align}

\begin{description}
  \item[$b$)] Noting that
\begin{align}
H(X|YE) = p(e=0)H(X|Y,e=0) + p(e=1)H(X|Y,e=1),
\end{align}
and that $H(X|Y,e=0) = 0$ (because $x= \hat x(y)$ is determined by $y$ when there is no error),
show that
\begin{align}
H(X|YE)\le P_e \log_2(d-1).
\end{align}

\item[$c$)] Finally, show that
\begin{align}
H(X|Y) \le H_2(P_e) + P_e\log_2(d-1),
\end{align}
which is {\em Fano's inequality}. 

\item[$d$)] Use Fano's inequality to derive eq.(\ref{eq:capacity-error}), hence completing the proof that the classical channel capacity $C$ is an upper bound on achievable rates for communication over a noisy channel with negligible error probability.
\end{description}

\item {\bf A quantum version of Fano's inequality} \label{ex:quantum-fano}

\begin{description}
\item[$a$)] In a $d$-dimensional system, suppose a density operator $\bfrho$ approximates the pure state $|\psi\rangle$ with fidelity
\begin{align}
F = \langle \psi|\bfrho|\psi\rangle = 1 - \varepsilon.
\end{align}
Show that
\begin{align}
H(\bfrho) \le H_2(\varepsilon) + \varepsilon \log_2(d-1).
\end{align}
{\bf Hint}: Recall that if a complete orthogonal measurement performed on the state $\bfrho$ has the distribution of outcomes $X$, then $H(\bfrho)\le H(X)$, where $H(X)$ is the Shannon entropy of $X$.

\item[$b$)] As in \S\ref{subsec:decoupling}, suppose that the noisy channel $\mathcal{N}^{A\to B}$ acts on the pure state $\psi_{RA}$, and is followed by the decoding map $\mathcal{D}^{B\to C}$. Show that
\begin{equation}
H(R)_\bfrho - I_c(R~\rangle B)_\bfrho\le 2H(RC)_\bfsigma,
\end{equation}
where
\begin{align}
\bfrho_{RB} = \mathcal{N}(\psi_{RA}), \quad\bfsigma_{RC} = \mathcal{D}\circ \mathcal{N}(\psi_{RA}). 
\end{align}
Therefore, if the decoder's output (the state of $RC$) is almost pure, then the coherent information of the channel ${\cal N}$ comes close to matching its input entropy. 
{\bf Hint}: Use the data processing inequality $I_c(R~\rangle C)_\bfsigma \le I_c(R~\rangle B)_\bfrho$ and the subadditivity of von Neumann entropy. It is convenient to consider the joint pure state of the reference system, the output, and environments of the dilations of $\mathcal{N}$ and $\mathcal{D}$. 

\item[$c$)] Suppose that the decoding map recovers the channel input with high fidelity, 
\begin{align}
F(\mathcal{D}\circ \mathcal{N}(\psi_{RA}),\psi_{RC}) = 1-\varepsilon.
\end{align}
Show that 
\begin{equation}
H(R)_\bfrho - I_c(R~\rangle B)_\bfrho \le 2 H_2(\varepsilon) + 2\varepsilon \log_2(d^2-1),
\end{equation}
assuming that $R$ and $C$ are $d$-dimensional. This is a quantum version of Fano's inequality, which we may use to derive an upper bound on the quantum channel capacity of $\mathcal{N}$.
\end{description}

\item {\bf Mother protocol for the GHZ state}

The mother resource inequality expresses an asymptotic resource conversion that can be achieved if Alice, Bob, and Eve share $n$ copies of the pure state $\phi_{ABE}$: by sending $\frac{n}{2}I(A;E)$ qubits to Bob, Alice can destroy the correlations of her state with Eve's state, so that Bob alone holds the purification of Eve's state, and furthermore Alice and Bob share $\frac{n}{2}I(A;B)$ ebits of entanglement at the end of the protocol; here $I(A;E)$ and $I(A;B)$ denote quantum mutual informations evaluated in the state $\phi_{ABE}$. 

Normally, the resource conversion can be realized with arbitrarily good fidelity only in the limit $n\to\infty$. But in this problem we will see that the conversion can be perfect if Alice, Bob and Eve share only $n=2$ copies of the three-qubit GHZ state
\begin{equation}
|\phi\rangle_{ABE} = \frac{1}{\sqrt{2}}\left(|000\rangle + |111\rangle\right)~.
\end{equation}
The protocol achieving this perfect conversion uses the notion of {\em coherent classical communication} defined in Chapter 4. 

\begin{description}

\item[$a$)] Show that in the GHZ state $|\phi\rangle_{ABE}$, $I(A;E)=I(A;B)=1$. Thus, for this state, the mother inequality becomes
\begin{equation}
\label{two-copy-mother}
2\langle \phi_{ABE} \rangle + [q\to q]_{AB} \ge [qq]_{AB} + 2\langle\phi'_{B_1E}\rangle ~.
\end{equation}

\item[$b)$] Suppose that in the GHZ state Alice measures the Pauli operator $X$, gets the outcome $+1$ and broadcasts her outcome to Bob and Eve. What state do Bob and Eve then share? What if Alice gets the outcome $-1$ instead?

\item[$c$)] Suppose that Alice, Bob, and Eve share just one copy of the GHZ state $\phi^{ABE}$. Find a protocol such that, after one unit of {\em coherent classical communication} from Alice to Bob, the shared state becomes $|\phi^+\rangle_{AB}\otimes |\phi^+\rangle_{BE}$, where $|\phi^+\rangle=\left(|00\rangle+|11\rangle\right)/\sqrt{2}$ is a maximally entangled Bell pair.

\item[$d$)] Now suppose that Alice, Bob, and Eve start out with two copies of the GHZ state, and suppose that Alice and Bob can borrow an ebit of entanglement, which will be repaid later, to catalyze the resource conversion.  Use coherent superdense coding to construct a protocol that achieves the (catalytic) conversion eq.~(\ref{two-copy-mother}) {\em perfectly}.

\end{description}

\item {\bf Degradability of amplitude damping and erasure} \label{ex:degradable}

 The qubit amplitude damping channel $\mathcal{N}_{\rm a.d.}^{A\to B}(p)$ discussed in \S 3.4.3 has the dilation $\bfU^{A\to BE}$ such that
\begin{align}
\bfU: &|0\rangle_A \mapsto |0\rangle_B \otimes |0\rangle_E,\notag\\
&|1\rangle_A \mapsto \sqrt{1-p}~|1\rangle_B \otimes |0\rangle_E + \sqrt{p}~|0\rangle_B \otimes |1\rangle_E;\notag
\end{align}
a qubit in its ``ground state'' $|0\rangle_A$ is unaffected by the channel, while a qubit in the ``excited state'' $|1\rangle_A$ decays to the ground state with probability $p$, and the decay process excites the environment. Note that $\bfU$ is invariant under interchange of systems $B$ and $E$ accompanied by transformation $p \leftrightarrow (1-p)$. Thus the channel complementary to $\mathcal{N}_{\rm a.d.}^{A\to B}(p)$ is $\mathcal{N}_{\rm a.d.}^{A\to E}(1-p)$. 
\begin{description}
\item[$a$)] Show that $\mathcal{N}_{\rm a.d.}^{A\to B}(p)$ is degradable for $p\le 1/2$. Therefore, the quantum capacity of the amplitude damping channel is its optimized one-shot coherent information. {\bf Hint}: It suffices to show that
\begin{align}
\mathcal{N}_{\rm a.d.}^{A\to E}(1-p) = \mathcal{N}_{\rm a.d.}^{B\to E}(q) \circ \mathcal{N}_{\rm a.d.}^{A\to B}(p),
\end{align}
where $0\le q\le 1$. 
\end{description}

The {\em erasure channel} $\mathcal{N}_{\rm erase}^{A\to B}(p)$ has the dilation $\bfU^{A\to BE}$ such that
\begin{align}
\bfU: |\psi\rangle_A \mapsto \sqrt{1-p}~|\psi\rangle_B \otimes |e\rangle_E + \sqrt{p}~|e\rangle_B \otimes |\psi\rangle_E;
\end{align}
Alice's system passes either to Bob (with probability $1-p$) or to Eve (with probability $p$), while the other party receives the ``erasure symbol'' $|e\rangle$, which is orthogonal to Alice's Hilbert space. Because $\bfU$ is invariant under interchange of systems $B$ and $E$ accompanied by the transformation $p \leftrightarrow (1-p)$, the channel complementary to $\mathcal{N}_{\rm erase}^{A\to B}(p)$ is $\mathcal{N}_{\rm erase}^{A\to E}(1-p)$. 
\begin{description}
\item[$b$)] Show that $\mathcal{N}_{\rm erase}^{A\to B}(p)$ is degradable for $p\le 1/2$. Therefore, the quantum capacity of the erasure channel is its optimized one-shot coherent information. {\bf Hint}: It suffices to show that
\begin{align}
\mathcal{N}_{\rm erase}^{A\to E}(1-p) = \mathcal{N}_{\rm erase}^{B\to E}(q) \circ \mathcal{N}_{\rm erase}^{A\to B}(p),
\end{align}
where $0\le q\le 1$. 

\item[$c$)] Show that for $p \le 1/2$ the quantum capacity of the erasure channel is 
\begin{align}
Q(\mathcal{N}_{\rm erase}^{A\to B}(p)) = (1 - 2p) \log_2 d,
\end{align}
where $A$ is $d$-dimensional, 
and that the capacity vanishes for $1/2 \le p \le 1$.
\end{description}

\item {\bf Quantum Singleton bound}

As noted in chapter 7, an $[[n,k,d]]$ quantum error-correcting code ($k$ protected qubits in a block of $n$ qubits, with code distance $d$) must obey the constraint
\begin{align}
n - k \ge 2 (d-1),
\end{align}
the {\em quantum Singleton bound}. This bound is actually a corollary of a stronger statement which you will prove in this exercise. 

Suppose that in the pure state $\phi_{RA}$ the reference system $R$ is maximally entangled with a code subspace of $A$, and that $E_1$ and $E_2$ are two disjoint correctable subsystems of system $A$ (erasure of either $E_1$ or $E_2$ can be corrected). You are to show that
\begin{align}\label{eq:quantum-singleton-general}
\log |A| - \log |R| \ge \log |E_1| + \log |E_2|.
\end{align}
Let $E^c$ denote the subsystem of $A$ complementary to $E_1E_2$, so that $A= E^c E_1 E_2$. 

\begin{description}
\item[$a$)] Recalling the error correction conditions $\bfrho_{RE_1} = \bfrho_R\otimes \bfrho_{E_1}$ and $\bfrho_{RE_2} = \bfrho_R\otimes \bfrho_{E_2}$, show that $\phi_{RE^cE_1E_2}$ has the property
\begin{align}\label{eq:quantum-singleton-almost}
H(R) =H(E^c) -\frac{1}{2} I(E^c; E_1) -\frac{1}{2} I(E^c;E_2).
\end{align}
\item[$b$)] Show that eq.(\ref{eq:quantum-singleton-almost}) implies eq.(\ref{eq:quantum-singleton-general}). 
\end{description}

\item {\bf Noisy superdense coding and teleportation}. 

\begin{description}
\item[$a$)] By converting the entanglement achieved by the mother protocol into classical communication, prove the noisy superdense coding resource inequality:
\begin{align}
Noisy~SD: \quad \langle \phi_{ABE} \rangle + H(A)[q\to q] \ge I(A;B) [c\to c].
\end{align}
Verify that this matches the standard noiseless superdense coding resource inequality when $\phi$ is a maximally entangled state of $AB$.

\item[$b$)] By converting the entanglement achieved by the mother protocol into quantum communication, prove the noisy teleportation resource inequality:
\begin{align}
Noisy~TP: \quad \langle \phi_{ABE} \rangle + I(A;B)[c\to c] \ge I_c(A\rangle B) [q\to q].
\end{align}
Verify that this matches the standard noiseless teleportation resource inequality when $\phi$ is a maximally entangled state of $AB$.

\end{description}

\item  {\bf The cost of erasure} \label{ex:erasure}

{\em Erasure} of a bit is a process in which the state of the bit is reset to 0. Erasure is {\em irreversible} --- knowing only the final state 0 after erasure, we cannot determine whether the initial state before erasure was 0 or 1. 
This irreversibility implies that erasure incurs an unavoidable thermodynamic cost. According to {\em Landauer's Principle}, erasing a bit at temperature $T$ requires work $W \ge kT \log 2$.  In this problem you will verify that a particular procedure for achieving erasure adheres to Landauer's Principle. 

Suppose that the two states of the bit both have zero energy. We erase the bit in two steps. In the first step, we bring the bit into contact with a reservoir at temperature $T> 0$, and wait for the bit to come to thermal equilibrium with the reservoir. In this step the bit ``forgets'' its initial value, but the bit is not yet erased because it has not been reset. 

We reset the bit in the second step, by slowly turning on a control field $\lambda$ which splits the degeneracy of the two states. For $\lambda \ge 0$, the state 0 has energy $E_0 = 0$ and the state 1 has energy $E_1 = \lambda$. After the bit thermalizes in step one, the value of $\lambda$ increases gradually from the initial value $\lambda=0$ to the final value $\lambda = \infty$; the increase in $\lambda$ is slow enough that the qubit remains in thermal equilibrium with the reservoir at all times. As $\lambda$ increases, the probability $P(0)$ that the qubit is in the state 0 approaches unity --- {\em i.e.}, the bit is reset to the state 0, which has zero energy.
\begin{description}
\item{$(a)$} For $\lambda \ne 0$, find the probability $P(0)$ that the qubit is in the state 0 and the probability $P(1)$ that the qubit is in the state 1.

\item{$(b)$} How much work is required to increase the control field from $\lambda$ to $\lambda + d\lambda$?

\item{$(c)$} How much work is expended as $\lambda$ increases slowly from $\lambda=0$ to $\lambda = \infty$? (You will have to evaluate an integral, which can be done analytically.) 
\end{description}

\item {\bf Proof of the decoupling inequality} \label{ex:decoupling}

In this problem we complete the derivation of the decoupling inequality sketched in \S \ref{subsec:decoupling-proof}. 

\begin{description}
\item[$a$)] Verify eq.(\ref{eq:swap-in-trace-trick}).

\end{description}
To derive the expression for $\mathbb{E}_{\bfU}\left[ \bfM_{AA'}(\bfU)\right]$ in eq.(\ref{eq:exp-value-2-2-poly}), we first note that the invariance property eq.(\ref{eq:expectation-unitary-invariance}) implies that $\mathbb{E}_{\bfU}\left[ \bfM_{AA'}(\bfU)\right]$ commutes with $\bfV\otimes \bfV$ for any unitary $\bfV$. Therefore, by Schur's lemma, $\mathbb{E}_{\bfU}\left[ \bfM_{AA'}(\bfU)\right]$ is a weighted sum of projections onto irreducible representations of the unitary group. The tensor product of two fundamental representations of $\bfU(d)$ contains two irreducible representations --- the symmetric and antisymmetric tensor representations. Therefore we may write
\begin{align}
\mathbb{E}_{\bfU}\left[ \bfM_{AA'}(\bfU)\right]= c_{\rm sym}~ \bfpi^{(\rm sym)}_{AA'} +c_{\rm anti}~\bfpi^{(\rm anti)}_{AA'};
\label{eq:integral-symmetric-antisymmetric}
\end{align}
here $\bfpi^{(\rm sym)}_{AA'}$ is the orthogonal projector onto the subspace of $AA'$ symmetric under the interchange of $A$ and $A'$, $\bfpi^{(\rm anti)}_{AA'}$ is the projector onto the antisymmetric subspace, and   $c_{\rm sym}$, $c_{\rm anti}$ are suitable constants. Note that
\begin{align}
& \bfpi^{(\rm sym)}_{AA'}=  \frac{1}{2} \left(\bfI_{AA'} + \bfS_{AA'}\right),\notag\\
& \bfpi^{(\rm anti)}_{AA'}= \frac{1}{2} \left(\bfI_{AA'} - \bfS_{AA'}\right), 
\end{align}
where $\bfS_{AA'}$ is the swap operator, and that the symmetric and antisymmetric subspaces have dimension $\frac{1}{2}|A|\left(|A|+1\right)$ and dimension $\frac{1}{2}|A|\left(|A|-1\right)$ respectively. 

Even if you are not familiar with group representation theory, you might regard eq.(\ref{eq:integral-symmetric-antisymmetric}) as obvious. We may write $\bfM_{AA'}(\bfU)$ as a sum of two terms, one symmetric and the other antisymmetric under the interchange of $A$ and $A'$. The expectation of the symmetric part must be symmetric, and the expectation value of the antisymmetric part must be antisymmetric. Furthermore, averaging over the unitary group ensures that no symmetric state is preferred over any other. 

\begin{description}

\item[$b$)] To evaluate the constant $c_{\rm sym}$, multiply both sides of eq.(\ref{eq:integral-symmetric-antisymmetric}) by $\bfpi^{(\rm sym)}_{AA'}$ and take the trace of both sides, thus finding
\begin{align}
c_{\rm sym} = \frac{|A_1| + |A_2|}{|A|+1}.
\end{align}

\item[$c$)] To evaluate the constant $c_{\rm anti}$, multiply both sides of eq.(\ref{eq:integral-symmetric-antisymmetric})) by $\bfpi^{(\rm anti)}_{AA'}$ and take the trace of both sides, thus finding
\begin{align}
c_{\rm anti} = \frac{|A_1| - |A_2|}{|A|-1}.
\end{align}

\item[$d$)] Using
\begin{align}
c_\bfI = \frac{1}{2}\left(c_{\rm sym} + c_{\rm anti}\right), \quad c_\bfS = \frac{1}{2}\left(c_{\rm sym}- c_{\rm anti}\right)
\end{align}
prove eq.(\ref{eq:c_bfI-c_bfS}). 
\end{description}

\end{exercises}


\begin{thebibliography}{99}

\bibitem{wilde} M.~M. Wilde, {\em Quantum Information Theory} (Cambridge, 2013). 
\bibitem{cover-thomas} T.~M. Cover and J.~A. Thomas, {\em Information Theory} (Wiley, 1991).
\bibitem{shannon} C.~E Shannon and W. Weaver, {\em The Mathematical Theory of Communication} (Illinois, 1949).
\bibitem{nielsen-chuang} M.~A. Nielsen and I.~L. Chuang, {\em Quantum Computation and Quantum Information} (Cambridge, 2000). 
\bibitem{wehrl} A. Wehrl, General properties of entropy, Rev. Mod. Phys. 50, 221 (1978).
\bibitem{lieb-ruskai} E.~H. Lieb and M.~B. Ruskai, A fundamental property of quantum-mechanical entropy, Phys. Rev. Lett. 30, 434 (1973).
\bibitem{hayden-ssa} P. Hayden, R. Jozsa, D. Petz, and A. Winter, Structure of states which satisfy strong subadditivity with equality, Comm. Math. Phys. 246, 359-374 (2003). 
\bibitem{nielsen-kempe} M.~A. Nielsen and J. Kempe, Separable states are more disordered globally than locally, Phys. Rev. Lett. 86, 5184 (2001). 
\bibitem{bekenstein} J. Bekenstein, Universal upper bound on the entropy-to-energy ration of bounded systems, Phys. Rev. D 23, 287 (1981).
\bibitem{casini} H. Casini, Relative entropy and the Bekenstein bound, Class. Quant. Grav. 25, 205021 (2008).
\bibitem{berta-uncertainty} P.~J. Coles, M. Berta, M. Tomamichel, S. Wehner, Entropic uncertainty relations and their applications, arXiv:1511.04857 (2015).
\bibitem{maassen} H. Maassen and J. Uffink, Phys. Rev. Lett. 60, 1103 (1988).
\bibitem{schumacher-compression} B. Schumacher, Quantum coding, Phys. Rev. A 51, 2738 (1995).
\bibitem{schumacher-jozsa} R. Jozsa and B. Schumacher, A new proof of the quantum noiseless coding theory, J. Mod. Optics 41, 2343-2349 (1994). 
\bibitem{bennett-concentration} C.~H. Bennett, H.~J. Bernstein, S. Popescu, and B. Schumacher, Concentrating partial entanglement by local operations, Phys. Rev. A 53, 2046 (1996).
\bibitem{horodecki-review} R. Horodecki, P. Horodecki, M. Horodecki, and K. Horodecki, Quantum entanglement, Rev. Mod. Phys. 81, 865 (2009).
\bibitem{brandao-locc} F.~G.~S.~L. Brand\~ao and M.~B. Plenio, A reversible theory of entanglement and its relation to the second law, Comm. Math. Phys. 295, 829-851 (2010).
\bibitem{hayashi-stein} M. Hayashi and H. Yamasaki, Generalized quantum Stein's lemma and second law of quantum resource theories, arXiv:2408.02722.
\bibitem{lami-stein} L. Lami, A solution of the generalised quantum Stein's lemma, arXiv:2408.06410.
\bibitem{christandl-winter-squashed} M. Christandl and A. Winter, ``Squashed entanglement'': an additive entanglement measure, J. Math. Phys. 45, 829 (2004).
\bibitem{koashi-winter-squashed} M. Koashi and A. Winter, Monogamy of quantum entanglement and other correlations, Phys. Rev. A 69, 022309 (2004).
\bibitem{brandao-squashed} F.~G.~S.~L. Brand\~ao, M. Cristandl, and J. Yard, Faithful squashed entanglement, Comm. Math. Phys. 306, 805-830 (2011).
\bibitem{doherty} A.~C. Doherty, P.~A. Parrilo, and F.~M. Spedalieri, Complete family of separability criteria, Phys. Rev. A 69, 022308 (2004).
\bibitem{holevo-bound} A.~S. Holevo, Bounds for the quantity of information transmitted by a quantum communication channel, Probl. Peredachi Inf. 9, 3-11 (1973).
\bibitem{peres-wootters} A. Peres and W.~K. Wootters, Optimal detection of quantum information, Phys. Rev. Lett 66, 1119 (1991).
\bibitem{holevo-capacity} A.~S. Holevo, The capacity of the quantum channel with general signal states, arXiv: quant-ph/9611023.
\bibitem{schumacher-westmoreland-capacity} B. Schumacher and M.~D. Westmoreland, Sending classical information via noisy quantum channels, Phys. Rev. A 56, 131-138 (1997).
\bibitem{hastings-superadditive} M.~B. Hastings, Superadditivity of communication capacity using entangled inputs, Nature Physics 5, 255-257 (2009).
\bibitem{horodecki-entanglement-breaking} M. Horodecki, P.~W. Shor, and M.~B. Ruskai, Entanglement breaking channels, Rev. Math. Phys. 15, 629-641 (2003).
\bibitem{shor-entanglement-breaking} P.~W. Shor, Additivity of the classical capacity for entanglement-breaking quantum channels, J. Math. Phys. 43, 4334 (2002).
\bibitem{schumacher-nielsen} B. Schumacher and M.~A. Nielsen, Quantum data processing and error correction, Phys. Rev. A 54, 2629 (1996).
\bibitem{schumacher-coherent} B. Schumacher, Sending entanglement through noisy quantum channels, Phys. Rev. A 54, 2614 (1996).
\bibitem{barnum-coherent} H. Barnum, E. Knill, and M.~A. Nielsen, On quantum fidelities and channel capacities, IEEE Trans. Inf. Theory 46, 1317-1329 (2000). 
\bibitem{lloyd-coherent} S. Lloyd, Capacity of the noisy quantum channel, Phys. Rev. A 55, 1613 (1997).
\bibitem{shor-quantum-capacity} P.~W. Shor, unpublished (2002).
\bibitem{devetak-quantum-capacity} I. Devetak, The private classical capacity and quantum capacity of a quantum channel, IEEE Trans. Inf. Theory 51, 44-55 (2005).
\bibitem{devetak-winter-hashing} I. Devetak and A. Winter, Distillation of secret key and entanglement from quantum states, Proc. Roy. Soc. A 461, 207-235 (2005).
\bibitem{schumacher-westmoreland-approximate} B. Schumacher and M.~D. Westmoreland, Approximate quantum error correction, Quant. Inf. Proc. 1, 5-12 (2002).
\bibitem{horodecki-merging} M. Horodecki, J. Oppenheim, and A. Winter, Quantum state merging and negative information, Comm. Math. Phys. 269, 107-136 (2007).
\bibitem{hayden-decoupling} P. Hayden, M. Horodecki, A. Winter, and J. Yard, Open Syst. Inf. Dyn. 15, 7-19 (2008).
\bibitem{abeyesinghe-decoupling} A. Abeyesinge, I. Devetak, P. Hayden, and A. Winter, Proc. Roy. Soc. A, 2537-2563 (2009). 
\bibitem{lubkin} E. Lubkin, Entropy of an $n$-system from its correlation with a $k$-reservoir, J. Math. Phys. 19, 1028 (1978).
\bibitem{lloyd-entropy} S. Lloyd and H. Pagels, Complexity as thermodynamic depth, Ann. Phys. 188, 186-213 (1988)
\bibitem{page} D.~N. Page, Average entropy of a subsystem, Phys. Rev. Lett. 71, 1291 (1993).
\bibitem{devetak-mother-father} I. Devetak, A.~W. Harrow, and A. Winter, A family of quantum protocols, Phys. Rev. Lett. 93, 230504 (2004).
\bibitem{devetak-mother-father-long} I. Devetak, A.~W. Harrow, and A. Winter, A resource framework for quantum Shannon theory, IEEE Trans. Inf. Theory 54, 4587-4618 (2008).
\bibitem{devetak-shor-degradable} I. Devetak and P.~W. Shor, The capacity of a quantum channel for simultaneous transmission of classical and quantum information, Comm. Math. Phys. 256, 287-303 (2005).
\bibitem{bennett-entanglement-assisted} C.~H. Bennett, P.~W. Shor, J.~A. Smolin, and A.~V. Thapliyal, Entanglement-assisted classical capacity of noisy quantum channels, Phys. Rev. Lett. 83, 3081 (1999). 
\bibitem{bennett-entanglement-assisted-more} C.~H. Bennett, P.~W. Shor, J.~A. Smolin, and A.~V. Thapliyal, Entanglement-assisted classical capacity of a quantum channel and the reverse Shannon theorem, IEEE Trans. Inf. Theory 48, 2637-2655 (2002).
\bibitem{shor-smolin} P.~W. Shor and J.~A. Smolin, Quantum error-correcting codes need not completely reveal the error syndrome, arXiv:quant-ph/9604006.
\bibitem{divincenzo-superadditive} D.~P. DiVincenzo, P.~W. Shor, and J.~A. Smolin, Quantum channel capacity of very noisy channels, Phys. Rev. A 57, 830 (1998).
\bibitem{smith-yard} G. Smith and J. Yard, Quantum communication with zero-capacity channels, Science 321, 1812-1815 (2008).
\bibitem{renner-landauer} L. del Rio, J. Aberg, R. Renner, O. Dahlsten, and V. Vedral, The thermodynamic meaning of negative entropy, Nature 474, 61-63 (2011).
\bibitem{hayden-preskill} P. Hayden and J. Preskill, Black holes as mirrors: quantum information in random subsystems, JHEP 09, 120 (2007).
\bibitem{sekino-susskind} Y. Sekino and L. Susskind, Fast scramblers, JHEP 10, 065 (2008).

\end{thebibliography}
\end{document}